\documentclass[fleqn,usenatbib,twocolumn]{mnras}

\usepackage{newtxtext,newtxmath,hyperref}

\usepackage[T1]{fontenc}
\usepackage{algorithm2e}
\usepackage{longtable,verbatim}

\DeclareRobustCommand{\VAN}[3]{#2}
\let\VANthebibliography\thebibliography
\def\thebibliography{\DeclareRobustCommand{\VAN}[3]{##3}\VANthebibliography}

\usepackage{graphicx}	
\usepackage{amsmath}	
\usepackage{amssymb}	
\usepackage[theme=default-plain, usecourier=false]{jlcode}

\title[Derivatives of TTVs]{A differentiable N-body code for transit timing and dynamical modeling. I. Algorithm and derivatives.}

\author[E. Agol et al.]{
Eric Agol,$^{1,2,3}$\thanks{E-mail: agol uw.edu (EA)}
David M. Hernandez,$^{4,5}$
\& Zachary Langford.$^{1}$
\\
$^{1}${Astronomy Department, University of Washington, Seattle, WA 98195, USA}\\
$^{2}${Institut d'Astrophysique de Paris, 98bis Boulevard Arago, Paris 75014, France}\\
$^{3}${Guggenheim Fellow} \\
$^{4}${Harvard--Smithsonian Center for Astrophysics, 60 Garden St., MS 51, Cambridge, MA 02138, USA} \\
$^{5}${Physics and Kavli Institute for Astrophysics and Space Research, Massachusetts Institute of Technology,  77 Massachusetts Ave., Cambridge, MA 02139, USA}
}

\date{Accepted XXX. Received YYY; in original form ZZZ}

\pubyear{2020}

\begin{document}
\label{firstpage}
\pagerange{\pageref{firstpage}--\pageref{lastpage}}

\maketitle

\begin{abstract}
When fitting $N$-body models to astronomical data -- including transit times, radial velocity, and astrometric positions at
observed times -- the derivatives of the model
outputs with respect to the initial conditions can help with model optimization and posterior sampling.  Here we describe a general-purpose symplectic integrator for
arbitrary orbital architectures{, including those with close encounters,} which we have recast to maintain numerical stability
and precision for small step sizes. We compute the derivatives of the $N$-body 
coordinates and velocities as a function of time with respect to the initial conditions
and masses by propagating the Jacobian along with the $N$-body integration.   For the 
first time we obtain the derivatives of the transit times with respect to the initial
conditions and masses using the chain rule, which is quicker and more accurate than using finite differences or automatic differentiation.   We implement this algorithm in an open source package, \texttt{NbodyGradient.jl}, written in the Julia language,  which has been used in the optimization and error analysis of
transit-timing variations in the TRAPPIST-1 system.
We present tests of the
accuracy and precision of the code, and show that it compares favorably in speed to other
integrators which are written in C.
\end{abstract}




\section{Introduction}
The gravitational $N$-body problem refers to the integration of the positions
and velocities of a set of $N$ point-particles forward or backward in time using Newton's
equations, after specifying their masses and initial phase-space coordinates.  
The solution of the $N$-body problem can be put to many uses, for example,
matching observational data on a set of astronomical bodies,  estimating the
long-term stability or sensitivity to initial conditions of a model system, or
determining the outcome of interactions between bodies.  

For each of these
applications, it {can be} beneficial to be able to compute the derivatives of the
state of the system at a given time with respect to the initial conditions
and masses. {As the N-body problem is highly non-linear, non-linear optimizers are needed to find the model parameters with the maximum a posteriori probability (MAP), or maximum likelihood estimate. 
Derivative-free optimization, such as Nelder-Mead, can be slow to converge, and typically becomes less efficient as the number of dimensions grows.  Hence, derivatives can significantly speed up the process of optimization.   Once the MAP is found, the Hessian can be computed with derivatives to estimate the uncertainties on parameters.  Then, each parameter can be varied along a fixed grid, and the non-linear optimization can be re-run to trace out the likelihood profile.  Finally, the posterior can be sampled using Bayesian techniques which take advantage of derivatives to improve the efficiency of the sampling in high dimensions.}

{Finite-difference derivatives can be easy to compute numerically; however, finite-differences are computationally expensive and limited by numerical precision.  The computation of derivatives along with N-body integration can yield higher precision with less computation time, enabling more effective application of non-linear optimization and parameter uncertainty estimation.} This calculation can be laborious, 
involving propagating derivatives
through each time step of an integration, but the result can be much
more computationally efficient and accurate relative to computing derivatives
with finite differences.  

The calculation of derivatives of the $N$-body problem has been investigated
in prior work.  \citet{Mikkola1999} and \citet{Rein2015} derive the variational
equations of the symplectic integrator of \cite{Wisdom1991} {, the ``WH" method,} to obtain the tangent map 
of an $N$-body system as a function of time, from which the positional variations
may be derived as a function of variations in the initial phase-space coordinates.  {This is implemented in the {\sc WHFast} integrator in the {\sc REBOUND} package based on the Wisdom-Holman method \citep{Rein2015}.}  
Second-order variational equations were derived by \citet{Rein2016} for a high-order
integrator {({\sc ias15}) which is} assumed to exactly solve the $N$-body equations.  \citet{Pal2010} used 
a Lie-integration scheme, including derivatives with respect to the initial
orbital elements and masses, to fit for planet-planet
perturbations in radial-velocity detected systems.

\subsection{Algorithm}

The purpose of this paper is to implement first-order derivatives
in a symplectic integrator, including the mass derivatives, and allowing
for a system hierarchy which is more general than standard symplectic integrators, and which includes derivatives of the transit times with respect to the initial conditions, which is currently absent in the literature.  Instead of using the variational equations,
which assume exact solution of the $N$-body problem for obtaining the derivatives,
we compute the derivatives of the $N$-body symplectic map, with the goal of yielding a more precise
result for the Jacobian of the state of the system at a given time with respect
to the specified initial conditions.
Although the \citet{Rein2016} integrator could have been put to use for this
problem, we are interested in developing a complementary code which trades
generality and precision for potentially more speed.  

The basic integrator we use has been described in two prior papers:
\citet{Hernandez2015} and \citet{Dehnen2017}.  The novel aspect underlying the integrator is to allow all bodies to be treated on
equal footing.  A universal Kepler solver \citep{Wisdom2015} is used
to integrate pairs of bodies with Keplerian drifts forwards in time, interspersed with 
constant-velocity corrections which are negative in time, while using operator 
splitting to create a symplectic and time-symmetric integrator 
out of the original concept proposed by \citet{GoncalvesFerrari2014}.
A potential advantage of this approach is the adaptability to different problems
with various geometries, such as hierarchical triples, pairs of binaries,
or other more complex hierarchies \citep{Hamers2016}.  The popular 
Wisdom--Holman method, and its variants, which use different coordinates 
and Hamiltonian splittings \citep{Hernandez2017} assume that there 
is a dominant mass and widely separated planets.  For general applications, these 
assumptions {can be} too constraining.

A drawback of this integrator is the potential for numerical cancellation errors
to accumulate due to alternating negative and positive time steps which
are a necessary part of the algorithm.  In developing this code, we found that
these cancellations caused numerical errors which accrue in proportion to the number of time steps.  This becomes more significant when the
time steps are short, as more steps are required for a given integration time.  We have rectified this problem by combining the
negative and positive time steps into a single step, and cancelling
the terms analytically, which we find reduces the numerical
errors significantly.  Thus, another goal of this paper is to describe this improved integrator.

\subsection{Motivation: TTVs and photodynamics}
The particular application we have in mind is the detection and 
characterization of exoplanet systems.
Planetary interactions become important when data are of high precision, or
if integrations are carried out on long timescales to study system
stability.  The first example of non-Keplerian
interactions being important was the pulsar exoplanet system PSR 1257+12 
\citep{Wolszczan1992}.  As had been predicted, the interactions of the planets were detected
in the pulsar timing, and then used to confirm the
planetary nature of the system, as well as measure the inclinations
and masses of the planets by breaking the mass-inclination degeneracy which accompanies Doppler shifts \citep{Rasio1992,Malhotra1992,Peale1993,
Wolszczan1994}.  
Second, high-precision radial-velocity measurements of
exoplanet systems also require accounting for planet-planet interactions.
An early example of this is GJ 876, which required an $N$-body integration
to match the observed stellar radial velocity instead of treating
the radial velocity signal as a sum of unperturbed Keplerians 
\citep{Laughlin2001}.  Third, the {\em Kepler} spacecraft
yielded sufficient precision of the times of transit of exoplanets
to produce a novel means of detecting and characterizing exoplanets:
transit-timing variations \citep[TTVs;][]{Holman2005,Agol2005}.
The Kepler-9 planet system showed strong anti-correlated variations
in the times of transit relative to a fixed ephemeris, which allows
for measurement of the planet masses \citep{Holman2010,Freudenthal2018,Borsato2019}.
Currently, several planets have been detected with TTVs, while hundreds
have been characterized \citep[see][and references therein]{Agol2017,JontofHutter2019}.

Transit-timing variations are entirely due to non-Keplerian motion of the
planetary orbits.  In the Newtonian two-body problem, transits occur at
regular intervals, and so the transit times are uniformly spaced in
time with the orbital period of the system.  When three or more bodies
interact, each pair of bodies no longer follows a Keplerian orbit,
but is perturbed by the other bodies in the system.  In the planetary
case, the perturbations of the times of transit by other planets
are typically small compare with the orbital period of the planet.
TTVs are defined as the residuals of a linear fit to the times of
transit \citep{Agol2005}, and so by definition TTVs are imparted by 
non-Keplerian motion. Consequently, the presence of TTVs typically requires an $N$-body 
model for the computation of the times of transit.

The advent of the detection of TTVs spurred theoretical models for
short-term planetary dynamics. Analytic prescriptions exist for transit-timing
variations \citep[e.g.][]{Agol2005,Nesvorn2010,Lithwick2012,Nesvorn2014,
Deck2015,Agol2016,Deck2016,Nesvorn2016,Hadden2016}.  
However, the dynamics of multi-planet
systems is sufficiently complex that any analytic prescription is only 
accurate in a confined region of parameter space and/or limited timescales, and generally needs
to be checked against numerical integration since it is unknown beforehand
whether these restrictions apply to the masses and orbital
elements of a particular system \citep[e.g.][]{Deck2015,JontofHutter2016,
Hadden2017,Linial2018,Yoffe2021}.  On the other hand, numerical integration 
can be much more computationally expensive and can accrue numerical errors.

When optimizing a TTV model, the gradient of the likelihood is often 
required to find the direction in which the variation of the initial conditions
will improve the likelihood.  The likelihood gradient in turn
requires the gradient of each transit time with respect to the initial
conditions.  Often finite-differences are used to estimate this gradient;
however, finite difference derivatives are limited by the numerical
accuracy of the integration; see \citet{Rein2016} regarding the drawbacks of finite-difference derivatives.  This can cause
difficulty in optimizing numerical TTV model fits.   In addition, finite-differences are expensive to compute as at least 
two integrations are required for each parameter, and truncation or round-off errors can limit the precision.  For planetary systems, this requires $14N$ integrations for $N$ planets. 

Computing the posterior
parameter distributions for observed systems requires numerous evaluations
of the likelihood, which becomes difficult to explore for high-dimensional
planetary systems due to the ``curse of dimensionality" making grid-based
and Markov-chain based integrations prohibitive.  This can be ameliorated
by Hamiltonian Markov chains, which require computation of the derivatives
of the likelihood function \citep{Girolami2011}.

Dynamical interactions have also been measured in systems where
multiple stars are present.  The triple-star KOI-126 was characterized
with a ``photodynamical" model \citep{Carter2011}, which requires coupling
an $N$-body code to a photometric model.  The architecture of this system
prohibits the use of a standard Wisdom-Holman type symplectic integrator to describe the
dynamics as there is a binary star in orbit about a more massive star.  
Similarly, circumbinary planets, such as Kepler-16b, have been 
found which require a photodynamical model of the stars and planets
\citep{Doyle2011}.  One can imagine more complex geometries, such as
a planet-moon pair orbiting a binary star, which would also require
a photodynamical code to model.  The computational expense of each of these models is significant, so that obtaining converged posterior parameters is a challenge. {An advantage of photodynamical models is that the covariance between transit parameters and orbital parameters may be computed directly from the light curve.}

Given these observational modeling problems, the motivation for this code is to 
provide derivatives of a general $N$-body integrator for short-term
integrations to model stellar and planetary systems with arbitrary
hierarchy, and to compute the derivatives of the model with respect
to the initial conditions to allow for better optimization of
the model likelihood and to explore parameter space more efficiently
with Hamiltonian Markov chain Monte Carlo.

In \S \ref{sec:symplectic_integrator} we describe the original symplectic integrator algorithm, and discuss its numerical instability.  In \S \ref{sec:drift_kepler}
we give the modifications to the algorithm we have made to prevent numerical cancellation. In \S \ref{sec:differentiation_symplectic} we
introduce the derivatives of the algorithm.  In \S \ref{sec:implementation} we describe the implementation and precision of the algorithm.  In \S \ref{sec:comparison} we compare the algorithm with other $N$-body integrators.  Finally, in \S \ref{sec:conclusions} we conclude.


\section{Overview of symplectic integrator} \label{sec:symplectic_integrator}

We carry out the $N$-body integration with a symplectic integrator \citep{Channell1991} 
which uses Kepler steps to integrate pairs of bodies, interspersed with constant
velocity corrections, thus treating each and every body in an identical 
manner \citep{Hernandez2015}. The
advantage of this approach {relative to the {\sc WH} method} is that the integrator can be used as an
all-purpose integrator for studying systems with a range of architectures.  
The integrator is especially powerful when binaries at any scale are present.
A fourth-order corrector gives higher precision to this integrator without
much additional computational cost \citep{Dehnen2017}; hence, we refer
to the algorithm as DH17\footnote{{In \cite{Dehnen2017}, this algorithm 
was called `DH16'.  We have used `DH17' to reflect the publication year.}} in 
what follows.  {DH17 is mathematically written by eq. (30) in \cite{Dehnen2017}, 
using $\alpha = 0$.  In Section \ref{sec:Jacobian} and Algorithm \ref{alg:AHL21_algorithm} we 
present a generalization of the method, described mathematically by eq. (40) in 
\cite{Dehnen2017}.  This generalization is also referred to as DH17 as  
\cite{Dehnen2017} also called both methods the same name.}  The methods described here are all time-reversible and time-symmetric \citep{Hairer2006,Hernandez2018}. 
We give an overview of DH17, along with transit-time finding, in algorithm \ref{alg:DH17_algorithm},
which uses a fixed time step, $h$, from initial time $t_0$ over a duration $t_\mathrm{max}$.

Unfortunately the DH17 algorithm is numerically unstable.  
Consequently, we have modified the DH17 algorithm, and present a modified algorithm, which we will refer to as
{\sc AHL21}, in which we combine pairs of steps of the DH17 algorithm into a single
step.  The {\sc AHL21} algorithm is mathematically identical to the DH17 algorithm;
however, thanks to carrying out the cancellation analytically rather than numerically, it more numerically stable, as we describe
in \S \ref{sec:drift_kepler}.  But first we start by outlining the DH17 algorithm and its drawbacks.

\subsection{DH17 algorithm summary}

The original DH17 algorithm is given in Algorithm \ref{alg:DH17_algorithm}.
The algorithm is derived from splitting the Hamiltonian into
pairwise Keplerian terms,
\begin{eqnarray}\label{eqn:Hamiltonian_splitting}
H &=& T + V,\cr
  &=& T + \sum_i \sum_{j> i} V_{ij},\cr
  &=& T + \sum_i \sum_{j> i} \left(K_{ij}-T_{ij}\right),
\end{eqnarray}
where $T$ is the kinetic energy, $V$ is the total {potential}
energy, while $T_{ij}$, $V_{ij}$, and $K_{ij}$ are the 
kinetic, potential, and total energy of a pair of bodies $i$
and $j$.  The $K_{ij}$ term is the two-body Hamiltonian,
whose solution amounts to a Keplerian orbit {whose center of mass moves at a constant velocity}, hence the
notation ``$K$" \citep{Hernandez2015}.  Note that the
minus sign in front of the kinetic energy term indicates 
a \emph{backward} drift in time.

{A symplectic integration is achieved by successively solving each component of the Hamiltoninan, using a Kepler solver and a simple drift, to give the positions and velocities at the start of the next component \citep{Dehnen2017}.} The creation of a second-order map from this splitting of the 
Hamiltonian involves division of each time-step of duration $h$
into two sub-steps of duration $h/2$.  In each of the substeps 
the order of application of the terms is reversed
to cancel first-order error terms.  In addition, a fourth-order
velocity corrector is added in the middle of the time step,
which amounts to applying tidal accelerations to the velocities
which are neglected in the two-body elements of the Hamiltonian, 
yielding much higher precision without much additional computational
effort; this results in Algorithm \ref{alg:DH17_algorithm}.

\begin{algorithm}
  \KwData{Initial Cartesian coordinates and masses at time $t=t_0$.}
 \KwResult{Integration of $N$-body system over time $t_\mathrm{max}$, and resulting times of transit and derivatives.}
 \For{$t-t_0 < t_\mathrm{max}$}{
  Drift all particles for time $h/2$\;
  \For{all pairs of particles $(i,j)$}{
   Drift particles $i$ and $j$ for time $-h/2$\;
   Apply a Kepler solver to advance the relative coordinates of $i$ and $j$ by $h/2$\;
   Advance center of mass coordinates of $i$ and $j$ by $h/2$\;
  }
  Apply fourth-order velocity correction to all particles over time step $h$\;
  \For{reversed pairs of particles $(i,j)$}{
   Apply a Kepler solver to advance the relative coordinates of $i$ and $j$ by $h/2$\;
   Advance center of mass coordinates of $i$ and $j$ by $h/2$\;
   Drift particles $i$ and $j$ for time $-h/2$\;
  }
  Drift all particles for time $h/2$\;
  \If{transit has occurred for particles $i$ and $j$}{
  Refine transit time, and save.}
  Increment time $t$ by $h$.
 }
 \caption{Transit times with DH17 symplectic integration}
 \label{alg:DH17_algorithm}
\end{algorithm}

As mentioned, there is an unfortunate drawback to the DH17 algorithm, which is
the \emph{negative} time step.  In cases in which the potential
energy term is small, the $K_{ij}$ and $-T_{ij}$ terms nearly
cancel.  What this means is that the motion induced by these
terms in the mapping can be nearly equal and opposite, causing
numerical cancellation which leads to roundoff
errors which accumulate with time.  This has two different causes.
First, the center-of-mass portion of
these Hamiltonians is identical, and thus cancels exactly
\citep{Dehnen2017}.  Second, if
the acceleration is weak, or the time step is short enough
that the acceleration does not result in a significant change
in velocity, then the Keplerian step is nearly inertial, and 
so entire Kepler step is very nearly equal and opposite to
the negative drift.  These two sources of cancellation can
lead to numerical errors when implementing the algorithm. 
We found these errors to be
severe for long integrations, for weakly interacting bodies (e.g.
pairs of planets), or for very short time steps, which compounds
the error more rapidly.  We present a solution to this
issue in the next section, which is our first main result. 

{As a side-note, the DH17 algorithm differs in its accuracy from symplectic integrators typically used for planetary systems.  Planetary sytems with a dominant mass are described by a Hamiltonian,
\begin{eqnarray}
\label{eq:hamilt}
H = H_A + \epsilon H_B,
\end{eqnarray}
where $\epsilon \ll 1$ in which the terms $H_A$ and $H_B$ are typically chosen to be integrable.  For this Hamiltonian, the Wisdom--Holman method (WH) \citep{Wisdom1991} has been developed which carries an energy error of $\mathcal O(\epsilon h^2)$\footnote{Different conventions are used for the scalings with $\epsilon$.  In the convention of \cite{Hernandez2017}, the error scales as $\mathcal O(\epsilon^2 h^2)$.  In this convention, all scalings get an extra factor of $\epsilon$.}, where $h$ is the step size.  In contrast, the DH17 algorithm is 4th order and its error is $\mathcal O(\epsilon h^4)$.  However, while WH assumes $\epsilon \ll 1$, DH17 does not require this assumption. } 

\section{The more accurate {\sc AHL21} algorithm}\label{sec:drift_kepler}

In this paper we present a modified version of the DH17 algorithm
in which the negative drifts ($-T_{ij}$) and Keplerian steps ($K_{ij}$) 
are combined algebraically, so that leading order terms are
cancelled by hand. This exact cancellation prevents the accumulation 
of round-off and truncation errors which occur when implementing the 
DH17 algorithm.
We find that this approach gives a higher precision numerical
algorithm yielding results that obey the expected $h^4$ scaling
of the algorithm down to machine precision on the timescales we have tested.

To describe this new approach, we first need to summarize the application
of these two sub-steps.

\subsection{Kinetic-energy drift}

The drift term is the most straightforward:  each particle (for $T$)
or pair of particles (for $T_{ij}$) simply drifts inertially,
{\begin{equation}\label{eqn:drift}
    \href{https://github.com/ericagol/NbodyGradient.jl/blob/0afa7a5fd387b307e3704ded05ec01838b01478c/src/integrator/ahl21/ahl21.jl\#L403-L422}{\mathbf{x}_{i}(t+h)} = \mathbf{x}_{i}(t) + h \mathbf{v}_{i}(t),
\end{equation}}
for a time-step $h${, where $\mathbf{x}_i(t)$ and $\mathbf{v}_i(t)$ are, respectively, the position and velocity vectors of the $i$th body at time $t$.} Again, note that $-T_{ij}$ in equation (\ref{eqn:Hamiltonian_splitting}) indicates that $h$ is negative.

\subsection{Universal Kepler step}\label{sec:universal_kepler}

To carry out the $K_{ij}$ mapping, we use a universal Kepler solver to compute the change in the relative position between the bodies \citep{Wisdom2015}.
The Kepler solver uses a universal Kepler equation based upon
the initial positions and velocities of a pair of bodies at the start of
a step.  The solution of Kepler's equation enables a mapping of the
initial phase-space coordinates to the final phase-space coordinates
after a time $h$ assuming pairwise Keplerian motion (i.e.\ neglecting
every other body in the system).


The equation of motion in Cartesian coordinates derived from the $K_{ij}$ Hamiltonian for each Kepler step is given by 
\begin{eqnarray}\label{eqn:xij}
    \ddot {\bf x}_{ij} = -\frac{k{\bf x}_{ij}}{r_{ij}^3},
\end{eqnarray}
where $k = Gm_{ij}$, 
{$m_{ij} = m_i + m_j$} is the sum of the masses of the {$i$th and $j$th} pair of bodies, $\href{https://github.com/ericagol/NbodyGradient.jl/blob/0afa7a5fd387b307e3704ded05ec01838b01478c/src/integrator/ahl21/ahl21.jl\#L809-L812}{{\bf x}_{ij}} = {\bf x}_i - {\bf x}_j$,
and $r_{ij} = \vert {\bf x}_{ij}\vert$.  The universal solver
transforms the time dependence to an independent variable, $s$,
defined by $\dot s = ds/dt = r^{-1}$, where $r$ is the distance between the bodies, which simplifies the equations of motion. In the rest of this section
we drop the subscript $ij$ from the mass, coordinate and velocity vectors, i.e. $m=m_{ij}$, $r=r_{ij}$, $\mathbf{x}\equiv \mathbf{x}_{ij}$, and $\mathbf{v}\equiv \mathbf{v}_{ij}$. We will refer to the Cartesian 
coordinates for the Keplerian as $(\mathbf{x}_0,\mathbf{v}_0)$ 
at the start of a step and $(\mathbf{x},\mathbf{v})$, a time $h$ later.

A Kepler step uses the fact 
that in the two-body problem angular momentum is conserved;  thus 
the final relative positions and velocities of the two bodies, $\mathbf{x}$
and $\mathbf{v}$, are in the same 
plane as the initial relative positions and velocities, $\mathbf{x}_0$
and $\mathbf{v}_0$, while the center-of-mass velocity remains constant and the center-of-mass position drifts at a constant rate as there are no external perturbers.  This means that 
the final relative positions and velocities can be expressed as a linear 
combination of the initial relative velocities and positions,
\begin{eqnarray}
\mathbf{x} &=& f \mathbf{x}_0 + g \mathbf{v}_0\cr
\mathbf{v} &=& \dot f \mathbf{x}_0 + \dot g \mathbf{v}_0,
\end{eqnarray}
where $f$ and $g$ are Gauss's functions, which we define in
more detail below as a function of $\mathbf{x}_0$,
$\mathbf{v}_0$, $h$, and $k$, where $k=Gm$ is the central
force constant.

Then, the equations describing the initial and final states are given
by \citet{Wisdom2015}, based on \citet{Mikkola1999}.  We define
{$r_0=\left|\mathbf{x}_0\right|$ is the initial separation, $v_0 =\left|\mathbf{v}_0\right|$ is the initial relative speed, 
and $r =\left|\mathbf{x}\right|,v = \left|\mathbf{v}\right|$} are the separation and relative speed at the end of the step.
We define {two} additional quantities,
{\begin{eqnarray}
\href{https://github.com/ericagol/NbodyGradient.jl/blob/0afa7a5fd387b307e3704ded05ec01838b01478c/src/integrator/ahl21/ahl21.jl\#L905-L909}{\eta_0} &=& 
 \mathbf{x}_0 \cdot \mathbf{v}_0,
\end{eqnarray}
\begin{eqnarray}\label{eqn:beta}
\href{https://github.com/ericagol/NbodyGradient.jl/blob/0afa7a5fd387b307e3704ded05ec01838b01478c/src/integrator/ahl21/ahl21.jl\#L896}{\beta} &=& \frac{2k}{r_0}-v_0^2,\\
 &=& \frac{2k}{r}-v^2.
\end{eqnarray}}

Expressing the final positions and velocities in terms of the initial
values requires the Gauss $f$ and $g$ functions, which are given by
\begin{eqnarray}
f &=& 1-\frac{k}{r_0}G_2,\\
g &=& r_0 G_1 + \eta_0 G_2,
\end{eqnarray}
and their derivatives
\begin{eqnarray}
\dot f &=& -\frac{k}{rr_0}G_1,\\
\dot g &=& \frac{1}{r}\left(r_0G_0 + \eta_0 G_1\right),
\end{eqnarray}
where $G_i(\beta,s)$ are four functions whose definitions depend on the sign of $\beta$ for $i=0,...,3$ (Table \ref{tab:G_functions}).
With these definitions, \citet{Wisdom2015} show that
\begin{eqnarray}\label{eqn:r_of_s}
\href{https://github.com/ericagol/NbodyGradient.jl/blob/0afa7a5fd387b307e3704ded05ec01838b01478c/src/integrator/ahl21/ahl21.jl\#L986}{r(s)} = r_0 G_0 + \eta_0 G_1 + k G_2.
\end{eqnarray}
This equation may also be derived from conservation of angular
momentum,  requiring  $\mathbf{x}_0\times \mathbf{v}_0 =
\mathbf{x}\times \mathbf{v}$, which yields the condition
$f \dot g - g \dot f = 1$; this 
equation is equivalent to equation 
(\ref{eqn:r_of_s}).  {We transform these equations from $s$ to $\gamma = \vert \beta \vert^{1/2} s$, as $\gamma$ is dimensionless.}
Equation (\ref{eqn:r_of_s}) can be integrated over a time step, $h$, to give an
implicit Kepler's equation for {$\gamma$}, 
\begin{eqnarray} \label{eqn:keplers_equation}
h = r_0 G_1 + \eta_0 G_2 + k G_3,
\end{eqnarray}
which can be solved using Newton's method to find {$\gamma$} 
as a function of $h$, $r0$, $\eta_0$, $k$, and $\beta$.  
The functions $G_0, G_1, G_2$ and $G_3$ are defined in Table 
\ref{tab:G_functions} in terms of trigonometric and hyperbolic functions \citep{Wisdom2015}, 
which differ based upon whether the bodies are bound (elliptic) or unbound (hyperbolic).\footnote{This equations simplifies to a \href{https://github.com/ericagol/NbodyGradient.jl/blob/0afa7a5fd387b307e3704ded05ec01838b01478c/src/integrator/ahl21/ahl21.jl\#L913}{cubic} in the parabolic case when $\beta = 0$. {The solution of this cubic is also used as the starting guess for the Newton's solver in the elliptic and hyperbolic cases}.}  As the $G_i$ functions only depend upon {$\gamma$} 
and $\beta$, once {\href{https://github.com/ericagol/NbodyGradient.jl/blob/0afa7a5fd387b307e3704ded05ec01838b01478c/src/integrator/ahl21/ahl21.jl\#L949}{$\gamma$}} 
is found numerically, the remainder of the Kepler step simply involves algebraic computation.


\begin{table*}
\centering
\caption{Functions $G_i(s)$ used in solving Universal Kepler equation, and functions $H_i$ used later in the combined drift+Kepler step and its derivatives.}
\begin{tabular}{c|c|c|c}
variable & elliptic & parabolic & hyperbolic\\
\hline
$\beta$  & $>0$ & $=0$ & $<0$\\
$\gamma$ & $\sqrt{\beta}s$ & --- & $\sqrt{-\beta}s$\\
 \href{https://github.com/ericagol/NbodyGradient.jl/blob/0afa7a5fd387b307e3704ded05ec01838b01478c/src/integrator/ahl21/ahl21.jl\#L979}{$G_0(\beta,\gamma)$} & {$\cos{\gamma} = 1-\beta G_2$}   & 1 & {$\cosh{\gamma}=1-\beta G_2$} \\
   \href{https://github.com/ericagol/NbodyGradient.jl/blob/0afa7a5fd387b307e3704ded05ec01838b01478c/src/integrator/ahl21/ahl21.jl\#L977}{$G_1(\beta,\gamma)$}& {$\beta^{-1/2}\sin{\gamma} = 2\beta^{-1/2}\sin{\tfrac{1}{2}\gamma}\cos{\tfrac{1}{2}\gamma}$}   & $s$ & {$(-\beta)^{-1/2}\sinh{\gamma}= 2(-\beta)^{-1/2}\sinh{\tfrac{1}{2}\gamma}\cosh{\tfrac{1}{2}\gamma}$} \\
   \href{https://github.com/ericagol/NbodyGradient.jl/blob/0afa7a5fd387b307e3704ded05ec01838b01478c/src/integrator/ahl21/ahl21.jl\#L978}{$G_2(\beta,\gamma)$} & {$\beta^{-1}(1-\cos{\gamma}) = 2\beta^{-1}\sin^2{\tfrac{1}{2}\gamma}$}   & $\tfrac{1}{2}s^2$ & {$\beta^{-1}(1-\cosh{\gamma})= -2\beta^{-1}\sinh^2{\tfrac{1}{2}\gamma}$}\\
   \href{https://github.com/ericagol/NbodyGradient.jl/blob/0afa7a5fd387b307e3704ded05ec01838b01478c/src/utils.jl\#L124-L135}{$G_3(\beta,\gamma)$} & $\beta^{-1}(\gamma-\sin{\gamma})/\sqrt{\beta}$ & $\tfrac{1}{6}s^3$ & $\beta^{-1}(\gamma-\sinh{\gamma})/\sqrt{-\beta}$\\
   \href{https://github.com/ericagol/NbodyGradient.jl/blob/0afa7a5fd387b307e3704ded05ec01838b01478c/src/utils.jl\#L167-L178}{$H_1(\beta,\gamma)$}& $\beta^{-2}(2-2\cos{\gamma}-\gamma\sin{\gamma})$ & $\frac{1}{12} s^4$ & $\beta^{-2}(2-2\cosh{\gamma}+\gamma\sinh{\gamma})$\\
   \href{https://github.com/ericagol/NbodyGradient.jl/blob/0afa7a5fd387b307e3704ded05ec01838b01478c/src/utils.jl\#L211-L223}{$H_2(\beta,\gamma)$}& $\beta^{-3/2}(\sin{\gamma}-\gamma\cos{\gamma})$ & $\frac{1}{3}s^3$ & $(-\beta)^{-3/2}(-\sinh{\gamma}+\gamma\cosh{\gamma})$\\
   \href{https://github.com/ericagol/NbodyGradient.jl/blob/0afa7a5fd387b307e3704ded05ec01838b01478c/src/utils.jl\#L256-L269}{$H_3(\beta,\gamma)$} & $\beta^{-1}\left(4\sin{\gamma} - \sin{\gamma}\cos{\gamma}-3\gamma \right)/\sqrt{\beta}$ & $-\frac{1}{10}\beta s^5$ & 
   $\beta^{-1}\left(4\sinh{\gamma}-\sinh{\gamma}\cosh{\gamma}-3\gamma\right)/\sqrt{-\beta}$\\
   \href{https://github.com/ericagol/NbodyGradient.jl/blob/0afa7a5fd387b307e3704ded05ec01838b01478c/src/utils.jl\#L305-L318}{$H_5(\beta,\gamma)$} & $\beta^{-1}\left(3\sin{\gamma} - \gamma\cos{\gamma}-2\gamma \right)/\sqrt{\beta}$ & $-\frac{1}{60}\beta s^5$ & $\beta^{-1}\left(3\sinh{\gamma}-\gamma\cosh{\gamma}-2\gamma\right)/\sqrt{-\beta}$\\
   \href{https://github.com/ericagol/NbodyGradient.jl/blob/0afa7a5fd387b307e3704ded05ec01838b01478c/src/utils.jl\#L351-L362}{$H_6(\beta,\gamma)$} & $\frac{1}{2}\beta^{-2}\left(9-8\cos{\gamma} - \cos{2\gamma}-6\gamma\sin{\gamma} \right)$ & $\frac{1}{40}\beta s^6$ & $\frac{1}{2}\beta^{-2}\left(9-8\cosh{\gamma} - \cosh{2\gamma}+6\gamma\sinh{\gamma} \right)$\\
\hline
\end{tabular}
\label{tab:G_functions}
\end{table*}

\subsection{Combined Kepler and Drift step} \label{sec:kepler_drift}

\begin{figure*}
    \centering
    \includegraphics{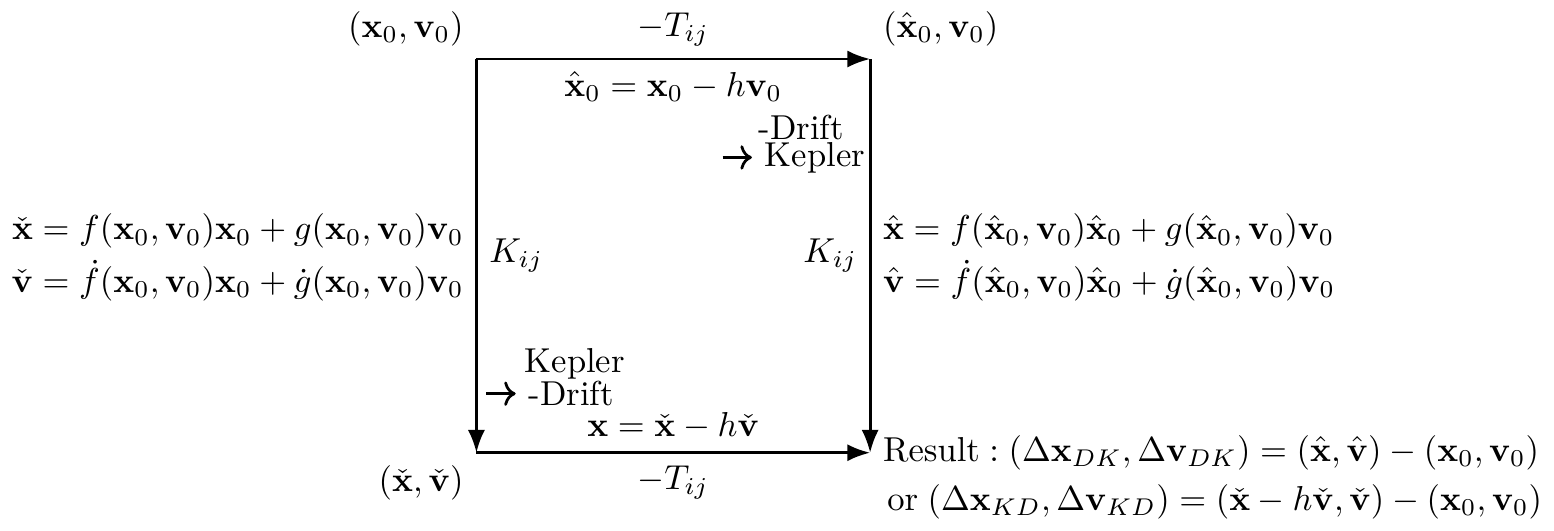}
    \caption{The order of the combined substeps (from upper left corner to lower right corner) has two sequences:
    first a negative drift followed by a Kepler step, then a Kepler step followed by a negative drift.  These two options need to be handled separately, and
    notation for the intermediate steps is summarized in this diagram.}
    \label{fig:kepler_drift}
\end{figure*}

In the {\sc AHL21} algorithm these two steps, $-T_{ij}$ and
$K_{ij}$, are combined in different
orders:  either a negative drift followed by a Kepler step, $-T_{ij}+K_{ij}$, or
a Kepler step followed by a negative drift, $K_{ij}-T_{ij}$ (Algorithm \ref{alg:AHL21_algorithm}).  As these operations
do not commute, we need to handle each one separately.
A diagram showing the order of these mappings is given in
Figure \ref{fig:kepler_drift}.  In each case, the position coordinate
takes on an intermediate value, which changes the nature of
the combined steps.

We describe these two options in the following subsections.

\subsubsection{Drift then Kepler {(DK)}}\label{sec:drift_then_kepler}

In the first case, the negative drift is taken first, yielding {an intermediate position}
\begin{eqnarray}\label{eqn:drift_first}
    \hat{\mathbf{x}}_0 &=& \mathbf{x}_0 - h \mathbf{v}_0.
\end{eqnarray}
With this modified value of the initial position, the Gauss
$f$, $g$, $\dot{f}$, and $\dot{g}$ functions need to be computed from $(\hat{\mathbf{x}}_0,
\mathbf{v}_0,k,h)$ {after solution of Kepler's equation for $\gamma$}, so we indicate these functions with a hat,
e.g.\ $\hat{f} \equiv f(\hat{\mathbf{x}}_0,\mathbf{v}_0,k,h)$.  
In addition, we
would like to find the difference between the final and initial
coordinates, $\Delta \mathbf{x}_{DK} = \hat{\mathbf{x}}-\mathbf{x}_0$ and
$\Delta \mathbf{v}_{DK} = \hat{\mathbf{v}}-\mathbf{v}_0$;  this allows for a more accurate computation
of these quantities when the step sizes are small.  In the combined step, -Drift+Kepler (which we indicate with ``DK"), the resulting term is
\begin{eqnarray}\label{eqn:dxv_drift_kepler}
\href{https://github.com/ericagol/NbodyGradient.jl/blob/0afa7a5fd387b307e3704ded05ec01838b01478c/src/integrator/ahl21/ahl21.jl\#L1022-L1024}{\Delta \mathbf{x}_{\mathrm{DK}}} &=& (\hat{f}-1) \mathbf{x}_0 + (\hat{g}-h\hat{f}) \mathbf{v}_0,\cr
\href{https://github.com/ericagol/NbodyGradient.jl/blob/0afa7a5fd387b307e3704ded05ec01838b01478c/src/integrator/ahl21/ahl21.jl\#L1026-L1028}{\Delta \mathbf{v}_{\mathrm{DK}}} &=& \dot{\hat{f}} \mathbf{x}_0 + (\dot{\hat{g}}-h\dot{\hat{f}}-1) \mathbf{v}_0,
\end{eqnarray}
where, again, $\hat{f}$, $\hat{g}$, $\dot{\hat{f}}$, and $\dot{\hat{g}}$ are all computed in terms
of $(\hat{\mathbf{x}}_0,\mathbf{v}_0,k,h)$.   {The scalar functions in equation \ref{eqn:dxv_drift_kepler} are given in Appendix \ref{sec:appendix1}.}


Note that in these equations the 1's are cancelled analytically; this yields more stable computation of the changes in the positions and velocities when these are small.

\subsubsection{Kepler then Drift {(KD)}}\label{sec:kepler_then_drift}

In the other case, a Kepler step is applied first, followed by a negative drift.  The Kepler step can be computed in terms of the initial coordinates, $\mathbf{x}_0$ and $\mathbf{v}_0$, yielding intermediate coordinates $(\check{\mathbf{x}},\check{\mathbf{v}})$, and then the negative drift is applied resulting in 
$\mathbf{x} = \check{\mathbf{x}}-h\check{\mathbf{v}}$ (Figure \ref{fig:kepler_drift}).

We combine these and take the difference with the initial coordinates, $\Delta \mathbf{x}_{KD} = \check{\mathbf{x}}-h\check{\mathbf{v}}-\mathbf{x}_0$ and $\Delta \mathbf{v}_{KD}=\check{\mathbf{v}}-\mathbf{v}_0$, to give the resulting difference vectors
\begin{eqnarray}\label{eqn:dxv_kepler_drift}
\href{https://github.com/ericagol/NbodyGradient.jl/blob/0afa7a5fd387b307e3704ded05ec01838b01478c/src/integrator/ahl21/ahl21.jl\#L1022-L1024}{\Delta \mathbf{x}_{\mathrm{KD}}} &=& (f-h\dot f-1) \mathbf{x}_0 + (g-h\dot g) \mathbf{v}_0,\cr
\href{https://github.com/ericagol/NbodyGradient.jl/blob/0afa7a5fd387b307e3704ded05ec01838b01478c/src/integrator/ahl21/ahl21.jl\#L1026-L1028}{\Delta \mathbf{v}_{\mathrm{KD}}} &=& \dot f \mathbf{x}_0 + (\dot g-1) \mathbf{v}_0,
\end{eqnarray}
where  $f$, $g$, $\dot f$, and $\dot g$ are all computed in terms
of $(\mathbf{x}_0,\mathbf{v}_0,k,h)$, and the ``KD" indicates that the Kepler step precedes the negative drift, Kepler-Drift.  {The scalar functions in equation \ref{eqn:dxv_kepler_drift} are given in Appendix \ref{sec:appendix1}.}

We take care that these functions are evaluated in a numerically-stable manner to avoid round-off error due to cancellations between terms at small time steps.  With the combination of the drift and Kepler steps, it turns out that we no longer 
need the drift of the center-of-mass coordinates in each Kepler step as these cancel 
exactly.  

Thus, the DH17 algorithm simplifies significantly at the expense of making the substeps slightly
more complicated.  This new combined algorithm we dub ``{\sc AHL21}", which is given in 
Algorithm \ref{alg:AHL21_algorithm}.  The fourth-order correction is the same as
that given in \citet{Dehnen2017}, and is summarized in \S \ref{sec:correction}.  
The transit-time finding is described below in \S \ref{sec:transit_times}.  
An alternate
version of the algorithm in which the combined drift and Kepler steps are replaced
by a kick for some pairs of bodies is given in \S \ref{sec:kicks}.
\begin{algorithm}
 \KwData{Initial Cartesian coordinates and masses at time $t=t_0$.}
 \KwResult{Integration of $N$-body system over time $t_\mathrm{max}$, and resulting times of transit and derivatives.}
 \For{$t-t_0 < t_\mathrm{max}$}{
  {Kick particles in $A^C$ for time $h/6$\; Drift all particles for time $h/2$\;}
  \For{pairs of particles ($i$, $j$) in $A$}{
   Apply a combined -Drift+Kepler step for bodies $i$ and $j$ over a time $h/2$ to
   give the changes in position and velocity of $\Delta \mathbf{x}_{DK}$ and
   $\Delta \mathbf{v}_{DK}$\, and update {$\mathbf{x}_{ij}$ and $\mathbf{v}_{ij}$};
  }
  Apply velocity correction and a kick, both multiplied by $2/3$, to particles in $A^C$\;
  Apply velocity correction to particles in $A$\;
  \For{reversed pairs of particles $(i,j)$ in $A$}{
   {Apply a combined Kepler-Drift step for bodies $i$ and $j$ over a time $h/2$ to
   give the changes in position and velocity of $\Delta \mathbf{x}_{KD}$ and
   $\Delta \mathbf{v}_{KD}$\, and update {$\mathbf{x}_{ij}$ and $\mathbf{v}_{ij}$};}
  }
  {Drift all particles for time $h/2$\; Kick particles in $A^C$ for time $h/6$\;}
  \If{transit has occurred for particles $i$ and $j$}{
  Refine transit time, and save.}
  Increment time $t$ by $h$.
 }
 \caption{Transit times with {\sc AHL21} symplectic integration.}
 \label{alg:AHL21_algorithm}
\end{algorithm}

The primary goal of this paper is to describe the implementation and
differentiation of the {\sc AHL21} algorithm,
yielding the derivatives of the transit times with respect to the initial 
conditions.  Along the way we compute the derivatives of the state of
the system at each time step with respect to the initial conditions, which
may be used for other applications such as photodynamics, radial velocity,
astrometry, or computation of Lyapunov exponents.  Next we describe the derivative computation.

\section{Differentiation of symplectic integrator} \label{sec:differentiation_symplectic}

We divide the {differentiation of algorithm \ref{alg:AHL21_algorithm}} into a series of steps,
\begin{enumerate}
    \item Derivative of coordinates at end of a symplectic step with respect 
    to coordinates at the beginning.  This includes the Kepler step, drifts 
    and kicks (\S \ref{sec:kepler_step}-\S \ref{sec:kicks}).
    \item Derivatives of fourth-order velocity correction (\S \ref{sec:correction}).
    \item Propagation of Jacobians through each of these steps (\S \ref{sec:Jacobian}).
    \item Derivative of parameters output at specified times with respect 
    to the coordinates at the symplectic time grid. Here we give the example 
    of the derivatives of the transit times with respect to the initial conditions 
    (\S \ref{sec:transit_times}), but
    this could also include eclipse times, radial velocity at pre-specified time, 
    or relative positions of the bodies at times of observation.   This step 
    involves an {\sc AHL21} step with a fractional time duration.
\end{enumerate}

We describe 
each of these steps in turn{, after some preliminaries}.

\subsection{Notation conventions} \label{sec:notation}

The integration is carried out in inertial Cartesian coordinates \citep{Hernandez2015}, 
while the initial conditions of the $N$ bodies are specified in either of two forms:
Cartesian coordinates of $N$ bodies, or orbital elements in a hierarchy of $N-1$
Keplerians (which we leave to a future paper).  The initial time of the start of 
the integration, $t_0$, requires 
a snapshot of the phase-space coordinates or orbital elements, which fully
specify the problem with the addition of the masses of the bodies, 
$\mathbf{m} = \{m_1,...,m_N\}$, which are constant in time. In this section we 
describe the phase-space coordinates.

\subsection{Code units}\label{sec:code_units}

We utilize units for masses in $M_\odot$, positions in AU, and
time in days.   Our gravitational constant is given by
$GM_\odot = 0.00029598$ AU$^3$day$^{-2} M_\odot^{-1}$.  The initial
conditions, then, simply need to be specified in terms of masses
in $M_\odot$, positions in AU, and velocities in
AU/day.

\subsection{Cartesian coordinates} \label{sec:cartesian_coordinates}

\begin{figure}
\center
	\includegraphics[width= .48\textwidth]{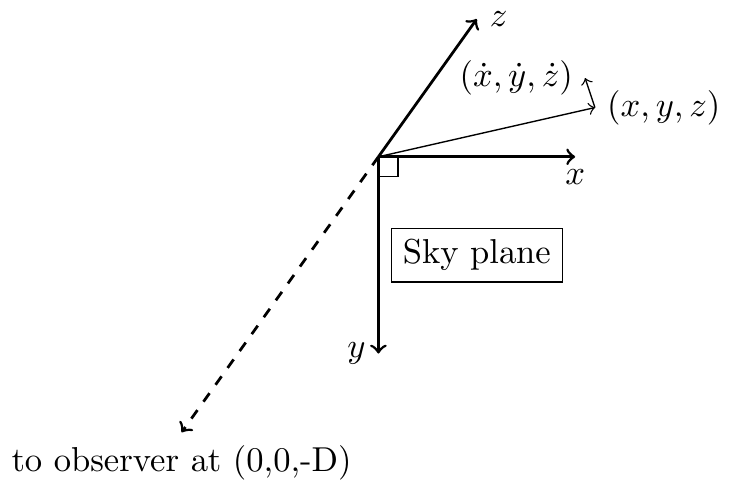}
\caption{
Cartesian coordinate system.  Body $i$ is at position 
${\bf x}_i=(x_i, y_i, z_i)$ with velocity ${\bf v}_i =(\dot x_i, \dot y_i, \dot z_i)$. 
}\label{fig:coordinates}  
\vspace{0.3cm}
\end{figure}

The Cartesian coordinates utilize a right-handed coordinate system for which the sky
plane is the $x-y$ plane, while the $z$ axis is along the line of sight,
increasing away from the observer (Figure \ref{fig:coordinates}). Positions for each body are denoted with a
vector ${\bf x}_i(t)=\left(x_i(t),y_i(t),z_i(t)\right)$, while velocities 
are denoted with ${\bf v}_i(t)=\left(\dot x_{i}(t),\dot y_{i}(t),\dot 
z_{i}(t)\right)$, with subscript $i=1,..,N$ labelling each body, and 
$\dot c = \frac{d c}{d t}$ indicates time derivative of
variable $c$.  The observer is located at ${\bf x}_{obs} = \left(0,0, 
-d\right)$, where $d$ is the distance of the observer to the center of mass 
of the system.  

The initial conditions are completely specified via ${\bf q}(t_0)$, where 
${\bf q}(t) = \{{\bf x}_i(t),{\bf v}_i(t), m_i; i=1,...,N\}$.  The vector
${\bf q}(t)$ has $7N$ elements, where the $7(i-1)+j$th element refers to planet
$i$ and the $j$th element of the vector 
\begin{equation}\label{eqn:coordinates_body}
{\bf q}_i(t) =\{x_i(t),y_i(t),z_i(t),\dot x_i(t),\dot y_i(t), \dot z_i(t), m_i\}
\end{equation}
where $j=1,...,7$.  
Note that we take the origin of the coordinates to be the
center of mass of the system, so that a constraint on the initial conditions is
$\sum_i m_i q_{i,j}(t_0) = 0$ for $j=1,...,6$, where $q_{i,j}$ denotes the
$j$th element of of ${\bf q}_i(t)$.\footnote{In general, the center-of-mass is
allowed to move at a constant velocity, which is not implemented in our initial
conditions, but could be if required.} 

The coordinate system is right-handed,  with the $x$-axis pointing to the 
right on the sky, then the $y$-axis points downwards, so that 
$\hat x \times \hat y = \hat z$ points away from the observer, for unit 
vectors $\{\hat x, \hat y, \hat z\}$ (Figure \ref{fig:coordinates}).

\subsection{Derivative of a combined Drift and Kepler step} \label{sec:kepler_step}

The building block of this integrator is the universal Kepler solver for
integrating pairs of bodies \citep{Wisdom2015}, which we combined with
a negative drift, before or after, described in \S \ref{sec:kepler_drift}.  
Standard Wisdom-Holman $N$-body symplectic 
integrators \citep{Wisdom1991} use an elliptic (bound) Kepler solver
for the `unperturbed' motion, while the weaker interactions between
low-mass or distant bodies are treated as impulses or kicks alternating with
the Kepler drifts.  In the case of the DH17 integrator, a hyperbolic step
is needed for the pairwise Keplerian integration of bodies that are unbound;
this is used as an alternative to kicks.  In this section we summarize 
the computation of the the Jacobian of the final relative coordinates with respect 
to the initial coordinates over a time step with duration $h$,  and
then the transformation to the coordinates of the individual bodies.


The variational equations for the Cartesian coordinates depends on
the ordering of the negative drift and Kepler step.  For the combined negative drift
followed by a Kepler step, the change in relative position and velocity is
\begin{eqnarray} 
\href{https://github.com/ericagol/NbodyGradient.jl/blob/0afa7a5fd387b307e3704ded05ec01838b01478c/src/integrator/ahl21/ahl21.jl\#L1108-L1120}{\delta \Delta \mathbf{x}_\mathrm{DK}} &=& (\hat{f}-1) \delta \mathbf{x}_0 + (\hat{g}-h\hat{f}) \delta\mathbf{v}_0 \cr &+& \delta (\hat{f}-1) \mathbf{x}_0 + \delta (\hat{g}-h\hat{f}) \mathbf{v}_0,\label{eqn:dxDK}\\
\href{https://github.com/ericagol/NbodyGradient.jl/blob/0afa7a5fd387b307e3704ded05ec01838b01478c/src/integrator/ahl21/ahl21.jl\#L1162-L1174}{\delta \Delta \mathbf{v}_\mathrm{DK}} &=& \dot{\hat{f}} \delta \mathbf{x}_0 + (\dot{ \hat{g}}-h\dot{\hat{f}}-1) \delta\mathbf{v}_0 \cr &+& \delta \dot{\hat{f}} \mathbf{x}_0 + \delta (\dot{\hat{g}}-h\dot{\hat{f}}-1) \mathbf{v}_0,\label{eqn:dvDK}
\end{eqnarray}
while for a combined Kepler step followed by a negative drift, the variation
of the change in relative coordinates is
\begin{eqnarray}
\href{https://github.com/ericagol/NbodyGradient.jl/blob/0afa7a5fd387b307e3704ded05ec01838b01478c/src/integrator/ahl21/ahl21.jl\#L1268-L1279}{\delta \Delta \mathbf{x}_\mathrm{KD}} &=& (f-h\dot f -1) \delta \mathbf{x}_0 + (g-h\dot g) \delta\mathbf{v}_0 \cr &+& \delta (f-h\dot f-1) \mathbf{x}_0 + \delta (g-h\dot g) \mathbf{v}_0,\label{eqn:dxKD}\\
\href{https://github.com/ericagol/NbodyGradient.jl/blob/0afa7a5fd387b307e3704ded05ec01838b01478c/src/integrator/ahl21/ahl21.jl\#L1310-L1321}{\delta \Delta \mathbf{v}_\mathrm{KD}} &=& \dot f \delta \mathbf{x}_0 + (\dot g-1) \delta\mathbf{v}_0 \cr &+& \delta \dot f \mathbf{x}_0 + \delta (\dot g-1) \mathbf{v}_0,\label{eqn:dvKD}
\end{eqnarray}
where we have taken the differential of equations (\ref{eqn:dxv_drift_kepler}) and (\ref{eqn:dxv_kepler_drift}).  The first line of each equation we have already computed, while the differentials of the Gauss functions in the second lines remain to be computed.

Each of the differential Gauss function terms involve the basis
$(\mathbf{x}_0,\mathbf{v}_0,k,h)$, while each of these functions 
is defined in terms of $G_i(\beta,\gamma)$, $\beta, \gamma, r_0, r$ and $\eta_0$
(or $\hat{G}_i(\hat{\beta},\hat{\gamma})$, $\hat{\beta}, \hat{\gamma}, \hat{r}_0, \hat{r}$ and $\hat{\eta}_0$).
Note that if $h$ is varied as a function of phase space, symplecticity is lost; however we accept a small symplecticity violation at a single timestep when searching for transit times.  A similar choice was made in \citet{Deck:2014}.  Thus, we first need to compute the differential of these Gauss 
function terms with respect to these intermediate quantities, and 
then propagate through these differentials using the chain rule to 
obtain the derivatives with respect to the basis.  There is an extra 
step involved in the drift-first case (DK):  since the functions on 
the right hand side are defined in terms of 
$\hat{\mathbf{x}}_0 = \mathbf{x}_0-h \mathbf{v}_0$, we also need to 
apply the chain rule to $\mathbf{\hat x}_0$ to transform the derivatives 
to the basis. 

{The differentials of intermediate quantities are given in Appendix \ref{sec:appendix2}.}

\subsubsection{Differential of drift-then-Kepler step}

The differential of the scalar quantities $\delta(\hat{f}-1)$, $\delta(\hat{g}-h\hat{f}),$
$\delta(\dot{\hat{f}})$, and $\delta (\dot{\hat{g}}-h\dot{\hat{f}} -1)$ should also be scalars, and can be expressed in terms similar to the $\delta \gamma$ and
$\delta r$ terms given {in Apppendix \ref{sec:appendix2}}. 
Note, however, that as the Kepler step takes place \emph{after} the negative drift, all of these functions are to be computed in terms of $\hat{\mathbf{x}}_0$ substituted for $\mathbf{x}_0$, and so we need to add an extra step in the derivation to find the differentials in terms of $\mathbf{x}_0$ in lieu of $\hat{\mathbf{x}}_0$. 
{The differential of these functions in
terms of intermediate scalar quantities is given in appendix \ref{sec:appendix2}.}

Substituting these differentials into equations (\ref{eqn:dxDK}) and
(\ref{eqn:dvDK}), we arrive at expressions for
$\partial{\Delta \mathbf{x}_\mathrm{DK}}/\partial{\mathbf{x}_0}$,
$\partial{ \Delta \mathbf{x}_\mathrm{DK}}/\partial{\mathbf{v}_0}$,  
$\partial{ \Delta \mathbf{x}_\mathrm{DK}}/\partial{k}$,
$\partial{ \Delta \mathbf{v}_\mathrm{DK}}/\partial{\mathbf{x}_0}$,
$\partial{ \Delta \mathbf{v}_\mathrm{DK}}/\partial{\mathbf{v}_0}$, and
$\partial{ \Delta \mathbf{v}_\mathrm{DK}}/\partial{k}$.  We also know
$\partial k/\partial \mathbf{x}_0$ = 0, 
$\partial k/\partial \mathbf{v}_0$ = 0, and $\partial k/\partial k = 1$,
which we insert into a Jacobian, $\mathbf{J}_\mathrm{kep}$, which
is a 7$\times$7 matrix.
In addition, we need the time derivatives of the coordinates with
respect to the time step, $h$, 
$\partial{ \Delta \mathbf{x}_\mathrm{DK}}/\partial{h}$,
$\partial{ \Delta \mathbf{v}_\mathrm{DK}}/\partial{h}$,
to obtain the time derivatives of the transit times with respect to
the initial coordinates 

This completes the {summary} 
of the Jacobian of the combined
negative drift then Kepler step for the variation of the relative
coordinates of bodies $i$ and $j$ (recall that we dropped the $ij$ subscripts in this section).  In the next subsection we {discuss}
the results for the Kepler step followed by a negative drift.

\subsubsection{Differential of Kepler then drift step}

The Kepler step followed by a negative drift is slightly simpler
as the Gauss functions can be expressed in terms of $\mathbf{x}_0$
rather than $\hat{\mathbf{x}}_0$.   {The differentials of the scalar
functions are given in Appendix \ref{sec:appendix2}.}

As with the prior combined step, the terms in these differentials may be inserted into a Jacobian,
$\mathbf{J}_\mathrm{kep}$, as well give the derivatives with respect to $h$.  This completes the description of the Jacobians computed for the combined drift and Kepler steps for the change in the relative coordinates between bodies $i$ and $j$.   This needs to be translated into the variations of the positions and velocities of the individual bodies $i$ and $j$, which we describe next.

\subsection{Jacobian of combined Kepler drift step} \label{sec:kepler_drift_jacobian}

The foregoing computation gives the variation in the \emph{relative}
difference between the positions and velocities of bodies $i$ and $j$.
This translates into variations in the positions of bodies $i$ and $j$
given by
\begin{eqnarray} \label{eqn:kepler_derivatives_ij}
\href{https://github.com/ericagol/NbodyGradient.jl/blob/0afa7a5fd387b307e3704ded05ec01838b01478c/src/integrator/ahl21/ahl21.jl\#L839}{\Delta \mathbf{x}_{i,\mathrm{DK}}} &=& \frac{m_j}{m_i+m_j} \Delta \mathbf{x}_\mathrm{DK},\cr
\href{https://github.com/ericagol/NbodyGradient.jl/blob/0afa7a5fd387b307e3704ded05ec01838b01478c/src/integrator/ahl21/ahl21.jl\#L840}{\Delta \mathbf{x}_{j,\mathrm{DK}}} &=& - \frac{m_i}{m_i+m_j} \Delta \mathbf{x}_\mathrm{DK},\cr
\href{https://github.com/ericagol/NbodyGradient.jl/blob/0afa7a5fd387b307e3704ded05ec01838b01478c/src/integrator/ahl21/ahl21.jl\#L843}{\Delta \mathbf{v}_{i,\mathrm{DK}}} &=&  \frac{m_j}{m_i+m_j} \Delta \mathbf{v}_\mathrm{DK},\cr
\href{https://github.com/ericagol/NbodyGradient.jl/blob/0afa7a5fd387b307e3704ded05ec01838b01478c/src/integrator/ahl21/ahl21.jl\#L844}{\Delta \mathbf{v}_{j,\mathrm{DK}}} &=&  - \frac{m_i}{m_i+m_j} \Delta \mathbf{v}_\mathrm{DK},
\end{eqnarray}
and likewise for DK $\rightarrow$ KD, where $\mathbf{x}_{i}(t+h) = 
\mathbf{x}_i(t)+\Delta \mathbf{x}_{i,\mathrm{DK}}$ {is carried out with compensated summation}.

The Jacobian may be found by differentiating these equations with respect
to the initial conditions of the Kepler-drift step, which is \href{https://github.com/ericagol/NbodyGradient.jl/blob/0afa7a5fd387b307e3704ded05ec01838b01478c/src/integrator/ahl21/ahl21.jl\#L847-L854}{straightforward}
for the position, velocity, and time step derivatives. However,
since this equation involves the masses $m_i$ and $m_j$, the mass derivative of 
a combined Kepler/drift step involves an additional term, where
\begin{eqnarray} \label{eqn:kepler_derivatives_mass}
\href{https://github.com/ericagol/NbodyGradient.jl/blob/0afa7a5fd387b307e3704ded05ec01838b01478c/src/integrator/ahl21/ahl21.jl\#L855-L868}{\frac{\partial\Delta \mathbf{x}_{i,\mathrm{DK}}}{\partial m_i}} &=& -\frac{m_j}{(m_i+m_j)^2} \Delta \mathbf{x}_\mathrm{DK} + \frac{G m_j}{m_i+m_j} \frac{\partial\Delta \mathbf{x}_\mathrm{DK}}{\partial k},\cr
\href{https://github.com/ericagol/NbodyGradient.jl/blob/0afa7a5fd387b307e3704ded05ec01838b01478c/src/integrator/ahl21/ahl21.jl#L855-L868}{\frac{\partial\Delta\mathbf{x}_{i,\mathrm{DK}}}{\partial m_j}} &=& \frac{m_i}{(m_i+m_j)^2} \Delta \mathbf{x}_\mathrm{DK} + \frac{G m_j}{m_i+m_j} \frac{\partial \Delta \mathbf{x}_\mathrm{DK}}{\partial k},
\end{eqnarray}
and the same equations apply for $\mathbf{x} \rightarrow \mathbf{v}$, $i \leftrightarrow j$, and $\mathrm{DK} \rightarrow \mathrm{KD}$.
The first of these two equations has a cancellation due to the difference in sign between the two terms on the right hand side. Specifically,
$\Delta \mathbf{x}_\mathrm{DK} \propto k$, so there is a term in the derivative, $\partial \Delta x_\mathrm{DK}/\partial k$, which equals $\Delta \mathbf{x}_\mathrm{DK}/k$,
which exactly cancels the first term in the equation. 
We carry out this cancellation algebraically, thus avoiding
roundoff errors which can occur when this term is much larger than the others in $\partial\Delta \mathbf{x}_\mathrm{DK}/\partial k$.
The second equation has both terms with the same sign, so this cancellation does not occur when the derivatives are with
respect to the mass of the other body.
Here we give the resulting derivatives:
\begin{eqnarray} \label{eqn:dx_DK_mass}
\href{https://github.com/ericagol/NbodyGradient.jl/blob/685014cfa11f9876c498d1e6e17b1be050ba4932/src/integrator/ahl21/ahl21.jl\#L1119}{\frac{\partial\Delta \mathbf{x}_{i,\mathrm{DK}}}{\partial m_i}} &=& \frac{G^2 m_j}{\hat \beta \hat r\hat r_0^2} \left[J_1 \mathbf{x}_0-J_2 \mathbf{v}_0\right],
\end{eqnarray}

\begin{eqnarray}\label{eqn:dv_DK_mass}
\href{https://github.com/ericagol/NbodyGradient.jl/blob/685014cfa11f9876c498d1e6e17b1be050ba4932/src/integrator/ahl21/ahl21.jl\#L1173}{\frac{\partial\Delta\mathbf{v}_{i,\mathrm{DK}}}{\partial m_i}} &=& \frac{G^2 m_j}{\hat \beta \hat r^3 \hat r_0^2} \left[ J_3 \mathbf{x}_0 + J_4 \mathbf{v}_0\right],
\end{eqnarray}
{where $J_1-J_4$ are functions given in Appendix \ref{sec:appendix3}.}
Note that the derivatives of $\Delta \mathbf{x}_{j,\mathrm{DK}}$ and $\Delta \mathbf{v}_{j,\mathrm{DK}}$
with respect to $m_j$ look identical save for replacing $m_j$ with $-m_i$.

Similarly, the mass derivatives in the Kepler followed by drift step are given as:
\begin{eqnarray}\label{eqn:dx_KD_mass}
\href{https://github.com/ericagol/NbodyGradient.jl/blob/685014cfa11f9876c498d1e6e17b1be050ba4932/src/integrator/ahl21/ahl21.jl\#L1278}{\frac{\partial\Delta \mathbf{x}_{i,\mathrm{KD}}}{\partial m_i}} &=& \frac{G^2 m_j}{\beta r^3 r_0^2} \left[J_5 \mathbf{x}_0 + J_6 \mathbf{v}_0\right],
    \end{eqnarray}
    \begin{eqnarray} \label{eqn:dv_KD_mass}
\href{https://github.com/ericagol/NbodyGradient.jl/blob/0afa7a5fd387b307e3704ded05ec01838b01478c/src/integrator/ahl21/ahl21.jl\#L1320}{\frac{\partial\Delta\mathbf{v}_{i,\mathrm{KD}}}{\partial m_i}} &=& \frac{G^2 m_j}{\beta r^3 r_0^2} \left[J_7 \mathbf{x}_0 + r_0 J_8 \mathbf{v}_0\right],
\end{eqnarray}
{where $J_5-J_8$ are given in Appendix \ref{sec:appendix3}.}
As above, the derivatives of $\Delta \mathbf{x}_{j,\mathrm{KD}}$ and $\Delta \mathbf{v}_{j,\mathrm{KD}}$
with respect to $m_j$ look identical save for replacing $m_j$ with $-m_i$.  {We place all of these derivatives into a Jacobian matrix for each drift+Kepler substep, $\Delta\mathbf{J}_{DK,ij}$ or $\Delta\mathbf{J}_{KD,ij}$.}

This completes the computation of the Jacobian of the drift plus Keplerian
evolution of bodies $i$ and $j$ with respect to one another.  Next, we describe the derivatives of the drift step applied at the start and end of each time step.

\subsection{Drift}\label{sec:drift}

The drift of an individual body is given by
\begin{eqnarray}
    \mathbf{x}_i(t+h) &=& \mathbf{x}_i(t) + h \mathbf{v}_i(t),\\
    \mathbf{v}_i(t+h) &=& \mathbf{v}_i(t).
\end{eqnarray}
This has the straightforward differential of
\begin{eqnarray}
    \delta \mathbf{x}_i(t+h) &=& \delta \mathbf{x}_i(t) + h \delta \mathbf{v}_i(t) + \delta h \mathbf{v}_i(t),\\
    \delta \mathbf{v}_i(t+h) &=& \delta \mathbf{v}_i(t).
\end{eqnarray}

There are two stages at which the drifts are applied: all particles drift
at the start and end of each {\sc AHL21} step with a duration $h/2$ (see algorithm 
\ref{alg:AHL21_algorithm}). We refer to this as {$\mathbf{I}+\Delta\mathbf{J}_\mathrm{D}(h)$} for 
drifting all of the planets.  In some cases it proves to be faster and sufficiently accurate to use instantaneous kicks between pairs of bodies rather than solving the Kepler problem;  we now turn to describing this option.

\subsection{Derivative of kicks} \label{sec:kicks}

\citet{Hernandez2015} show that for some pairs of particles (typically distant
or unbound), sufficient accuracy may be obtained by applying a gravitational 
kick between particles, rather than a Keplerian step and negative drift.  Letting $A$ be the set of pairs {$(i,j)$} advanced with drift+Kepler steps, then  $A^C$ is the complementary set which receives pairwise kicks {such that $A \cap A^C = \varnothing$}.  {Note that if all pairs are in $A^C$, the integrator becomes leapfrog.}  

Algorithm \ref{alg:AHL21_algorithm}
implements this method by applying the pairwise kicks {(to set $A^C$) for a time step $h/6$ before} the initial drifts  , 
then after the combined drift-Kepler is applied to set $A$, a second set of kicks is applied for a time step $2h/3$ along with separate correction terms for the pairs in $A$ and $A^C$, and then after the {second } Kepler-drift step is applied to $A$ {in reverse order}, there is a final set of pairwise kicks applied to $A^C$ for a time $h/6$ {after} the final drifts (on set $A^C$).

For a pair of particles $i$ and $j$, the kicks applied over a time step $h$ are given by:
\begin{eqnarray}\label{eqn:kick}
\href{https://github.com/ericagol/NbodyGradient.jl/blob/0afa7a5fd387b307e3704ded05ec01838b01478c/src/integrator/ahl21/ahl21.jl\#L447}{\Delta \mathbf{v}_i} &=& -h\frac{G m_j \mathbf{x}_{ij}}{r_{ij}^3},\cr
\href{https://github.com/ericagol/NbodyGradient.jl/blob/0afa7a5fd387b307e3704ded05ec01838b01478c/src/integrator/ahl21/ahl21.jl\#L449}{\Delta \mathbf{v}_j} &=& h\frac{G m_i \mathbf{x}_{ij}}{r_{ij}^3},
\end{eqnarray}
where, as above, ${\bf x}_{ij}
= {\bf x}_i-{\bf x}_j$, $r_{ij} = \vert {\bf x}_{ij}\vert$, which has derivatives given by
\begin{eqnarray}\label{eqn:kick_deriv}
\href{https://github.com/ericagol/NbodyGradient.jl/blob/0afa7a5fd387b307e3704ded05ec01838b01478c/src/integrator/ahl21/ahl21.jl\#L456}{\delta \Delta \mathbf{v}_i} &=& -h\frac{G m_j}{r_{ij}^5} \left[{\bf x}_{ij} \frac{\delta m_j}{m_j} r_{ij}^2+\mathbf{w}_{ij}\right],\\
\href{https://github.com/ericagol/NbodyGradient.jl/blob/0afa7a5fd387b307e3704ded05ec01838b01478c/src/integrator/ahl21/ahl21.jl\#L458}{\delta \Delta \mathbf{v}_j} &=& h\frac{G m_i}{r_{ij}^5} \left[{\bf x}_{ij} \frac{\delta m_i}{m_i} r_{ij}^2+\mathbf{w}_{ij}\right],\\
\href{https://github.com/ericagol/NbodyGradient.jl/blob/0afa7a5fd387b307e3704ded05ec01838b01478c/src/integrator/ahl21/ahl21.jl\#L460-L472}{\mathbf{w}_{ij}} &=& \delta {\bf x}_{ij} r_{ij}^2 -  3 {\bf x}_{ij} {\bf x}_{ij} \cdot \delta {\bf x}_{ij}.
\end{eqnarray}
These differentials yield a Jacobian for the kicks between all pairs of bodies {$(i,j) \in A^C$}.  We now move on to describing the derivatives of the {fourth}-order correction which is used to improve the order of the algorithm.

\subsection{Derivative of correction} \label{sec:correction}

\citet{Dehnen2017} reduced the error in the \citet{Hernandez2015} mapping,
obtaining a symplectic integrator accurate to $h^4$.  We incorporate this
correction into our integrator, using $\alpha=0$ (eq. (40) of Dehnen and Hernandez 2017) which only requires one
call of the corrector in between the sequences of binary drift-Kepler and Kepler-drift steps (in the middle of algorithm \ref{alg:AHL21_algorithm}). {Two corrections need to be computed:  one for the pairs in $A$, and one for those in $A^C$.  We describe these in the next two subsections.}

{\subsubsection{Drift+Kepler pairs correction ($A$)}}

The {first} correction is applied to the velocities {of the bodies which are treated with the Kepler+drift splitting}, with an impulse
term for the $i$th body {in $A$} of
\begin{eqnarray}\label{eqn:correction}
\href{https://github.com/ericagol/NbodyGradient.jl/blob/0afa7a5fd387b307e3704ded05ec01838b01478c/src/integrator/ahl21/ahl21.jl#L722-L729}{
\Delta {\bf v}_i} &=& \frac{h^3}{24} \sum_{{i,j \in A}} \frac{G m_j}{r_{ij}^5} {\bf T}_{ij},
\end{eqnarray}
\begin{eqnarray}\label{eqn:tij}
{\bf T}_{ij} &=& {\bf x}_{ij} \left(\frac{2 Gm}{r_{ij}} + 3 {\bf a}_{ij} \cdot {\bf x}_{ij}\right)
-r_{ij}^2{\bf a}_{ij},
\end{eqnarray}
where $m=m_i+m_j$, ${\bf a}_{ij} = {\bf a}_i-{\bf a}_j$,  and
\begin{eqnarray}\label{eqn:correct_ai}
\href{https://github.com/ericagol/NbodyGradient.jl/blob/0afa7a5fd387b307e3704ded05ec01838b01478c/src/integrator/ahl21/ahl21.jl#L676-L677}{
{\bf a}_i} = -\sum_{{i,j \in A}} \frac{G m_j}{r_{ij}^3} {\bf x}_{ij}.
\end{eqnarray}
{Note that the sum is only taken over the particles in $A$, and no} correction is required for the positions.  We will define a constant $\href{https://github.com/ericagol/NbodyGradient.jl/blob/0afa7a5fd387b307e3704ded05ec01838b01478c/src/integrator/ahl21/ahl21.jl\#L660}{C} = 
(Gh^3)/24$ in what follows.

The derivative of this correction term can be computed in two steps,
first computing the derivative of ${\bf a}_i$, and then the derivative
of $\Delta {\bf v}_i$,
\begin{eqnarray}\label{eqn:correction_derivative}
\href{https://github.com/ericagol/NbodyGradient.jl/blob/0afa7a5fd387b307e3704ded05ec01838b01478c/src/integrator/ahl21/ahl21.jl\#L663-L700}
{\delta {\bf a}_i} = -\sum_{{i,j \in A}} \frac{G m_j}{r_{ij}^5} 
\left[ {\bf x}_{ij} \frac{\delta m_j}{m_j} r_{ij}^2 + \delta {\bf x}_{ij} r_{ij}^2
-  3 {\bf x}_{ij} {\bf x}_{ij} \cdot \delta {\bf x}_{ij}\right],
\end{eqnarray}
\begin{eqnarray}
\href{https://github.com/ericagol/NbodyGradient.jl/blob/0afa7a5fd387b307e3704ded05ec01838b01478c/src/integrator/ahl21/ahl21.jl\#L733-L792}
{\delta \Delta {\bf v}_i} &=& C \sum_{{i,j \in A}} \left[\frac{\delta m_j}{r_{ij}^5} 
- \frac{m_j}{r_{ij}^6}\frac{5{\bf x}_{ij}\cdot \delta {\bf x}_{ij}}{r_{ij}}
\right]{\bf T}_{ij}\notag \\
&+& C \sum_{{i,j \in A}} \frac{m_j}{r_{ij}^5} \delta {\bf T}_{ij},
\end{eqnarray}
with
\begin{eqnarray}
\href{https://github.com/ericagol/NbodyGradient.jl/blob/0afa7a5fd387b307e3704ded05ec01838b01478c/src/integrator/ahl21/ahl21.jl\#L744-L792}{\delta {\bf T}_{ij}} &=& \delta {\bf x}_{ij} \left(\frac{2Gm}{r_{ij}}+
3{\bf a}_{ij}\cdot{\bf x}_{ij}\right)\notag \\
&+& \frac{2Gm{\bf x}_{ij}}{r_{ij}}\left(\frac{\delta m_i+\delta m_j}{m} - \frac{{\bf x}_{ij}\cdot 
\delta {\bf x}_{ij}}{r_{ij}^2}\right)\notag\\
&+&  {\bf x}_{ij}\left(3 {\bf x}_{ij}\cdot \delta {\bf a}_{ij} + 3 {\bf a}_{ij}\cdot \delta {\bf x}_{ij}\right)\notag \\
&-& 2({\bf x}_{ij}\cdot \delta {\bf x}_{ij}){\bf a}_{ij} - r_{ij}^2 \delta {\bf a}_{ij}.
\end{eqnarray}
When implementing these equations as computer code, we pre-compute and store the dot products
to save computational time.

{\subsubsection{Correction for fast-kick pairs ($A^C$)}}

{The pairs in $A^C$ also require a correction, but with a slightly simpler relation:} 

{\begin{eqnarray}\label{eqn:correction_fastkick}
\href{}{\Delta {\bf v}_i} &=& \frac{h^3}{36} \sum_{i,j \in A^C} \frac{G m_j}{r_{ij}^5} \left[3\mathbf{x}_{ij}(\mathbf{a}_{ij}\cdot \mathbf{x}_{ij})-\mathbf{a}_{ij}r_{ij}^2\right],
\end{eqnarray}}
{where the sum is taken only over pairs in $A^C$.  The derivatives are computed in a manner similar to that described in the prior sub-section.}

The overall Jacobian for this step is given by {$\mathbf{I}+\Delta{\bf J}_\mathrm{4th}(h)$},
which is the identity matrix for the position and mass component,
and is given by $\partial{\Delta \mathbf{v}_i}/\partial \mathbf{x}_j$
for the offdiagonal components relating bodies $i$ and $j$. {The time derivatives are straightforward as they involve derivatives with respect to $h^3$, and so involve the same formulae multiplied by $3/h$.}
With the Jacobians now defined for each component of a time step, we next describe how we combine these into the Jacobian of a full time step.



\subsection{Jacobian of a time step} \label{sec:Jacobian}

With the Jacobian transformations computed at each step of algorithm 
\ref{alg:AHL21_algorithm}, we can now compute the complete derivative of each transit 
time with respect to the initial conditions, keeping track of the product of
Jacobians throughout \ref{alg:AHL21_algorithm}.  Now, in each case we compute the
change in the coordinates over a time step, and so the Jacobian of each substep
has the form:
\begin{eqnarray}
    \mathbf{J}_\mathrm{substep} = \mathbf{I} + \Delta \mathbf{J}_\mathrm{substep},
\end{eqnarray}
where $\Delta \mathbf{J}_\mathrm{substep}$ is the Jacobian of the change in
coordinates at the end of the substep with respect to the coordinates
at the beginning of the substep.  Consequently, the Jacobian can be written
as:
\begin{eqnarray}\label{eqn:substep_jacobian}
    \href{https://github.com/ericagol/NbodyGradient.jl/blob/0afa7a5fd387b307e3704ded05ec01838b01478c/src/utils.jl\#L25-L46}{\mathbf{J}_\mathrm{current}} &=& (\mathbf{I} + \Delta \mathbf{J}_\mathrm{substep}) \mathbf{J}_\mathrm{prior}\cr
    &=& \mathbf{J}_\mathrm{prior} + \Delta \mathbf{J}_\mathrm{substep} \mathbf{J}_\mathrm{prior}.
\end{eqnarray}
The propagation of the Jacobian involves adding
terms to the prior Jacobian as a function of each substep.  Now, when
the timestep is small, this involves very small additions to the Jacobian
which can increase the impact of round-off error during the propagation
of the derivatives.  To mitigate the impact of this, we use compensated
summation \citep{Kahan1965}.

As an example, in the 3-body case, an individual
{\sc AHL21} step looks like:
\begin{eqnarray}
   \mathbf{J}_\mathrm{AHL21}(h) &=& (\mathbf{I}+\Delta \mathbf{J}_\mathrm{D}(\tfrac{h}{2}))(\mathbf{I}+\Delta {\bf J}_{\mathrm{DK},12}(\tfrac{h}{2}))\cr
   && (\mathbf{I}+\Delta {\bf J}_{\mathrm{DK},13}(\tfrac{h}{2})) (\mathbf{I}+\Delta \mathbf{J}_{\mathrm{DK},23}(\tfrac{h}{2}))\cr
   && (\mathbf{I}+\Delta {\bf J}_\mathrm{4th}(h))\cr
   && (\mathbf{I}+\Delta \mathbf{J}_{\mathrm{KD},23}(\tfrac{h}{2}))(\mathbf{I}+\Delta \mathbf{J}_{\mathrm{KD},13}(\tfrac{h}{2}))\cr
   && (\mathbf{I}+\Delta \mathbf{J}_{\mathrm{KD},12}(\tfrac{h}{2}))(\mathbf{I}+\Delta \mathbf{J}_\mathrm{D}(\tfrac{h}{2})).
\end{eqnarray}
In this example we do not use fast kicks for any pair of bodies.
Note that each Jacobian in this product is computed with the updated
state of the system from the prior substep.  

We have also implemented a version of the integrator which allows 
the drift + Kepler interactions to be replaced with kicks for
some subset of pairs bodies (\S \ref{sec:kicks}).  For this
version of the integrator an additional three
Jacobians must be multiplied.

After taking $n+1$ steps whereby the first transit occurs in between
steps $n$ and $n+1$, a substep is taken to find the intermediate
time, $\Delta t=t-nh-t_0$,  which minimizes the sky separation between
the two bodies at time $t$.  To compute the derivatives of the times of transit
requires computing the Jacobian of a step with intermediate time
$\Delta t$, which is used to compute the derivatives of the time of
transit with respect to the initial conditions, which we describe next.

\subsection{Time derivative of a step}\label{sec:time_derivative}

The derivative of the positions and coordinates as a function of the
time step duration, $\Delta t$, requires a propagation of the time step derivative
through all of the substeps of a single time step.  The involves
applying the chain rule through each of the sub-steps in algorithm \ref{alg:AHL21_algorithm}.  Let {$\mathbf{q}_\mathrm{current} = \Delta \mathbf{q} + \mathbf{q}_\mathrm{prior}$} be the coordinates and velocities of all bodies {after applying} one component of a substep.  Then, 
\begin{eqnarray}
    \href{https://github.com/ericagol/NbodyGradient.jl/blob/0afa7a5fd387b307e3704ded05ec01838b01478c/src/integrator/ahl21/ahl21.jl\#L19-L20}{\frac{d\mathbf{q}_\mathrm{current}}{d 
    \Delta t}} = (\mathbf{I} + \Delta \mathbf{J}_\mathrm{substep}) \frac{d \mathbf{q}_\mathrm{prior}}{d \Delta t} + \left.\frac{\partial \Delta \mathbf{q}}{\partial \Delta t}\right|_{\mathbf{q}_\mathrm{prior}},
\end{eqnarray}
where $\frac{\partial \Delta \mathbf{q}}{\partial \Delta t}$ is the partial derivative of a particular sub-step with respect to the time step $\Delta t$.  Note that in the 
{\sc AHL21} algorithm (\ref{alg:AHL21_algorithm}), the combined drift and Kepler steps take place over a time $\Delta t  = h/2$, which introduces a factor of 1/2 in the partial derivatives with respect to $h$.

At the end of the step we refer to the derivative over the time step with respect to $\Delta t$ as:
\begin{eqnarray}\label{eqn:dstepdt}
    \frac{d \mathbf{q}(t)}{d \Delta t},
\end{eqnarray}
where $t$ is the total simulation time upon completion of the time step of duration $\Delta t$.
This may then be used to compute the transit times and their derivatives, as described next.

\subsection{Derivative of transit times} \label{sec:transit_times}

We define the times of transit between bodies $i$ and $j$ as the point in time where the sky-projected separation is at a minimum, and body $i$ is in front of body $j$ \citep{Fabrycky2010}.   Since multiple transits can occur between two bodies, we count these with a third index, $k$, so that the set of transit times during the time integration is given by $\left\{t_{ijk}; \forall i,j,k \right\}$.  At a transit time, the sky-velocity between the two bodies must be perpendicular to their sky separation, where the ``sky plane" is the $x-y$ plane;  this guarantees an extremum of the sky separation between the bodies.  The dot product of the relative sky separation and sky velocity of the two bodies equals zero at the time of transit, and is negative/positive just before/after transit.
So, transit times are computed from the constraints 
\begin{eqnarray}\label{eqn:gsky}
\href{https://github.com/ericagol/NbodyGradient.jl/blob/cb2fabd04b1009d9e36d5b144ea544e6f282f1e2/src/transits/timing.jl\#L139-L144}{g_{\mathrm{sky},ij}(t_{ijk})} &=& (x_i-x_j) 
(v_{x,i}-v_{x,j}) + (y_i-y_j) (v_{y,i} - v_{y,j}) = 0 \cr
z_i &<& z_j \cr
\frac{d g_{\mathrm{sky},ij}}{d t} &>& 0,
\end{eqnarray}
where $i$ is the index of the planet, and $j$ is the index of the star 
\citep{Fabrycky2010}, and $k$ is an index for the number of transits between
the bodies.  

Throughout the time-integration of a system, transits between a planet and star (or any pair of bodies) are checked for by identifying when 
$g_{\mathrm{sky},ij}(t)$ changes sign from negative to positive between two time steps, 
and the planet (or occultor) is nearer to the observer than the star.  
Once a transit time has 
been identified as occurring between time steps $n$ and $n+1$, where $t_n = t_0 + nh$, by the condition  $g_{\mathrm{sky},ij}(t_n) < 0$ 
and $g_{ski,ij} (t_{n+1}) > 0$ and $z_i(t_n) < z_j(t_n)$, then the time of transit is solved for with Newton's 
method,which makes use of our Jacobian calculation.  Newton's method is applied to obtain the time $\href{https://github.com/ericagol/NbodyGradient.jl/blob/cb2fabd04b1009d9e36d5b144ea544e6f282f1e2/src/transits/timing.jl\#L83}{t_{ijk}} = t_n + \Delta t$, where $\Delta t$ is the time after $t_n$ at which $g_{\mathrm{sky},i,j}= 0$, which is taken as the time of transit.  The initial guess for the time of transit, $\Delta t_\mathrm{init}$, is found by linear interpolation, 
\begin{eqnarray}
\href{https://github.com/ericagol/NbodyGradient.jl/blob/cb2fabd04b1009d9e36d5b144ea544e6f282f1e2/src/transits/timing.jl\#L20}{\Delta t_\mathrm{init}} =-\frac{g_{\mathrm{sky},ij}(t_n)h}{g_{\mathrm{sky},ij}(t_{n+1})-g_{\mathrm{sky},ij}(t_n)}.
\end{eqnarray}

To implement Newton's method, the system is integrated in between these time steps with a single {\sc AHL21} step, 
but with a time $\Delta t < h$, instead of $h$, giving ${\bf q}(t_n+\Delta t)$.  
From these coordinates, 
$g_{\mathrm{sky}}(t_n+\Delta t)$ is computed between the two bodies, and refined using
\begin{eqnarray}
\href{https://github.com/ericagol/NbodyGradient.jl/blob/cb2fabd04b1009d9e36d5b144ea544e6f282f1e2/src/transits/timing.jl\#L62}{\delta \Delta t} = -g_{\mathrm{sky}} \left(\frac{d g_{\mathrm{sky}}}{d t_{ijk}}\right)^{-1},
\end{eqnarray}
where 
\begin{eqnarray}
\href{https://github.com/ericagol/NbodyGradient.jl/blob/cb2fabd04b1009d9e36d5b144ea544e6f282f1e2/src/transits/timing.jl\#L146-L149}{\frac{d g_{\mathrm{sky}}}{d t_{ijk}}} &=& x_{ij}\left(\frac{d v_{x,i}}{d \Delta t}-\frac{d v_{x,j}}{d \Delta t}\right)
+ y_{ij}\left(\frac{d v_{y,i}}{d \Delta t}-\frac{d v_{y,j}}{d \Delta t}\right)\cr 
&+& 
v_{x,ij}\left(\frac{d x_i}{d \Delta t}-\frac{d x_j}{d \Delta t}\right)
+ v_{y,ij}\left(\frac{d y_i}{d \Delta t}-\frac{d y_j}{d \Delta t}\right),
\end{eqnarray}
and $x_{ij} = x_i-x_j$, etc., and the time derivatives with respect to $\Delta t$ are computed with equation (\ref{eqn:dstepdt}).  Note that in practice since the integration time step, $h$, is fixed, for the transit time derivatives $\delta t_{ijk} = \delta (\Delta t)$.


Once a transit time is found, how does it vary with the initial conditions?  We focus
on the initial conditions just before transit at time $t_n$, $\mathbf{q}_n=\mathbf{q}(t_n)$.
If $\mathbf{q}_n$ is perturbed slightly, then the time of transit will
change, but the new time of transit must still satisfy $g_{\mathrm{sky},ij}(t_{ijk} + \delta \Delta t) = 0$, where the new time of transit is at $t_n + \Delta t 
+ \delta (\Delta t)$.  So,
\begin{eqnarray}
    \frac{\partial g_{\mathrm{sky}}}{\partial \mathbf{q}_n}  \delta (\mathbf{q}_n)+ \frac{d g_{\mathrm{sky}}}{d t} \delta (\Delta t) = 0,
\end{eqnarray}
where we have dropped the $i,j,k$ subscripts from $t$ in this equation.  

Thus the gradient of each transit time with respect to the state, $\mathbf{q}_n=\mathbf{q}(t_n)$,
at the beginning of the $n$th time step just preceding the transit 
is given by
\begin{eqnarray}
    \href{https://github.com/ericagol/NbodyGradient.jl/blob/cb2fabd04b1009d9e36d5b144ea544e6f282f1e2/src/transits/timing.jl\#L168-L169}{\frac{d (\Delta t)}{d \mathbf{q}_n}} &=& -\left(\frac{d g_{\mathrm{sky}}}
    {d t_{ijk}}\right)^{-1}\left[\left(\frac{\partial x_i}
    {\partial \mathbf{q}_n}-\frac{\partial x_j}
    {\partial \mathbf{q}_n}\right)(v_{x,i}- v_{x,j})\right.\notag \\
    &+&\left(\frac{\partial v_{x,i}}{\partial \mathbf{q}_n}-\frac{\partial v_{x,j}}
    {\partial \mathbf{q}_n}\right)(x_i-x_j)\notag \\
    &+&\left(\frac{\partial y_i}{\partial \mathbf{q}_n}-\frac{\partial y_j}{\partial \mathbf{q}_n}\right)(v_{y,i}-v_{y,j})\notag \\
    &+&\left. \left(\frac{\partial v_{y,i}}{\partial \mathbf{q}_n}-\frac{\partial v_{y,j}}{\partial \mathbf{q}_n}\right)(y_i-y_j)
    \right],
\end{eqnarray}
where the gradients are computed over the partial time step, so that,
for example, $\frac{\partial x_j}{\partial \mathbf{q}_n}$ is the component
of  $\mathbf{J}_\mathrm{AHL21}(\Delta t)$ associated with the $x$-component
of body $j$. Note again that the transit time is 
$t_{ijk} = t_0 + nh + \Delta t$, but since
$h$ and $t_0$ are fixed, $d t_{ijk}/d \mathbf{q}_n = d \Delta t/
d \mathbf{q}_n$.

In addition to the transit time, $t_{ijk}$, it is also useful to compute
the sky velocity, $v_{\mathrm{sky},ijk}$, and the impact parameter squared, $b^2_{\mathrm{sky},ijk}$,
at the time of transit.  These can be used to compute transit light curves, as
well as measure the variation of the impact parameter and duration as
a function of time as additional dynamical constraints on a system.  
These two quantities are given by:
\begin{eqnarray}
\href{https://github.com/ericagol/NbodyGradient.jl/blob/cb2fabd04b1009d9e36d5b144ea544e6f282f1e2/src/transits/timing.jl\#L152}{v_{\mathrm{sky},ijk}(\mathbf{q}_n,t_{ijk})} &=& \sqrt{(v_{x,i}-v_{x,j})^2+(v_{y,i}-v_{y,j})^2},\\
\href{https://github.com/ericagol/NbodyGradient.jl/blob/cb2fabd04b1009d9e36d5b144ea544e6f282f1e2/src/transits/timing.jl\#L151}{b^2_{\mathrm{sky},ijk}(\mathbf{q}_n,t_{ijk})} &=& (x_{i}-x_{j})^2+(y_{i}-y_{j})^2,
\end{eqnarray}
where there is a direct dependence upon $\mathbf{q}_n$ which is
propagated to the time of transit within the timestep, $\Delta t$,
and there is an indirect dependence upon $\mathbf{q}_n$ through
the fact that these are evaluated at the time of transit, $t_{ijk}(\mathbf{q}_n)$.

Taking the derivative of these with respect to $\mathbf{q}_n$ gives:
\begin{eqnarray}
\href{https://github.com/ericagol/NbodyGradient.jl/blob/cb2fabd04b1009d9e36d5b144ea544e6f282f1e2/src/transits/timing.jl\#L185}{\frac{d v_{\mathrm{sky},ijk}}{d \mathbf{q}_n}} &=& \frac{\partial v_{\mathrm{sky},ijk}}{\partial \mathbf{q}_n} + \frac{d v_{\mathrm{sky},ijk}}{d \Delta t} \frac{\partial \Delta t}{\partial \mathbf{q}_n},\\
%
\href{https://github.com/ericagol/NbodyGradient.jl/blob/cb2fabd04b1009d9e36d5b144ea544e6f282f1e2/src/transits/timing.jl\#L185}{\frac{\partial v_{\mathrm{sky},ijk}}{\partial \mathbf{q}_n}} &=& v_{\mathrm{sky},ijk}^{-1}
\Bigg[(v_{x,i}-v_{x,j})\left(\frac{\partial v_{x,i}}{\partial \mathbf{q}_n}-\frac{\partial v_{x,j}}{\partial \mathbf{q}_n}\right)\cr 
&+& (v_{y,i}-v_{y,j})\left(\frac{\partial v_{y,i}}{\partial \mathbf{q}_n}-\frac{\partial v_{y,j}}{\partial \mathbf{q}_n}\right)\Bigg],\\
%
\href{https://github.com/ericagol/NbodyGradient.jl/blob/cb2fabd04b1009d9e36d5b144ea544e6f282f1e2/src/transits/timing.jl\#L178}{\frac{d v_{\mathrm{sky},ijk}}{d \Delta t}} &=& v_{\mathrm{sky},ijk}^{-1}
\Bigg[(v_{x,i}-v_{x,j})\left(\frac{d v_{x,i}}{d \Delta t}-\frac{d v_{x,j}}{d \Delta t}\right) \cr 
&+& (v_{y,i}-v_{y,j})\left(\frac{d v_{y,i}}{d \Delta t}-\frac{d v_{y,j}}{d \Delta t}\right)\Bigg],\\
\href{https://github.com/ericagol/NbodyGradient.jl/blob/cb2fabd04b1009d9e36d5b144ea544e6f282f1e2/src/transits/timing.jl#L187}{\frac{d b^2_{\mathrm{sky},ijk}}{d \mathbf{q}_n}} &=& 2
\Bigg[(x_{i}-x_{j})\left(\frac{\partial x_{i}}{\partial \mathbf{q}_n}-\frac{\partial x_{j}}{\partial \mathbf{q}_n}\right) \cr 
&+& (y_{i}-y_{j})\left(\frac{\partial y_{i}}{\partial \mathbf{q}_n}-\frac{\partial y_{j}}{\partial \mathbf{q}_n}\right)\Bigg].
%
\end{eqnarray}
  Note that we compute
$b^2_{\mathrm{sky}}$ rather than $b_{\mathrm{sky}}$ to avoid the problem that when the orbits are edge-on, the impact parameter is zero at
mid-transit, causing the derivative of $b_\mathrm{sky}$ to be divided by $b_\mathrm{sky}=0$, which results in a NaN.


This completes the computation of all of the Jacobians needed to propagate the derivatives
of the transit times, and sky velocity/impact parameter, through to the initial conditions, 
which we describe next.

\subsection{Jacobians of positions, velocities, transit times}\label{sec:jacobian_total}

With the Jacobians computed at each of the steps, we can recursively
compute the Jacobian at step $n$ with $t=t_0+nh$ as
\begin{eqnarray}
    \mathbf{J}_{n} = \mathbf{J}_\mathrm{AHL21}(h) \mathbf{J}_{n-1}.
\end{eqnarray}
Starting with the initial state $\mathbf{q}_0 = \mathbf{q}(t_0)$ and initial Jacobian
$\mathbf{J}_0= \frac{\partial \mathbf{q}_0}{\partial \mathbf{q}_0}
= \mathbf{I}$ (the identity matrix),
we iteratively compute the Jacobian at step $n$ with respect to the
state at initial time, $t_0$ ($n=0$), giving the
Jacobian transformation from $\mathbf{q}_0$ to $\mathbf{q}_n$,
\begin{eqnarray}
\mathbf{J}_{n} = \frac{\partial \mathbf{q}_n}{\partial \mathbf{q}_0}.
\end{eqnarray}\\
Then, the gradient of the transit times is given by
\begin{eqnarray}
    \frac{d \Delta t}{d \mathbf{q}(t_0)} = \frac{d \Delta t}
    {d \mathbf{q}(t_n)} \mathbf{J}_n.
\end{eqnarray}
We save this gradient for each transit time in an array that is pre-allocated
when calling the routine.   

{In our implementation, we do not compute $\mathbf{J}_\mathrm{AHL21}(h)$ for each step, directly;  instead, we iteratively multiply the current Jacobian by the Jacobian for each sub-step.}



{This completes the description of the algorithm and its derivatives. We now turn to the implementation and testing of the code.}

\section{Implementation and testing} \label{sec:implementation}
We have developed {\texttt{NbodyGradient.jl}}\footnote{\url{http://github.com/ericagol/NbodyGradient.jl}} in the Julia language for carrying out the foregoing computations. This involves the initialization of the algorithm, the $N$-body integration, the finding of transit times, and the Jacobian propagation.  Given the complicated nature of the calculations, we have written unit tests for each of the steps in the algorithm;  these were critical in developing the code for computing the derivatives, and helped to pinpoint inaccuracies in the DH17 algorithm which led to developing the {\sc AHL21} algorithm.  We have also created tests of the code as a whole, and carried out comparisons with other codes for both speed and accuracy, which are summarized here.

In this section we describe some aspects of the implementation  of the code (\S \ref{sec:julia}) and the tests we have carried out.   We test the $N$-body algorithm for accuracy by varying the step size and checking for conservation of energy and angular momentum (\S \ref{sec:energy_angmom}), while we check the transit-time algorithm for accuracy by  measuring the variation in transit times with step size (\S \ref{sec:tt_accuracy}).
We compare the $N$-body integrator with a C implementation to check for speed (\S \ref{sec:c_comparison}).
We check the numerical precision of the code by carrying out comparisons with extended precision (\S \ref{sec:transit_precision}), and we check the accuracy of the derivatives by comparing with finite-differences carried out in extended precision (\S \ref{sec:transit_precision}).
Most of our tests are carried out with integrations of the outer Solar System and of the TRAPPIST-1 system.

We start by describing the implementation of the algorithm in Julia.

\subsection{Julia implementation} \label{sec:julia}

We chose the Julia language to develop this code \citep{Bezanson2017},
given its several advantages. The high-level, 
interactive (``REPL") capability can make debugging code more straightforward.
The just-in-time compiler can make the code execution competitive with compiled
C, if attention is paid to memory allocation and type stability.  An advantage of Julia for testing code accuracy is that different numerical types can easily be changed which allows for straightforward computation at different precisions.  Julia also uses multiple dispatch which allows us to automatically select versions of functions that match in precision, and gives us control over computing gradients, transit times, and other outputs. Finally, Julia is open-source, and thus amenable to distribution and usage amongst scientists.

We have optimized the code keeping in mind several unique aspects of Julia.
Memory allocation and garbage collection were minimized by defining arrays
at higher levels which were then passed to subroutine functions to avoid
repeated allocation of large arrays.  Matrix multiplication can be sped up
by utilizing the BLAS linear algebra routines \citep{Blackford2002}, specifically \texttt{gemmv}, which
gave significant reduction in run-time for the multiplication of Jacobians at each
step.  For multiplication of the {$\mathbf{J}_{\mathrm{DK/KD},ij}$} Jacobian, we found
more efficiency by copying the portion of $\mathbf{J}_n$ (times the prior
substeps) relevant to bodies $i$ and $j$ to a new $14\times 7N$ matrix, then 
using the BLAS routine to carry out the multiplication, and then copying the result
back into $\mathbf{J}_n$.  For loops in which we access elements of arrays
successively, we try to step through elements which are adjacent in memory, and we also avoid index-checking to save time.
Finally, we try to avoid changing the types of variables, and we define the
types up front to make this explicit.  Thanks to these details, we find a favorable run-time comparison with C (\S \ref{sec:c_comparison}).

Another aspect of our implementation is that the code is simple to use and extendable. Here is an example of running the integrator and computing transit times interactively from the Julia prompt (\texttt{REPL}), or from within a Jupyter or Pluto notebook\footnote{Further details on running the code can be found in the documentation at \url{http://github.com/ericagol/NbodyGradient.jl}} (a slightly modified version of the script used in the comparisons with other codes in \S \ref{sec:comparison}):

{\Large
\begin{jllisting}[xleftmargin=0.0pt, xrightmargin=0.53\textwidth, frame=lines, rulecolor=\color{black}]
using NbodyGradient

# Set up initial conditions from file of orbital elements
# (Initial time, # of bodies, orbital elements file)
ic = ElementsIC(0.0, 8, "elements.txt") 
# Time step. Period of planet b / 100
h = ic.elements[2,2]/100 
# Set up integrator (time step, initial time, elapsed time)
intr = Integrator(h, 0.0, 4533.0) 
# Compute and store initial Cartesian coordinates
s = State(ic) 
# Allocate arrays for transit times and derivatives
tt = TransitTiming(intr.tmax, ic)  
# Run integration & compute transit times w/ derivatives.
intr(s, tt) 
\end{jllisting}}

\noindent{Here, the initial conditions are specified by a file containing rows of orbital elements, \texttt{elements.txt}\footnote{We leave discussion of initial conditions to future work, but for sake of completeness the elements file is set up as follows. Each row is given by the mass, period, time of initial transit, eccentricity vector components, Inclination, and Long. of Ascending Node for each body. The columns are delimited by a comma (','). The eccentricity vector is defined by ($e$cos($\varpi$), $e$sin($\varpi$)) where $e$ is the eccentricity and $\varpi$ is the long. of periastron.  The orbital elements are given in Jacobi coordinates which are converted to Cartesian coordinates to start the integration.}. An integration is triggered by passing a \texttt{State} type (\texttt{s} in the example) to an \texttt{Integrator} (\texttt{intr} in the example), along with any “output” type (\texttt{tt} in the example): \texttt{intr(s,tt)}. The \texttt{State} holds the Cartesian coordinates and Jacobian which are updated at each step. Passing the output structure \texttt{tt} of type \texttt{TransitTiming} tells the integrator to compute transit times of the system and store the results in the \texttt{tt} structure.  The transit times can be accessed within the structure as \texttt{tt.tt}, which is a two dimensional array of size $N$ by $N_{tt}$, which holds the transit times for each planet, each of which have a count in the vector \texttt{tt.count} with a maximum allowed value of $N_{tt}$.  The derivatives with respect to the initial Cartesian coordinates and masses are stored as \texttt{tt.dtdq0} which is a 4-dimensional array with the same first two dimensions as \texttt{tt.tt}, and the last two dimensions of sizes $7$ and $N$ which hold the derivatives with respect to $x$, $v$, and $m$ for each body ($\mathbf{q}_0$).
By utilizing multiple dispatch, adding functionality to the code consists of simply making a new \texttt{Integrator} method and a structure to hold the output.}



Next, we describe the accuracy of the $N$-body algorithm by checking conservation of energy and angular momentum.

\subsection{Energy and angular momentum conservation} \label{sec:energy_angmom}

To test the accuracy of the algorithm, 
we have carried out integrations of the outer Solar System.   We start with positions, velocities, and
masses given in \citet{Hairer2006}, only including the giant planets (Jupiter, Saturn,
Uranus and Neptune).  The mass of the Sun is added to the sum of the masses of the terrestrial
planets for a fifth inner body.

\begin{figure}
    \centering
    \includegraphics[width=0.49\textwidth]{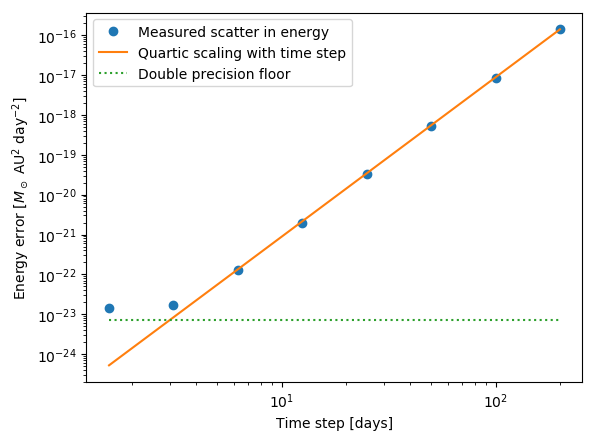}
    \includegraphics[width=0.49\textwidth]{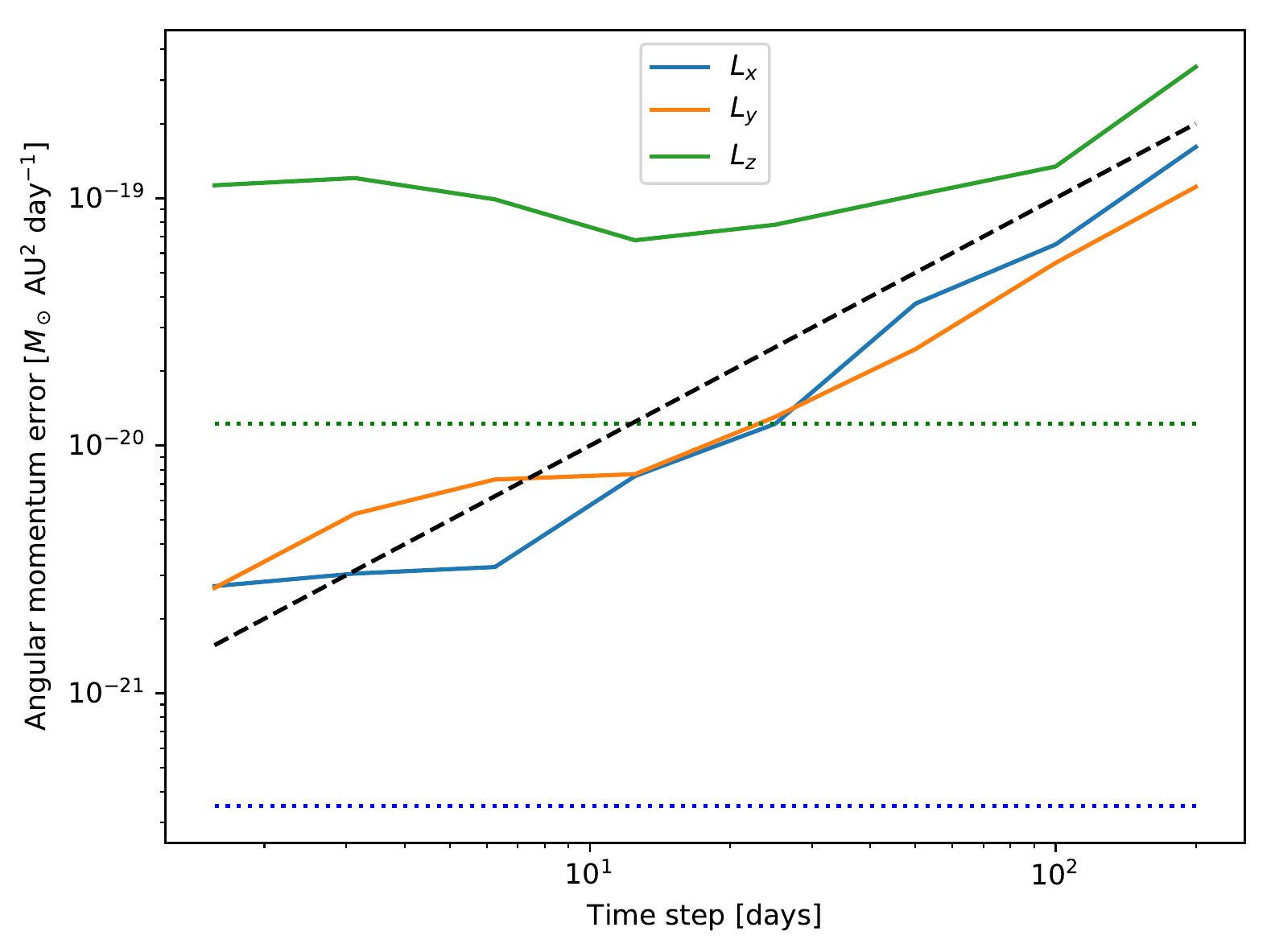}
    \caption{
    (Top) Conservation of energy.  Standard deviation of energy versus step size (blue dots).  Orange line shows $h^4$ scaling.  Green dotted line shows double-precision limit. (Bottom) Angular momentum conservation.  The standard deviation of each component is plotted versus step size.  
     The dotted lines shows the double-precision limit for angular momentum ($L_y$ is zero, so this is not shown).  The black dashed curve shows that angular momentum error scales $\propto h$ according to Brouwer's law. 
    }
    \label{fig:energy_conservation}
\end{figure}

We compute the total energy and angular momentum
of the system as a function of time.  We measured the RMS value for the energy
and all three components of the total angular momentum, and varied the time
steps by factors of two.  We expect the energy precision to scale with time 
step to the fourth power, $\propto h^4$. Figure \ref{fig:energy_conservation} shows that this
scaling holds over a range of two orders of magnitude in the time step.  We used time steps 
from 1.5625 to 200 days, and the RMS energy and angular momentum was measured
over $\approx 10^6$ time steps in each case.  The upper end of the time steps was set by
the requirement that the time step be smaller than 1/20 of the shortest orbital
period, in this case Jupiter.  At the lower end of this range (1.5625 and
3.125 days) we see a deviation from the $h^4$ scaling thanks to the limit of
double-precision representation of the energy.  This occurs at approximately
$2^{-52} = 2.2\times 10^{-16}$ of the absolute error value of each conserved quantity, which is plotted
as dotted lines in each panel in Figure \ref{fig:coordinates} (the energy
is $\approx -3\times10^{-8} M_\odot \mathrm{AU}^2 \mathrm{day}^{-2}$, while
the total angular momentum is $\approx 6\times 10^{-5} M_\odot \mathrm{AU}^2
\mathrm{day}^{-1}$).  The three
components of angular momentum show a better conservation precision which
is close to double-precision for all values of the time step.  Numerical errors 
accumulate with time step, and the expectation is that these scale as 
$\approx \epsilon_\mathrm{err} h^{1/2} t^{1/2} = \epsilon_\mathrm{err} h N_S^{1/2}$, where $N_S$ is the 
number of time steps and $\epsilon_\mathrm{err}$ is a random numerical error \citep{Hairer2008}.   
This is also borne out in Figure \ref{fig:energy_conservation} which {shows} a scaling of 
the error with time step $h$ ($N_S$ is held fixed in these integrations).

Our conclusion is that the numerical integration
is behaving as expected:  energy is conserved with an accuracy $\propto h^4$
above the double-precision limit, and angular momentum is conserved close to
double-precision, but grows according to Brouwer's law.   Note that the RMS relative error (defined as $\Delta E/E_0$, with $E_0$ the initial energy, and $\Delta E$ the change in this energy) measures an oscillation which can be orders of magnitude larger than the mean relative error over time. 

Given this evidence of accurate behavior of the $N$-body algorithm, we next ask:  how does the $N$-body implementation fare in computation time? 

\subsection{Comparing with C implementation} \label{sec:c_comparison}

To check that we have optimized the computational speed of the $N$-body integrator, 
we carried out a comparison of the Julia version of our code with a C
implementation without derivatives \citep{Hernandez2016}.  First, we compared the Kepler solver \citep{Wisdom2015} and found that our Julia implementation matches the C version.  Both versions take 0.15 $\mu$ sec per Kepler step for a bound orbit with $e=0.5$.\footnote{\label{cpu}These comparisons were made on a Macbook with a 2.8 GHz Intel Core i7 processor with Julia v1.6.  The C code was compiled with \texttt{cc -O3}.}  Note that in this comparison we used a version of the Kepler step which is not combined with a backward drift.

Next, we carried out an integration of the outer solar system (with 5 bodies, as described in the prior section) with the C implementation.  We found that the C implementation runs at the same speed as our Julia implementation of {\sc AHL21}.  With 50-day time steps, both versions take about 4.7 $\mu$sec per timestep.\footnote{See footnote \ref{cpu}.}  In this comparison, we use the same convergence criterion for the Kepler solver; when we include the 4th-order corrector in {\sc AHL21} it increases the run time by $\approx$ 10\% for the outer solar system problem. Thus, we conclude that the speed of the Julia implementation is comparable to compiled C.

When using the convergence that a fractional tolerance of $10^{-8}$ is reached for the solution to Kepler's equation -- we find a bias in the long-term energy conservation which causes it to drift with time.  If instead we use the convergence criterion that the eccentric-anomaly, $\gamma$, remains unchanged relative to one of the prior two iterations of Newton's solver -- then we find that this bias is significantly reduced.  This adds iterations to the Kepler-solver, typically 1-2, and thus causes the code to take about 10\% longer to run, but with the trade-off of better energy conservation.  Thus, using the 4th-order corrector and this criterion adds about 20\% to the overall run-time, amounting to 5.7 $\mu$sec per time-step compared with the example above.

Now that we have verified the speed and accuracy of the $N$-body algorithm, we next examine the accuracy and precision of the transit times as a function of step size.

\subsection{Transit-timing accuracy} 
\label{sec:tt_accuracy}

Since the numerical accuracy of the integration depends upon the step
size parameter, $h$, as the {\sc AHL21} integrator is fourth order
in $h$ (Fig. \ref{fig:energy_conservation}), we also expect that the accuracy of transit times should scale with $h^{4}$.
Further precision could be obtained if we were to use a corrector at the
start of the integration;  however, such a corrector is left for future work.  

\begin{figure}
    \centering
    \includegraphics[width=0.49\textwidth]{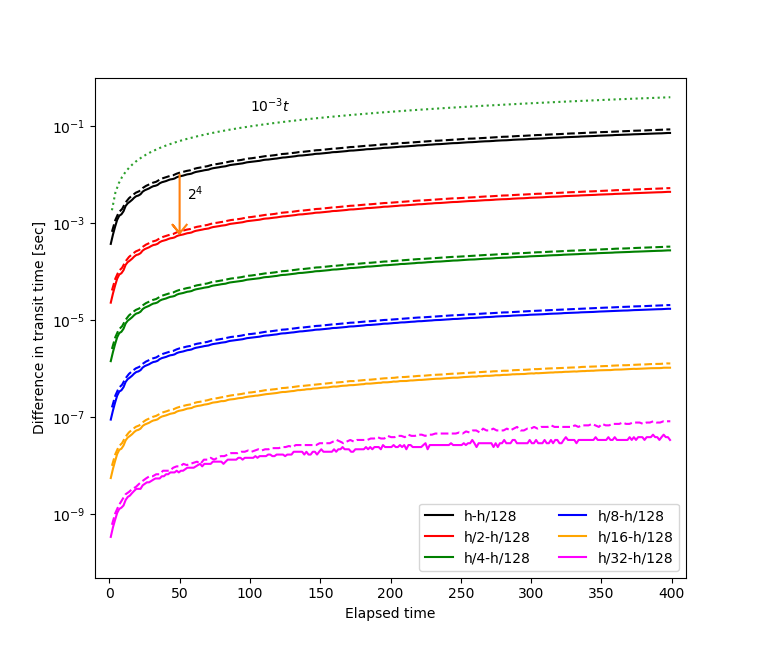}
    \includegraphics[width=0.49\textwidth]{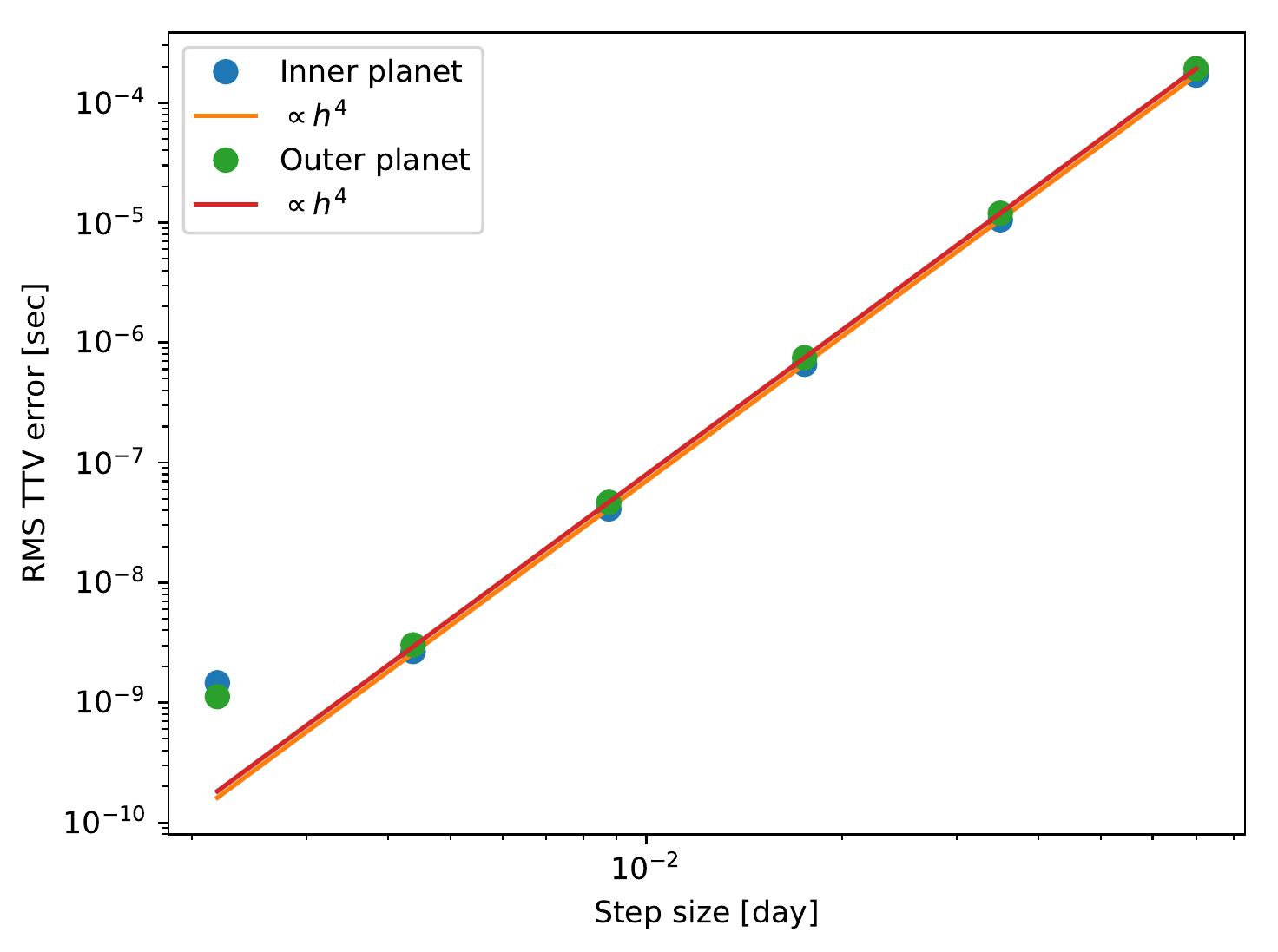}
    \caption{(Left) Variation in the transit times with step size.  The absolute value of the difference in times between the indicated step size and a step size of $h/128$ is plotted for the inner two planets of TRAPPIST-1 (solid and dashed, respectively) simulated over 400 days with a nominal step size of $h=0.06$ days.  The dotted curve shows a linear scaling, while the arrow indicates a $1/2^4$ decrease in timing difference when the step-size is halved. (Right) RMS precision of TTVs for the TRAPPIST-1 b/c two-planet system with six stepsizes, 
    compared with an integration which is $1/2$ of the shortest stepsize, after 
    removing a linear fit from the difference to isolate the comparison to the transit-timing variations.  For both planets the TTV errors scale as $h^4$ until double-precision is reached at $\approx 10^{-9}$ seconds.}
    \label{fig:timing_accuracy}
\end{figure}

Figure \ref{fig:timing_accuracy} shows the change in the transit times with stepsize for a simulation of TRAPPIST-1 b and c over 400
days.  Compared with the times computed with a very small step-size, the transit times drift with time.  This behavior is expected due to the difference between the symplectic Hamiltonian and the full Hamiltonian which contains high-frequency terms which cause the coordinates of the symplectic integrator to be offset from the real coordinates.  This offset causes the longitudes to drift with time, and due to the slight difference in orbital frequency, the drift grows linearly as shown in Figure \ref{fig:timing_accuracy}.  Since the AHL21 algorithm has order $h^4$, these offsets decrease with step-size as $h^4$ (Fig.\ \ref{fig:timing_accuracy}), and so at small stepsize the symplectic integrator better approximates the real system.  In practice these coordinate offsets are not expected to be important for transit-timing analyses as they will lead to very small differences between the inferred parameters and real parameters, even for large $h$.  We recommend that the user determine which step size is appropriate by checking the difference in transit times as a function of step size.

For the purposes of transit-timing variations, we
are primarily interested in the precision of the non-linear portion
of the transit times versus epoch.   So, to assess the TTV precision,
we subtract a linear fit from the difference between an integration
with large $h$ with an integration with small $h$, and then compute
the RMS of the residuals.  We expect the RMS to scale as $h^{4}$, and
figure \ref{fig:timing_accuracy} indeed shows that this is the case
for large step size for a system with two planets of periods 1.5 and 2.4 days, 
masses $3 \times 10^{-5}$ of the star, low eccentricity, and integrated
over 400 days (this approximates the inner two planets of the TRAPPIST-1 system).  In both cases the TTV precision reaches
a value that is $\approx 10^{-14}$ of each planet's orbital period.  We also find that this precision scales in proportion to the ratio of the masses of the planets to the star, as expected (see discussion under eq. \eqref{eq:hamilt}).

Having demonstrated that the accuracy of the transit times scales as expected, we next examine the numerical precision of the computed times, as well as their derivatives.

\subsection{Precision of transit times and their derivatives} \label{sec:transit_precision}

Given a fixed step size for the algorithm, we next ask the question:  how numerically precise are the transit times computed for that step size?   And, how precise are the derivatives computed as a function of the initial conditions?  These questions involve the control of truncation and round-off errors in the algorithm, which motivated the development of the {\sc AHL21} algorithm.

\begin{figure}
    \centering
    \includegraphics[width=0.49\textwidth]{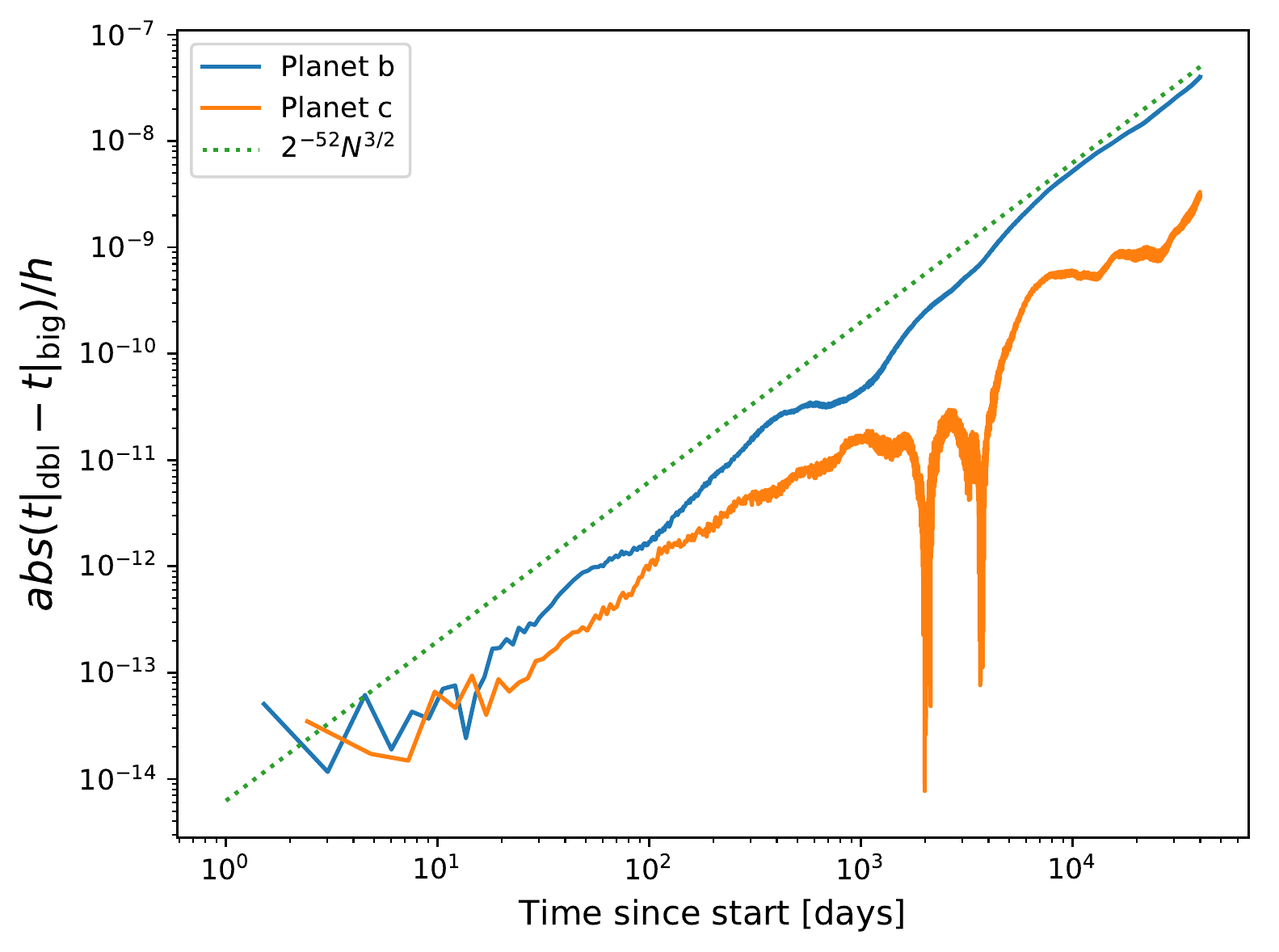}
        \includegraphics[width=0.49\textwidth]{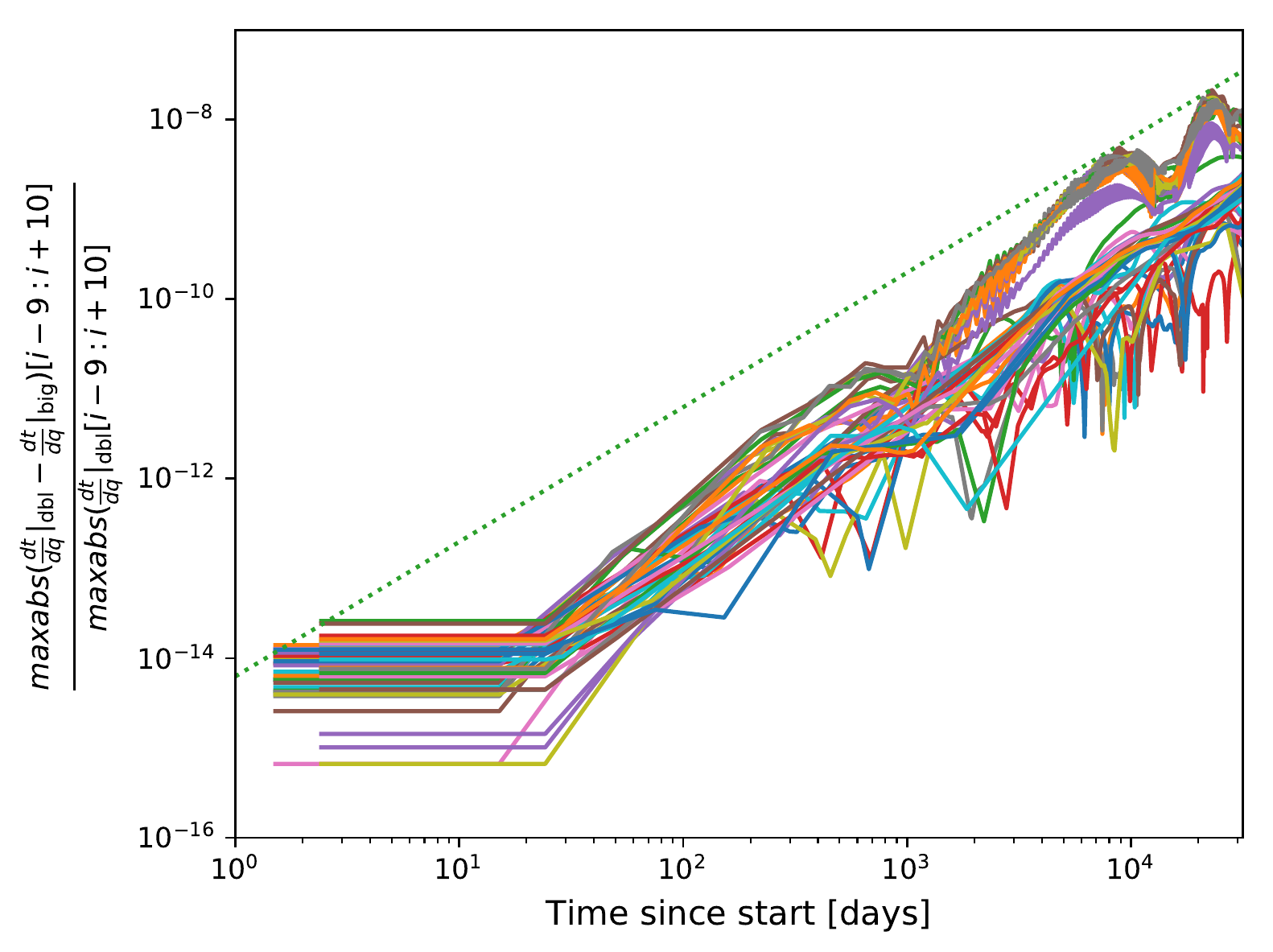}
    \caption{(Left) Fractional numerical error of transit times computed over 40,000 days for TRAPPIST-1 b and c computed from double and BigFloat integrations.  The error is plotted relative to the time ste, $h$.  (Right) Fractional numerical error on the transit time derivatives with respect to the initial cartesian coordinates and masses.  The dotted green lines in both panels scale as $2^{-52} N_S^{3/2}$, where $N_S$ is the number of time steps.  The maximum absolute derivative differences have been taken over 20 steps, and divided by the maximum absolute derivatives to give the fractional differences.  
    }
    \label{fig:timing_precision}
\end{figure}

We check the numerical precision of the algorithm by comparing the transit times and their derivatives computed at both double-precision and extended precision (using the double-precision \texttt{Float64} type with 64 bits, and the extended-precision \texttt{BigFloat} type with 256 bits in \texttt{Julia}).  Figure \ref{fig:timing_precision} shows the difference in the  times of transit
in the TRAPPIST-1 b and c case computed in double-precision relative to \texttt{BigFloat} precision.   We find that the computational errors grow at a rate that is bounded by $\approx 2^{-52}hN_S^{3/2}$, where $N_S$ is the number of elapsed time steps 
(Fig.\ \ref{fig:timing_precision}), as expected for phase-errors based on Brouwer's Law \citep{Brouwer1937}.  The computation was carried out for 400,000 days for an inner orbital period of 1.5 days, for a total of $\approx 10^7$ time steps.

Next, we carry out tests of the numerical precision of the Jacobians
at each substep in the calculation, as well as for the entire integration
interval, and for the transit time derivatives.  We do this in two ways:  1). by computing finite-difference derivatives in extended precision and 2). by comparing the derivatives in double precision with derivatives computed with extended precision arithmetic.  The finite-difference test checks that the formulas derived in \S \ref{sec:differentiation_symplectic} are valid, while the extended precision test checks that the numerical implementation is precise. 

To compute the finite-difference
derivatives, we carry out integrations for each parameter using \texttt{BigFloat},
and compute a finite difference approximation of the partial derivative with parameters perturbed just above and below the nominal
value,
\begin{eqnarray}\label{eqn:finite_diff}
    \frac{\partial t}{\partial q_{i,j}} \approx \frac{\Delta t}{\Delta q_{i,j}} = \frac{t(q_{i,j} (1+\varepsilon_\mathrm{diff})) - t(q_{i,j}(1-\varepsilon_\mathrm{diff}))}{2\varepsilon_\mathrm{diff} q_{i,j}},
\end{eqnarray}
where $t(q_{i,j})$ indicates the transit time evaluated at initial conditions $\mathbf{q}_0$ with the $i,j$th initial condition given by $q_{i,j}$.
Typically we use $\varepsilon_\mathrm{diff} = 10^{-18}$, and we find that the finite-difference derivative is insensitive to this value when rounded back to double-precision.
We used these finite differences in writing and debugging each
substep of the code, and we have created a suite of tests which can be used when
further modifying or developing the code.  We found that the finite difference derivatives agree with the derivatives computed from propagating the Jacobian, at a level close to the double-precision limit, which validates our implementation of the algorithm based on the formulae in \S \ref{sec:differentiation_symplectic}.  

Next, we estimate the fractional numerical errors on the derivatives of the transit times with respect to the initial conditions and masses propagated through the numerical integration (Fig.\ \ref{fig:timing_precision}, right panel) by comparing the derivatives computed at double precision with those computed at extended precision.  We find that these double-precision numerical errors also shows a growth which is bounded by
$2^{-52}N_S^{3/2}$, according to Brouwer's Law.  In this case we filtered the derivatives before computing the fractional error taking the maximum absolute value of each derivative over 20 transit times normalized by the maximum absolute derivative over the same 20 times to avoid the case in which the values of the derivatives approach zero.  We did not find that the Brouwer's law limit applied to be the DH17 algorithm - the errors significantly exceeded the Brouwer's law limit for long integrations - which motivated the development of the {\sc AHL21} algorithm.
We did not carry out longer integrations due to the high computational expense for \texttt{BigFloat} precision.

We conclude that based on these tests the code is performing as expected:  the $N$-body code is fast and as accurate as the algorithm allows; the transit times are precise; the derivative formulae are correct; and the derivatives are precise.  With these validations of the code completed, we now turn to compare our code with other publicly-available $N$-body and transit-timing codes.

\section{Comparison with other codes} \label{sec:comparison}

In this section we compare with two existing open-source $N$-body integrators which
have been used for transit-timing and $N$-body integration:  \texttt{TTVFast} and \texttt{REBOUND}.  Although other codes
are available, such as \texttt{SYSTEMIC} \citep{Meschiari2010} and 
\texttt{TRADES} \citep{Borsato2014}, as well as numerous proprietary
codes for modelling transit timing, \texttt{TTVFast} and \texttt{REBOUND} are both widely-used and open-source. These comparisons provide further validation of the accuracy of our code, as well as timing benchmarks of the relative speeds.


\subsection{Comparison of transit times with TTVFast and REBOUND}

The TTVFast approach uses a Wisdom-Holman integrator \citep{Wisdom1991} with a central dominant body, appropriate for planetary systems orbiting a single star (or planets orbiting a single star in a wide binary).   A third-order corrector is used at the start of each simulation to transform from real coordinates to symplectic coordiates.   Two versions of TTVFast have been developed in FORTRAN and C;  here we describe comparisons with the latter.\footnote{\url{https://github.com/kdeck/TTVFast}} The initial conditions may be specified in either Jacobi or heliocentric orbital elements, or heliocentric Cartesian coordinates.  We use the initial Cartesian coordinates from NbodyGradient transformed to heliocentric coordinates, and then rotated by 180$^\circ$ about the $y$ axis so that the observer is located along the +$z$ axis, the convention adopted in TTVFast. 

The TTVFast algorithm uses an approximate method to find times of transit.  When a transit time is found to occur for one of the planets between two timesteps, then two Keplerian integrations 
between the planet and star are integrated forwards and backwards from the prior and subsequent timesteps, and weighted to approximate the position of the planet relative to the star.  Newton's method is then used to find the time of transit in the same manner described above (\S \ref{sec:transit_times}).

\begin{figure*}
    \centering
    \includegraphics[width=0.95\textwidth]{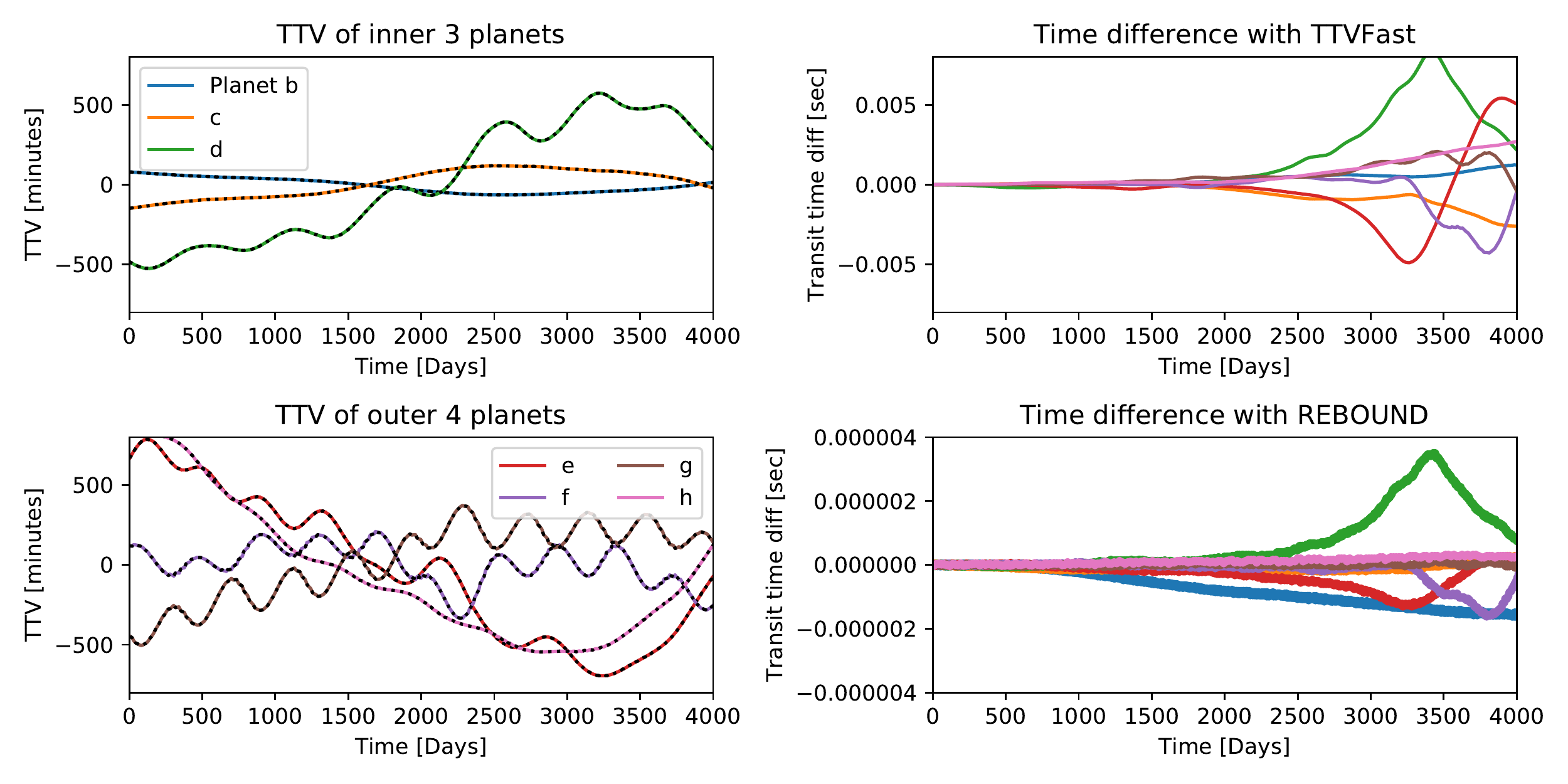}
    \caption{Transit-timing comparison for TRAPPIST-1.  (Left) TTVs from NbodyGradient (in color), and TTVFast and REBOUND (black dots) for the seven TRAPPIST-1 planets over 4000 days.  (Right) Timing differences in seconds between NbodyGradient and TTVFast (top) and REBOUND (bottom); the colors are same as in left hand panel. Note that the panel in the upper right has a range which has been expanded by  $6\times 10^6$ relative to the left panels, while the lower right panel is expanded by a factor of $1.2\times 10^{10}$. 
    }
    \label{fig:timing_comparison2}
\end{figure*}

We have made a comparison of the transit times from NbodyGradient with TTVFast using the best-fit initial conditions for the 7-planet TRAPPIST-1 system
\citep{Agol2021}.  For this comparison we use a time step for TTVFast which is 0.05\%\footnote{We also had to modify TTVFast to avoid accumulation of numerical errors which occur for such a small timestep.  Rather than adding the time step to the elapsed time every time step, we multiply the current number of steps by the time step to obtain the elapsed time.} of the orbital period of planet b (Figure \ref{fig:timing_comparison2}) to reduce the difference between the symplectic and real coordinates (we use a larger step of 0.1\% for NbodyGradient as this integrator is higher precision).  We find that over a timescale
of $\sim$4532 days (an estimate of the total time between initial and final TRAPPIST-1 observations over the lifetime of JWST), the difference between TTVFast and NbodyGradient is better than a few milliseconds for all seven planets, with better agreement for the inner planets than for the outer.  This agreement is quite good, and we attribute the remaining differences, which grow with time, as being due to differences between the initial mapping and real coordinates which cause phase errors to grow with time.

Although \texttt{REBOUND} is not primarily designed for transit-timing, there is a Python notebook in the REBOUND repository which gives an example of transit-time computation.\footnote{See \url{https://rebound.readthedocs.io/en/latest/ipython/TransitTimingVariations.html} for a description.}  We used the same initial cartesian coordinates as the NbodyGradient computation for TRAPPIST-1, and computed the transit times with a tolerance of $10^{-12}$ days for REBOUND.  We transform z$\rightarrow$-x, x$\rightarrow$y, and y$\rightarrow$-z to allow for the fact that the REBOUND computation places the observer along the x axis rather than along the -z axis (as assumed in NbodyGradient, Fig.\ \ref{fig:coordinates}).   Figure \ref{fig:timing_comparison2} shows that over 4000 days for TRAPPIST-1, the times agree between NbodyGradient and REBOUND at the  $<4\mu$sec level.  This was computed with a time step of 0.0015 days for NbodyGradient, about 1/1000 of the orbital period of the inner planet, TRAPPIST-1b, to reduce the difference between the symplectic and real coordinates.

Unfortunately TTVFast does not include derivatives, which was part of the
motivation for developing the NbodyGradient code.  However, given that the
transit times compare well, and that we have compared the NbodyGradient derivatives with finite-differences computed at high precision (\S \ref{sec:transit_precision}), this gives us confidence  that the NbodyGradient derivatives are also being computed accurately.  We have made scripts available for reproducing this comparison in the
NbodyGradient repository.


Next we compare the run-time of NbodyGradient with REBOUND, with and without gradients.

\subsection{Run-time Comparison with \texttt{REBOUND}}

\begin{figure}
    \centering
    \includegraphics[width=0.49\textwidth]{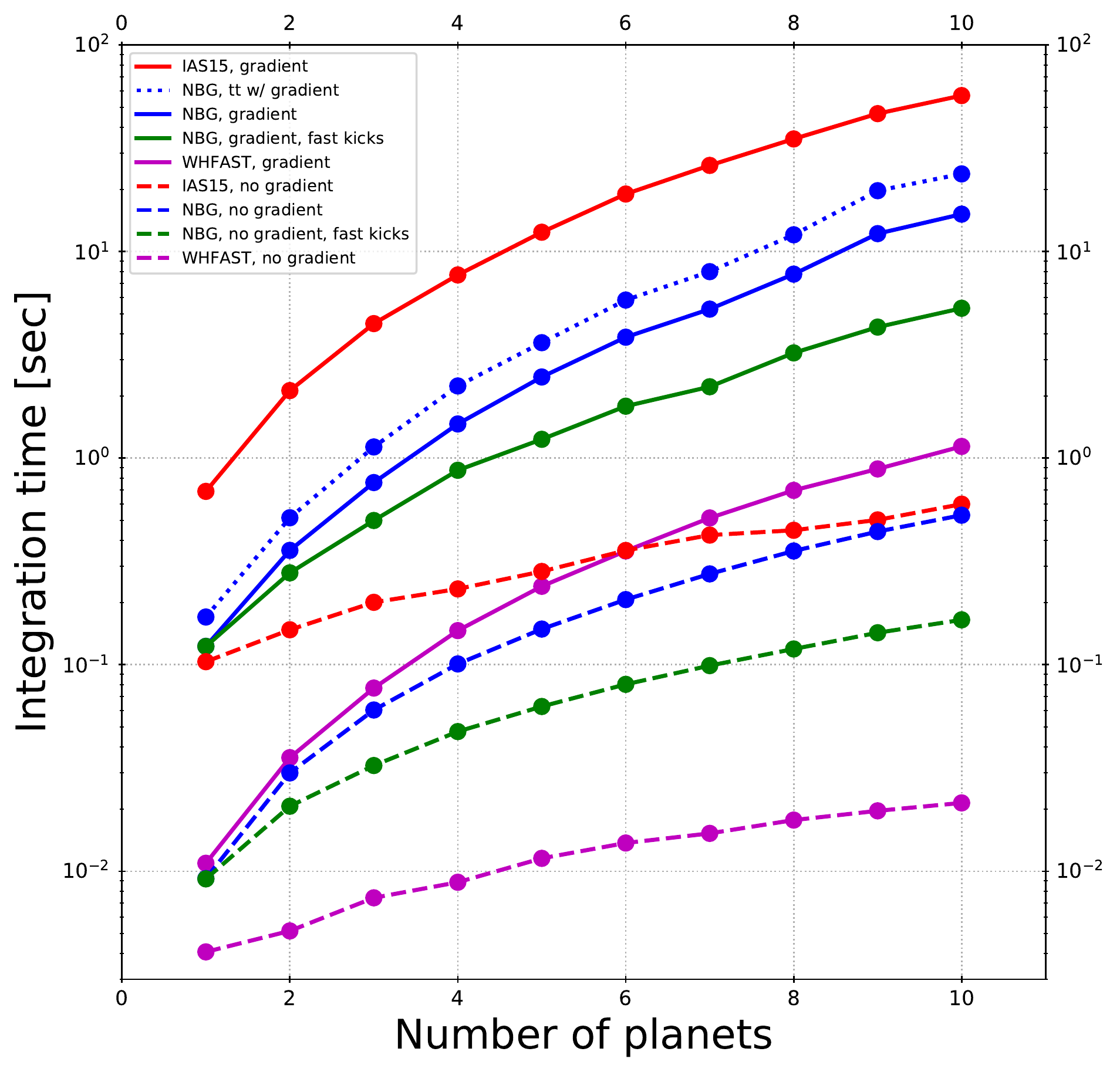}
    \caption{Computational time of {the NbodyGradient} code without (blue), with fast kicks for {planet pairs} (green), and with
    transit-time computation (magenta) versus \texttt{REBOUND} {\sc ias15} (red),
    with (solid) and without (dashed) computation of the gradient. {We also compare with the \texttt{REBOUND} WHFast algorithm (magenta), both with gradients (solid) and without (dashed).} 
    }
    \label{fig:rebound_vs_nbodygrad}
\end{figure}

The REBOUND integrator {\sc ias15} \citep{Rein2015} allows for the computation of
the variational equations, {and may be used to model systems with close encounters and an arbitrary architecture.} \footnote{\url{https://github.com/hannorein/rebound}}  Figure \ref{fig:rebound_vs_nbodygrad} compares the \texttt{REBOUND} {\sc ias15}
integrator computational speed with our code.   We spaced planets by
a ratio of semi-major axis of 1.8, and with initial orbital angles separated
by 1.4 radians.  For the {\sc AHL21} integration, we use a step size that is
1/20 of the orbital period of the inner planet and we integrate for
800 orbits of the inner planet.
We ran both codes with a range
of planets from one to ten, and we turned off the transit finding to
make a fair comparison.  We tried two different versions of the {\sc AHL21}
integrator:  with fast kicks for {pairs of planets}, and with
fast kicks turned off.  When the fast kicks are used for 
{planet pairs}, we find that the {\sc AHL21} code compares well to \texttt{REBOUND} {{\sc ias15} algorithm}
when no gradients are computed, either slightly faster or comparable
in wall clock time for 1-10 planets ({red and blue} dashed lines in Figure \ref{fig:rebound_vs_nbodygrad}).  However, when the gradient is computed,
our code takes a computational time that is $\approx 4-5$ times faster than \texttt{REBOUND} {{\sc ias15}}
for a large number of planets when fast kicks are turned off.  If the fast-kicks
are used for {planet pairs}, then NbodyGradient is an order of magnitude
faster than the {\sc ias15} integrator in \texttt{REBOUND}.  Note that both REBOUND gradients and NbodyGradient assume the Newtonian equations of motion when computing gradients.

{We also compare with the \texttt{WHFast} algorithm in \texttt{REBOUND}.  This algorithm is also symplectic, but requires a central dominant mass.  The \texttt{WHFast} algorithm is by far the speediest of the three: both
with and without gradients it is about an order of magnitude faster than either
{\sc AHL21} or {\sc ias15}.  However, as it is a second-order algorithm, it may require the use of a corrector, and/or shorter step-sizes, to obtain similar precision as the {\sc AHL21}
algorithm with is fourth-order;  this may come with extra computational cost, depending on the particular application.}  {From a theoretical standpoint, \emph{if} the Kepler solver function calls dominate the compute time, \texttt{WHFast} should only be twice as fast as 
{\sc AHL21} with kicks between planets in \texttt{NbodyGradient} (green dashed curve).  Since we have not achieved this, it may indicate we have not yet properly optimized our code.  Julia, \texttt{NbodyGradient}'s language, is believed to be able to achieve speeds comparable to C++, \texttt{WHFast}'s language.  We plan to continue to optimize \texttt{NbodyGradient.} }

Our primary goal in developing
this $N$-body code is for modeling observational data for which the uncertainties
are typically dominated by measurement errors rather than model accuracy.
Hence, we are willing to exchange some accuracy for computational speed by using a symplectic integrator with a large time step. {Thus, {\sc AHL21} may provide a useful compromise between {\sc ias15} and \texttt{WHFast}.}
Note that the integrator is still precise, but the symplectic Hamiltonian only approximates the real Hamiltonian, and we have not been able to derive a corrector to transform between symplectic and real coordinates; see the discussion
in \citet{Deck:2014} and references therein.

In {sum, the AHL21 algorithm in} NbodyGradient may gain
some computation time {for general N-body problems} with the tradeoff of interpreting the initial conditions as symplectic coordinates, not real coordinates.  In addition, currently \texttt{REBOUND} {(either {\sc ias15} or \texttt{WHFast}) }does not yet
implement the gradients of the transit times with respect to the initial
conditions, which NbodyGradient was designed to compute from the start.


\section{Summary and conclusions} \label{sec:conclusions}

The original goal of our development of this paper and code was to make possible the analytic computation of the derivatives of the times of transit
with respect to the initial conditions;  this is the first time this has been done in an $N$-body code to our knowledge.  We have accomplished this with a fast and robust code written in the Julia language.  Julia has the advantage of matching compiled C speeds if it is written in an optimum manner, which our benchmarks indicate we have achieved. Yet it allows for interactive usage and high-level coding which makes building and debugging the code more straightforward.  In addition, Julia easily allows changing of the variable types, so that higher-precision computations are available simply by calling the $N$-body functions with coordinates and masses initialized with high-precision variables; the functions will be automatically recompiled at the first time being called with different numeric types. 

We have also developed this code with generality in mind;  in particular, we would like to eventually apply it to hierarchical systems such as circumbinary planets or planets hosting moons, which is why we have based the symplectic splitting on the DH17 algorithm which does not assume a dominant body (or bodies).  {In addition, the popular Wisdom--Holman (WH) method assumes the perturbed approximation holds, which implies the method breaks down during close encounters.  The explanation, using error analysis, is that WH has two-body error terms.  However, when DH17 is used without kicks, it only has three-body terms which blow up during strong three-body encounters \citep{Dehnen2017}.  So our code can better handle close two-body encounters, as was shown by \cite{Dehnen2017}, who used it to simulate a stellar cluster.}

However, a drawback of the DH17 algorithm we found was the lack of precision caused by cancellations between the backward drifts and forward Kepler steps.  We have fixed this problem by carrying out analytic cancellation of these expressions with modified versions of Gauss's $f$ and $g$ functions.   This fix creates an algorithm which is both numerically stable and precise:  energy and angular momentum are conserved well for long integrations.  The algorithm is accurate to fourth order in time step, but even for large time steps it will integrate the non-Keplerian perturbations between the bodies with sufficient accuracy for observational data (and the time step can be decreased until the desired precision is reached;  this happens rapidly thanks to the fourth-order scaling of the algorithm with time step).  This accuracy is higher order than the Wisdom-Holman symplectic integrator or its versions with symplectic correctors, which have error terms scaling as $h^2$.  As with any symplectic integrator, the integration coordinates are offset slightly from the real coordinates \citep{Wisdom1996}, which causes a long-term phase shift.  However, this shift is small, even for large time steps, and  should not affect the interpretation of the state of multi-body dynamical systems.  In practice it results in a slight offset of the initial coordinates which is caused by introducing high-frequency terms in the physical Hamiltonian.

In order to compute the derivatives of the transit times with respect to the initial conditions, we have propagated  the Jacobian of the $N$-body positions and velocities  with respect to the initial conditions and masses throughout each  timestep of the $N$-body integration.  We have eliminated numerical cancellations in this expansion, and used series expansions for special functions when cancellation of the leading orders occurs.  This has given an algorithm which yields precise derivatives and which appears to adhere to Brouwer's law for up to $10^7$ timesteps for the problem we tested.  We also find that it compares favorably in run time to the variational equations integrated by {\sc ias15} \citep{Rein2016}, with a factor of 4-10 speed-up for long time steps in the comparison we tested.

We have found that the derivatives make possible the optimization
of the masses of planets in the TRAPPIST-1 planetary system, and the
results compare well with an analysis with the GENGA code \citep{Grimm2014}, as reported in \citet{Agol2021}. In particular,
we were able to use the derivatives to efficiently find the maximum likelihood
and to compute the Hessian at the maximum likelihood.  We have also used it to 
find the likelihood profile
as a function of the masses of the planets and orbital parameters, as well as to run a
Hamiltonian Markov Chain Monte Carlo computation in 35 dimensions to derive the posterior
distribution for the system parameters, which agree well
with a complementary analysis based on the code described in \citet{Grimm2018}.  Finally, the derivatives enabled an efficient search for an eighth planet, which required optimization over 40 free parameters;  no strong evidence for an eighth planet turned up in this search \citep{Agol2021}.

These analyses depend on initial conditions which were specified in terms of
orbital elements, which we plan to describe in subsequent work.  We are also continuing to develop the output options, API, and documentation, and we welcome community contributions to the code repository.\footnote{\url{http://github.com/ericagol/NbodyGradient.jl}}

There are some limitations to our work.  Due to its symplectic nature, our code does not allow for non-conservative effects to be included, such as tidal forces, drag, or general relativity.  However, these could be added in future work through the machinery described by \citet{Tamayo2020}.  We assume that the masses of the bodies are constant.  We do not compute second- or third- order derivatives, as has been implemented by \citet{Rein2016}.  Even so, we expect this code to find application in a wide range of dynamical problems related to observation of exoplanetary systems and beyond.  


\section*{Acknowledgments}

EA acknowledges support from the Guggenheim Foundation, from NSF grant AST-1615315, and from
the NASA Astrobiology Institute's Virtual Planetary Laboratory Lead Team,
funded through the NASA Astrobiology Institute under solicitation NNH12ZDA002C
and Cooperative Agreement Number NNA13AA93A. ZL acknowledges support from the Washington NASA Space Grant Consortium Summer Undergraduate Research Program. We thank Mos\'e Giordano for advice on optimizing Julia code, {and we thank Hanno Rein, Dan Tamayo, and the anonymous referee for comments on the submitted version which greatly improved the paper.}

\section*{Data Availability}

No new data were generated or analysed in support of this research.  Simulated data which was used for making the figures will be made available in a github repository.

\bibliographystyle{mnras}
\bibliography{bib}

\appendix

%
%
%
%

{\section{Derivation of {\sc AHL21} Kepler + Drift step} \label{sec:appendix1}}

{Here we give more detail for the derivation of the combined Kepler and Drift steps which were described in \S \ref{sec:kepler_drift}.   As a reminder, we start with two bodies $i$ and $j$ with relative coordinates $\mathbf{x}_{ij} = \mathbf{x}_i-\mathbf{x}_j$ and $\mathbf{v}_{ij} = \mathbf{v}_i-\mathbf{v}_j$. In this section we drop the subscript $ij$.  At the start of the time step, the coordinates are $\mathbf{x}_0$ and $\mathbf{v}_0$, while at the end of the time step the coordinates are $\mathbf{x}$ and $\mathbf{v}$.  We propagate these forward with either a negative drift followed by a Kepler step, yielding a change in position and velocity of $\Delta \mathbf{x}_\mathrm{DK}$ and $\Delta \mathbf{v}_\mathrm{DK}$, or a Kepler step followed by a negative drift, yielding $\Delta \mathbf{x}_\mathrm{KD}$ and $\Delta \mathbf{v}_\mathrm{KD}$.  In the process we need to solve Kepler's equation \ref{eqn:keplers_equation}.  The two pathways are shown in Figure \ref{fig:kepler_drift}.}

{\subsection{Drift then Kepler}\label{sec:drift_kepler_details}}

{As discussed in \S \ref{sec:drift_kepler}, the negative drift is taken first, yielding an intermediate position $\hat{\mathbf{x}}_0 = \mathbf{x}_0 - h \mathbf{v}_0.$  The resulting change in position and velocity over the time step is given by equation \ref{eqn:dxv_drift_kepler},
\begin{eqnarray}
\href{https://github.com/ericagol/NbodyGradient.jl/blob/0afa7a5fd387b307e3704ded05ec01838b01478c/src/integrator/ahl21/ahl21.jl\#L1022-L1025}{\Delta \mathbf{x}_{\mathrm{DK}}} =& \hat{\mathbf{x}}-\mathbf{x}_0 &= (\hat{f}-1) \mathbf{x}_0 + (\hat{g}-h\hat{f}) \mathbf{v}_0,\cr
\href{https://github.com/ericagol/NbodyGradient.jl/blob/0afa7a5fd387b307e3704ded05ec01838b01478c/src/integrator/ahl21/ahl21.jl\#L1026-L1028}{\Delta \mathbf{v}_{\mathrm{DK}}} =& \hat{\mathbf{v}}-\mathbf{v}_0 &=\dot{\hat{f}} \mathbf{x}_0 + (\dot{\hat{g}}-h\dot{\hat{f}}-1) \mathbf{v}_0,
\end{eqnarray}
where, again, $\hat{f}$, $\hat{g}$, $\dot{\hat{f}}$, and $\dot{\hat{g}}$ are all computed in terms
of $(\hat{\mathbf{x}}_0,\mathbf{v}_0,k,h)$.  This means that the scalar functions these depend on also need to be computed in terms of $\hat{\mathbf{x}}_0$, i.e., 
$\hat{r}_0 = \vert\hat{\mathbf{x}}_0\vert,$
$\hat{\beta} = 2k/\hat{r}_0-v_0^2,$
$\hat{\eta}_0 = \hat{\mathbf{x}}_0 \cdot \mathbf{v}_0,$ and
$\hat{G}_i = G_i (\hat{\beta},\hat{\gamma}),$
where $\hat{\gamma}$ can be computed with Newton's method from equation (\ref{eqn:keplers_equation}) evaluated using $\hat{r}_0$, $\hat{\eta}_0$,
and $\hat{G}_i$, and $\hat{r}$ can be computed from (\ref{eqn:r_of_s})
in the same manner.}

{The Gauss function terms in equation (\ref{eqn:dxv_drift_kepler}) are given as
\begin{eqnarray}\label{eqn:dxv_drift_kepler_redux}
\href{https://github.com/ericagol/NbodyGradient.jl/blob/0afa7a5fd387b307e3704ded05ec01838b01478c/src/integrator/ahl21/ahl21.jl\#L994}{\hat{f}-1} &=& -\frac{k}{\hat{r}_0} \hat{G}_2,\cr
\href{https://github.com/ericagol/NbodyGradient.jl/blob/0afa7a5fd387b307e3704ded05ec01838b01478c/src/integrator/ahl21/ahl21.jl\#L998}{\hat{g}-h\hat{f}} &=& k\left(\frac{h}{\hat{r}_0} \hat{G}_2 - \hat{G}_3\right),\cr
\href{https://github.com/ericagol/NbodyGradient.jl/blob/0afa7a5fd387b307e3704ded05ec01838b01478c/src/integrator/ahl21/ahl21.jl\#L991}{\dot{\hat{f}}} &=& -\frac{k}{\hat{r} \hat{r}_0} \hat{G}_1,\cr
\href{https://github.com/ericagol/NbodyGradient.jl/blob/0afa7a5fd387b307e3704ded05ec01838b01478c/src/integrator/ahl21/ahl21.jl\#L1014}{\dot{\hat{g}}-h \dot{\hat{f}}-1} &=& \frac{k}{\hat{r}}\left(\frac{h}{\hat{r}_0}\hat{G}_1 - \hat{G}_2\right),
\end{eqnarray}
where the 1's have been cancelled analytically for more accurate expressions at small time steps.}

{\subsection{Kepler then Drift}} 

{In the case of a Kepler step followed by a negative drift, the Kepler step is computed in terms of the initial coordinates, $\mathbf{x}_0$ and $\mathbf{v}_0$, yielding intermediate coordinates $(\check{\mathbf{x}},\check{\mathbf{v}})$, and then the negative drift is applied resulting in 
$\mathbf{x} = \check{\mathbf{x}}-h\check{\mathbf{v}}$ (Figure \ref{fig:kepler_drift}).  This yields the change in position and velocity of:
\begin{eqnarray}\label{eqn:dxv_kepler_drift_redux}
\href{https://github.com/ericagol/NbodyGradient.jl/blob/0afa7a5fd387b307e3704ded05ec01838b01478c/src/integrator/ahl21/ahl21.jl\#L1022-L1025}{\Delta \mathbf{x}_{\mathrm{KD}}} =& \check{\mathbf{x}}-h\check{\mathbf{v}}-\mathbf{x}_0 &= (f-h\dot f-1) \mathbf{x}_0 + (g-h\dot g) \mathbf{v}_0,\cr
\href{https://github.com/ericagol/NbodyGradient.jl/blob/0afa7a5fd387b307e3704ded05ec01838b01478c/src/integrator/ahl21/ahl21.jl\#L1026-L1028}{\Delta \mathbf{v}_{\mathrm{KD}}} =& \check{\mathbf{v}}-\mathbf{v}_0 &= \dot f \mathbf{x}_0 + (\dot g-1) \mathbf{v}_0,
\end{eqnarray}
where  $f$, $g$, $\dot f$, and $\dot g$ are all computed in terms
of $(\mathbf{x}_0,\mathbf{v}_0,k,h)$, and the ``KD" indicates that the Kepler step precedes the negative drift, Kepler-Drift.}

{These functions can also be expressed in terms of the Gauss functions as
\begin{eqnarray}
{f-h\dot f-1} &=& \frac{k}{r}\left(G_2 -\frac{k}{r_0}\left(G_2^2-G_1G_3\right)\right),\cr
g-h\dot g &=& \frac{k}{r} \left(r_0\left(G_1G_2-G_0G_3\right)+\eta_0\left(G_2^2-G_1G_3\right)\right),\cr
\href{https://github.com/ericagol/NbodyGradient.jl/blob/0afa7a5fd387b307e3704ded05ec01838b01478c/src/integrator/ahl21/ahl21.jl\#L991}{\dot f} &=&-\frac{k}{r r_0} G_1,\cr
\href{https://github.com/ericagol/NbodyGradient.jl/blob/0afa7a5fd387b307e3704ded05ec01838b01478c/src/integrator/ahl21/ahl21.jl\#L1017}{\dot g - 1} &=& - \frac{k}{r}G_2,
\end{eqnarray}
where we have used equations (\ref{eqn:keplers_equation}) and (\ref{eqn:r_of_s}) to transform
these equations.  Note that in this case each of these functions depends on $(\mathbf{x}_0,\mathbf{v}_0,k,h)$ as the Kepler step is applied \emph{before} the drift.}

{Unfortunately these equations can lead to numerical instability for small values of
$\gamma = \sqrt{\vert\beta\vert}s$.\footnote{Note that in the elliptic Kepler's 
equation, $\gamma$ is equal to the change in eccentric anomaly over the time step.}    
The offending terms involve difference of products of $G_i$ functions:  $G_1G_2-G_0G_3$ and 
$G_2^2-G_1G_3$.  These terms have a Taylor series expansion in which the leading 
order terms in $s$ cancel;  this is also true for the function $G_3$.}

{So, we define two new functions, 
\begin{eqnarray}\label{eqn:H1_H2}
H_1 &=& G_2^2-G_1 G_3,\cr
H_2 &=& G_1 G_2 - G_0 G_3,
\end{eqnarray} 
given in Table \ref{tab:G_functions}.  In terms of these functions we have
\begin{eqnarray}
\href{https://github.com/ericagol/NbodyGradient.jl/blob/0afa7a5fd387b307e3704ded05ec01838b01478c/src/integrator/ahl21/ahl21.jl\#L1004}{f-h\dot f-1} &=& \frac{k}{r}\left(G_2 -\frac{k}{r_0}H_1\right),\cr
\href{https://github.com/ericagol/NbodyGradient.jl/blob/0afa7a5fd387b307e3704ded05ec01838b01478c/src/integrator/ahl21/ahl21.jl\#L1007}{g-h\dot g} 
&=& \frac{k}{r} \left(r_0H_2 + \eta_0 H_1\right).
\end{eqnarray}
For large values of $\gamma$ we evaluate these with the special function definitions, summarized in 
Table \ref{tab:G_functions}, while for small values of $\gamma$, we evaluate 
$G_3, H_1$ and $H_2$ in terms of the following Taylor series:
\begin{eqnarray}
\href{https://github.com/ericagol/NbodyGradient.jl/blob/0afa7a5fd387b307e3704ded05ec01838b01478c/src/utils.jl\#L180-L209}{H_1(\gamma,\beta)} &=& \frac{2\gamma^4}{\beta^2} \sum_{n=0}^\infty \frac{ (\varepsilon \gamma ^2)^{n} (n+1)}{(2n+4)!},\label{eqn:H1}\\
\href{https://github.com/ericagol/NbodyGradient.jl/blob/0afa7a5fd387b307e3704ded05ec01838b01478c/src/utils.jl\#L225-L254}{H_2(\gamma,\beta)} &=& \frac{2 \gamma^3}{\vert \beta\vert^{3/2}} \sum_{n=0}^\infty \frac{ (\varepsilon\gamma^2)^{n} (n+1)}{(2n+3)!},\label{eqn:H2}\\
\href{https://github.com/ericagol/NbodyGradient.jl/blob/0afa7a5fd387b307e3704ded05ec01838b01478c/src/utils.jl\#L137-L165}{G_3(\gamma,\beta)} &=& \frac{\gamma^3}{\vert \beta\vert^{3/2}} \sum_{n=0}^\infty \frac{(\varepsilon\gamma^2)^{n}}{(2n+3)!}.\label{eqn:G3}
\end{eqnarray}
where $\varepsilon=-1$ for $\beta >0$ (elliptic) and $\varepsilon=1$ for $\beta<0$
(hyperbolic) cases.  Note that in evaluating these series expansions we compute each term recursively, and terminate the series expansion when the function matches one of the two prior partial sums (indicating that the series is converged to machine precision).}

{The fact that these functions have leading terms $\propto \gamma^3$ and $\propto \gamma^4$
is due to cancellation of lower order terms in the trigonometric representation.
This cancellation can lead to round-off errors for small values of $\gamma$, which
are commonly encountered when there are a wide range of orbital timescales in
a system.  We find that higher precision is obtained by evaluating the series 
expressions for $\gamma < 1/2$ out to $\approx 6$ terms in double-precision, 
while for $\gamma > 1/2$, higher precision is obtained from the full trigonometric
expression.}

{The other $G_i$ functions ($G_0, G_1$ and $G_2$) we 
evaluate with the stable trigonometric and hyperbolic function transforms discussed in \citet{Wisdom2015}, also given in Table \ref{tab:G_functions}.  With the combination of the drift and Kepler steps, it turns out that we no longer 
need the drift of the center-of-mass coordinates in each Kepler step as these cancel 
exactly.}

{\section{Derivation of derivatives of Kepler+Drift steps} \label{sec:appendix2}}

{In this appendix we give more detail on the derivation of the derivatives of the combined Kepler and drift steps, as well as formulae for the derivatives of the scalar quantities in equations \ref{eqn:dxDK}, \ref{eqn:dvDK}, \ref{eqn:dxKD}, and \ref{eqn:dvKD}.}

{\subsection{Differential of intermediate quantities}}

{The differentials of $r_0, \beta, \eta_0$ are given by
\begin{eqnarray}
\delta r_0 &=& \frac{\mathbf{x}_0 \cdot \delta \mathbf{x}_0  }{r_0},\\
\delta \beta &=& \frac{\delta k}{r_0} - \frac{k \delta \mathbf{x}_0 \cdot \mathbf{x}_0}{r_0^3}
- 2 \mathbf{v}_0 \cdot \delta \mathbf{v}_0,\\
\delta \eta_0 &=& \mathbf{v}_0 \cdot \delta \mathbf{x}_0 + \mathbf{x}_0 \cdot \delta \mathbf{v}_0.
\end{eqnarray}
Note that these quantities are constant over a time step, and so there is
no dependence upon $\delta h$.}

{Taking the differential of the Universal Kepler equation, 
\ref{eqn:keplers_equation}, we find
\begin{eqnarray}\label{eqn:gamma_D_diff}
     \vert \beta \vert^{-1/2} r \delta \gamma &=&
     \delta h + \delta k\left[\frac{D}{r_0}-G_3\right]\cr
     &-& \delta \mathbf{x}_0 \cdot \left[\left(k D + G_1 r_0^2\right)\frac{\mathbf{x}_0}{r_0^3} + G_2 \mathbf{v}_0\right] \cr &-&
     \delta \mathbf{v}_0 \cdot \left[D \mathbf{v}_0 + G_2 \mathbf{x}_0\right],\\
     \href{https://github.com/ericagol/NbodyGradient.jl/blob/0afa7a5fd387b307e3704ded05ec01838b01478c/src/integrator/ahl21/ahl21.jl\#L1062}{D} &=& \beta^{-1} \left[h + \eta_0 G_2 + 2 k G_3\right].
\end{eqnarray}
Note that the $\delta \mathbf{v}_0\cdot \mathbf{x}_0$ and $\delta \mathbf{x}_0\cdot \mathbf{v}_0$
have the same derivative terms;  this is due to both of these terms deriving from $\delta \eta_0$.}

{Taking the differential of the radial equation, (\ref{eqn:r_of_s}), we find
\begin{eqnarray}
\delta r &=& \frac{\delta \gamma}{\vert \beta \vert^{1/2}}(\eta_0 G_0 + \zeta G_1) + \frac{\delta k}{\beta r_0} \left(r_0-r-k G_2\right) \cr
&+& \delta \mathbf{x}_0 \cdot \left[\left(kr+k^2 G_2-\zeta r_0 G_0\right)\frac{\mathbf{x}_0}{\beta r_0^3}+G_1 \mathbf{v}_0\right]\cr
&+& \delta \mathbf{v}_0 \cdot \left[\left(\eta_0 G_1 + 2 k G_2\right)\frac{\mathbf{v}_0}{\beta} + G_1 \mathbf{x}_0\right],
\end{eqnarray}
where we define 
\begin{equation}\label{eqn:zeta}
\href{https://github.com/ericagol/NbodyGradient.jl/blob/0afa7a5fd387b307e3704ded05ec01838b01478c/src/integrator/ahl21/ahl21.j\#L900}{\zeta} = k - r_0 \beta.
\end{equation}}
{The functions $G_i(\beta,\gamma)$ have derivatives in terms of $\beta$ and $\gamma$, which 
we can combine with the foregoing differentials for these quantities to obtain the derivatives
with respect to the basis $(\mathbf{x}_0,\mathbf{v}_0,k,h)$.
These intermediate derivatives of $G_i$ (in both the elliptic and hyperbolic cases) are
given by
\begin{eqnarray} \label{eqn:dGidgamma}
\frac{\partial G_{i+1}}{\partial \gamma} &=& \frac{G_i}{\vert \beta \vert^{1/2}} \ \ (i \ge 0),\\
\frac{\partial G_0}{\partial \gamma} &=& - \frac{\beta G_1}{\vert \beta \vert^{1/2}},
\end{eqnarray}
and
\begin{eqnarray}\label{eqn:dGidbeta}
\frac{\partial G_n}{\partial \beta} &=& -\frac{n}{2\beta} G_n.
\end{eqnarray}}

{With these intermediate derivatives in hand, we can compute the full differentials
of the Gauss functions in the combined drift and Kepler terms.  Since these formulae
differ in the two cases, we consider each separately in turn in the following two
subsections.}

{\subsection{Differential of drift-then-Kepler step}} \label{app:drift_then_kepler}

{The differential of the scalar quantities $\delta(\hat{f}-1)$, $\delta(\hat{g}-h\hat{f}),$
$\delta(\dot{\hat{f}})$, and $\delta (\dot{\hat{g}}-h\dot{\hat{f}} -1)$ should also be scalars, and can be expressed in terms similar to the $\delta \gamma$ and
$\delta r$ terms given above.  Note, however, that as the Kepler step takes place \emph{after} the negative drift, all of these functions are to be computed in terms of $\hat{\mathbf{x}}_0$ substituted for $\mathbf{x}_0$, and so we need to add an extra step in the derivation to find the differentials in terms of $\mathbf{x}_0$ in lieu of $\hat{\mathbf{x}}_0$.}

{The differential of these functions in
terms of intermediate scalar quantities is given by}
{\begin{eqnarray}
    \frac{\delta(\hat{f}-1)}{\hat{f}-1} &=& \frac{\delta k}{k} - \frac{\delta \hat{r}_0}{\hat{r}_0} + \frac{\hat{G}_1}{ \hat{G}_2} \frac{\delta \hat{\gamma}}{\vert \hat{\beta}\vert^{1/2}} - \frac{\delta \hat{\beta}}{\hat{\beta}},
    \end{eqnarray}}
    {\begin{eqnarray}
    \delta(\hat{g}-h\hat{f}) &=& \frac{\delta k}{k}(\hat{g}-h\hat{f}) + k\Bigg[\left( \frac{\delta h}{\hat{r}_0}-\frac{h\delta \hat{r}_0}{{\hat{r}_0}^2}\right)\hat{G}_2 \cr
    &+&
    \frac{\delta \hat{\gamma}}{\vert \hat{\beta}\vert^{\tfrac{1}{2}}}\left(\frac{h}{\hat{r}_0}\hat{G}_1-\hat{G}_2\right)-\frac{\delta \hat{\beta}}{\hat{\beta}}\left( \frac{h}{\hat{r}_0}-\tfrac{3}{2}\hat{G}_3\right)\Bigg],\\
    \frac{\delta \dot{\hat{f}}}{\dot f} &=&  \frac{\delta k}{k} - \frac{\delta \hat{r}_0}{\hat{r}_0} -\frac{\delta \hat{r}}{\hat{r}} + \frac{\hat{G}_0}{\hat{G}_1} \frac{\delta \hat{\gamma}}{\vert \hat{\beta}\vert^{1/2}} - \frac{1}{2}\frac{\delta \hat{\beta}}{\hat{\beta}},\\
    \delta(\dot{\hat{g}}-h\dot{\hat{f}}-1) &=& \left[\frac{\delta k}{\hat{r}_0 \hat{r}}-\frac{k\delta \hat{r}_0}{{\hat{r}_0}^2 \hat{r}}-\frac{k\delta \hat{r}}{\hat{r}_0 {\hat{r}}^2}\right]\left(h \hat{G}_1 - \hat{r}_0 \hat{G}_2\right) \cr
    &+& \frac{k}{\hat{r}_0 \hat{r}}\left[\delta h \hat{G}_1  -\delta \hat{r}_0 \hat{G}_2 -\frac{1}{2}\frac{\delta \hat{\beta}}{\hat{\beta}}\left(h \hat{G}_1 -2\hat{r}_0 \hat{G}_2\right)\right.\cr
    &+&\left. \left(h \hat{G}_0-r_0\hat{G}_1\right)
    \frac{\delta \hat{\gamma}}{{\vert \hat{\beta} \vert^{1/2}}}\right].
\end{eqnarray}}
{In this equation we have used the `\,$\hat{}$\,' symbol to indicate that each of these quantities is a function of $\hat{\mathbf{x}}_0$ rather
than $\mathbf{x}_0$ (with the exception of $h$ and $k$).  Into
these differentials we can substitute our expressions for $\delta \hat{\beta}$, $\delta \hat{r}$, $\delta \hat{r}_0$, $\delta \hat{\eta}_0$, and $\delta \hat{\gamma}$ given above, keeping in mind that these need to be computed in terms of
$\delta \hat{\mathbf{x}}_0$ and $\hat{\mathbf{x}}_0$ substituted for 
$\delta \mathbf{x}_0$ and $\mathbf{x}_0$.  Then, we need to replace the differential
$\delta \hat{\mathbf{x}}_0$ by $\delta \mathbf{x}_0 - h \delta \mathbf{v}_0 - \mathbf{v}_0 \delta h$.  The $\mathbf{v}_0 \delta h$
term leads to dot products of $\mathbf{v}_0 \cdot \hat{\mathbf{x}}_0 = \hat{\eta}_0$ and $\mathbf{v}_0 \cdot \mathbf{v}_0 = \frac{2k}{\hat{r}_0} - \hat{\beta}$.  The algebraic computation of these operations was aided by Mathematica \citep{Mathematica}, yielding the following results:}
{\begin{eqnarray}
\href{https://github.com/ericagol/NbodyGradient.jl/blob/0afa7a5fd387b307e3704ded05ec01838b01478c/src/integrator/ahl21/ahl21.jl\#L1112-L1117}{\frac{\delta(\hat{f}-1)}{\hat{f}-1}} &=& \delta k \mathcal{A}_k +
 \delta h \mathcal{A}_h +\delta \mathbf{x}_0 \cdot \mathbf{x}_0 \mathcal{A}_{xx}
   \cr  &+&  \delta (\mathbf{x}_0 \cdot \mathbf{v}_0) \mathcal{A}_{xv} 
   + \delta \mathbf{v}_0 \cdot \mathbf{v}_0 \mathcal{A}_{vv},\cr 
   \href{https://github.com/ericagol/NbodyGradient.jl/blob/0afa7a5fd387b307e3704ded05ec01838b01478c/src/integrator/ahl21/ahl21.jl\#L1081}{\mathcal{A}_k} &=& \frac{1}{k}-\frac{2}{\beta  r_0}+\frac{c_1 G_1}{r r_0G_2}, \cr 
   \href{https://github.com/ericagol/NbodyGradient.jl/blob/0afa7a5fd387b307e3704ded05ec01838b01478c/src/integrator/ahl21/ahl21.jl\#L1080}{\mathcal{A}_h} &=& \frac{G_1}{r} \left(\frac{2 k}{r_0}-\beta
   +\frac{1}{G_2}\right)-c_{24} \eta_0, \cr
   \href{https://github.com/ericagol/NbodyGradient.jl/blob/0afa7a5fd387b307e3704ded05ec01838b01478c/src/integrator/ahl21/ahl21.jl\#L1075}{\mathcal{A}_{xx}} &=& c_{24},\cr
   \href{https://github.com/ericagol/NbodyGradient.jl/blob/0afa7a5fd387b307e3704ded05ec01838b01478c/src/integrator/ahl21/ahl21.jl\#L1076}{\mathcal{A}_{xv}} &=& - \left[c_{24} h+\frac{G_1}{r}\right],\cr
   \href{https://github.com/ericagol/NbodyGradient.jl/blob/0afa7a5fd387b307e3704ded05ec01838b01478c/src/integrator/ahl21/ahl21.jl\#L1079}{\mathcal{A}_{vv}} &=& \frac{1}{r}\left[\frac{k H_6}{\beta G_2} -r_0G_2 + h\left(2G_1+h r c_{24}\right)\right] ,\label{eqn:dlnfm1}  
   \end{eqnarray}}
{\begin{eqnarray}
   \href{https://github.com/ericagol/NbodyGradient.jl/blob/0afa7a5fd387b307e3704ded05ec01838b01478c/src/integrator/ahl21/ahl21.jl\#L1112-L1117}{\delta(\hat{g}-h\hat{f})} &=&\delta k \mathcal{B}_k +
 \delta h \mathcal{B}_h +\delta \mathbf{x}_0 \cdot \mathbf{x}_0 \mathcal{B}_{xx}
   \cr  &+&  \delta (\mathbf{x}_0 \cdot \mathbf{v}_0) \mathcal{B}_{xv} 
   + \delta \mathbf{v}_0 \cdot \mathbf{v}_0 \mathcal{B}_{vv},\cr 
   \href{https://github.com/ericagol/NbodyGradient.jl/blob/0afa7a5fd387b307e3704ded05ec01838b01478c/src/integrator/ahl21/ahl21.jl\#L1103}{\mathcal{B}_k} &=& -\frac{c_9 k}{\beta  r_0^2}+\frac{c_{1} c_{13} k}{r r_0^2}+\frac{G_2
   h-G_3 r_0}{r_0},\cr
   \href{https://github.com/ericagol/NbodyGradient.jl/blob/0afa7a5fd387b307e3704ded05ec01838b01478c/src/integrator/ahl21/ahl21.jl\#L1102}{\mathcal{B}_h} &=&
   \frac{kG_2}{r_0}+\frac{kc_{13}}{r r_0}+\frac{kG_2c_{13}}{rr_0}\left(\frac{2k}{r_0}-\beta\right)-\eta_0 c_{10},\cr
   \href{https://github.com/ericagol/NbodyGradient.jl/blob/0afa7a5fd387b307e3704ded05ec01838b01478c/src/integrator/ahl21/ahl21.jl\#L1093}{\mathcal{B}_{xx}} &=& c_{10},\cr
   \href{https://github.com/ericagol/NbodyGradient.jl/blob/0afa7a5fd387b307e3704ded05ec01838b01478c/src/integrator/ahl21/ahl21.jl\#L1094}{\mathcal{B}_{xv}} &=& -\left[c_{10}
   h+\frac{c_{13} G_2 k}{r r_0}\right],\cr
   \href{https://github.com/ericagol/NbodyGradient.jl/blob/0afa7a5fd387b307e3704ded05ec01838b01478c/src/integrator/ahl21/ahl21.jl\#L1100-L1101}{\mathcal{B}_{vv}} &=& \frac{2h k G_2 c_{13}}{r r_0}+h^2 c_{10}\cr 
  &+&\frac{k}{\beta r r_0}\left[r_0^2H_8-\beta h r_0G_2^2 + (h k+\eta_0 r_0)H_6\right], 
   \end{eqnarray}}
  {\begin{eqnarray}
   \href{https://github.com/ericagol/NbodyGradient.jl/blob/0afa7a5fd387b307e3704ded05ec01838b01478c/src/integrator/ahl21/ahl21.jl\#L1166-L1171}{\frac{\delta \dot{\hat{f}}}{\dot{\hat{f}}}} &=& \delta k \mathcal{C}_k +
 \delta h \mathcal{C}_h +\delta \mathbf{x}_0 \cdot \mathbf{x}_0 \mathcal{C}_{xx}
   \cr  
   &+&  \delta (\mathbf{x}_0 \cdot \mathbf{v}_0) \mathcal{C}_{xv} 
   + \delta \mathbf{v}_0 \cdot \mathbf{v}_0 \mathcal{C}_{vv},\cr 
   \href{https://github.com/ericagol/NbodyGradient.jl/blob/0afa7a5fd387b307e3704ded05ec01838b01478c/src/integrator/ahl21/ahl21.jl\#L1142}{\mathcal{C}_k} &=& \frac{1}{k}-\frac{1}{\beta r_0}-\frac{c_{17}}{\beta r r_0}-\frac{c_1(G_1c_2-G_0r)}{r^2r_0G_1},\cr 
   \href{https://github.com/ericagol/NbodyGradient.jl/blob/0afa7a5fd387b307e3704ded05ec01838b01478c/src/integrator/ahl21/ahl21.jl\#L1147}{\mathcal{C}_h}&=&c_{22}
   \left(\frac{2 k}{r_0}-\beta \right)-\frac{c_{2}}{r^2}+c_{21} \eta_0
   +\frac{G_0}{G_1 r},\cr 
   \href{https://github.com/ericagol/NbodyGradient.jl/blob/0afa7a5fd387b307e3704ded05ec01838b01478c/src/integrator/ahl21/ahl21.jl\#L1135}{\mathcal{C}_{xx}} &=& c_{21},\cr
    \href{https://github.com/ericagol/NbodyGradient.jl/blob/0afa7a5fd387b307e3704ded05ec01838b01478c/src/integrator/ahl21/ahl21.jl\#L1136}{\mathcal{C}_{xv}} &=& c_{22}-c_{21} h,\cr
   \href{https://github.com/ericagol/NbodyGradient.jl/blob/0afa7a5fd387b307e3704ded05ec01838b01478c/src/integrator/ahl21/ahl21.jl\#L1141}{\mathcal{C}_{vv}} &=&c_{34}- 2 h c_{22} +h^2 c_{21},
   \end{eqnarray}}
   {\begin{eqnarray}
   \href{https://github.com/ericagol/NbodyGradient.jl/blob/0afa7a5fd387b307e3704ded05ec01838b01478c/src/integrator/ahl21/ahl21.jl\#L1166-L1171}{\delta(\dot{\hat{g}} - h \dot{\hat{f}} -1)}&=& \delta k \mathcal{D}_k +
 \delta h \mathcal{D}_h +\delta \mathbf{x}_0 \cdot \mathbf{x}_0 \mathcal{D}_{xx}
   \cr  
   &+&  \delta (\mathbf{x}_0 \cdot \mathbf{v}_0) \mathcal{D}_{xv} 
   + \delta \mathbf{v}_0 \cdot \mathbf{v}_0 \mathcal{D}_{vv},\cr 
   \href{https://github.com/ericagol/NbodyGradient.jl/blob/0afa7a5fd387b307e3704ded05ec01838b01478c/src/integrator/ahl21/ahl21.jl\#L1156}{\mathcal{D}_k} &=& \frac{1}{rr_0}\Bigg[-\frac{k(c_{13}-G_2r_0)}{\beta r_0}+c_{13}\cr
   &-&\frac{kc_{13}c_{17}}{\beta rr_0}+\frac{kc_1c_{12}}{rr_0}-\frac{kc_1c_2c_{13}}{r^2r_0}\Bigg],\cr 
   \href{https://github.com/ericagol/NbodyGradient.jl/blob/0afa7a5fd387b307e3704ded05ec01838b01478c/src/integrator/ahl21/ahl21.jl\#L1161}{\mathcal{D}_h} &=&\frac{kG_1}{rr_0}+\frac{kc_{12}}{r^2r_0}-\frac{kc_{13}c_{2}}{r^3r_0}\cr&-&c_{26}
   \left(\frac{2 k}{r_0}-\beta \right)-c_{25} \eta_0, \cr 
   \href{https://github.com/ericagol/NbodyGradient.jl/blob/0afa7a5fd387b307e3704ded05ec01838b01478c/src/integrator/ahl21/ahl21.jl\#L1148}{\mathcal{D}_{xx}} &=& c_{25} ,\cr
   \href{https://github.com/ericagol/NbodyGradient.jl/blob/0afa7a5fd387b307e3704ded05ec01838b01478c/src/integrator/ahl21/ahl21.jl\#L1149}{\mathcal{D}_{xv}} &=&c_{26}-c_{25} h,\cr
   \href{https://github.com/ericagol/NbodyGradient.jl/blob/0afa7a5fd387b307e3704ded05ec01838b01478c/src/integrator/ahl21/ahl21.jl\#L1155}{\mathcal{D}_{vv}} &=& 
    c_{33}+c_{25} h^2-2 c_{26} h,
\end{eqnarray}}
{with auxiliary quantities defined as} 
{\begin{eqnarray}
  \href{https://github.com/ericagol/NbodyGradient.jl/blob/0afa7a5fd387b307e3704ded05ec01838b01478c/src/integrator/ahl21/ahl21.jl\#L1063}{c_1} &=& D-r_0G_3,\cr
  \href{https://github.com/ericagol/NbodyGradient.jl/blob/0afa7a5fd387b307e3704ded05ec01838b01478c/src/integrator/ahl21/ahl21.jl\#L1064}{c_2} &=& \eta_0 G_0+G_1\zeta,\cr
  \href{https://github.com/ericagol/NbodyGradient.jl/blob/0afa7a5fd387b307e3704ded05ec01838b01478c/src/integrator/ahl21/ahl21.jl\#L1065}{c_3}  &=& D k+G_1 r_0^2,\cr
  \href{https://github.com/ericagol/NbodyGradient.jl/blob/0afa7a5fd387b307e3704ded05ec01838b01478c/src/integrator/ahl21/ahl21.jl\#L1066}{c_4} &=& \eta_0 G_1+2G_0r_0,\cr
  \href{https://github.com/ericagol/NbodyGradient.jl/blob/0afa7a5fd387b307e3704ded05ec01838b01478c/src/integrator/ahl21/ahl21.jl\#L1122}{c_5} &=& \frac{r_0-kG_2}{r G_1},\cr
  \href{https://github.com/ericagol/NbodyGradient.jl/blob/0afa7a5fd387b307e3704ded05ec01838b01478c/src/integrator/ahl21/ahl21.jl\#L1123}{c_6} &=& \frac{r_0G_0-kG_2}{\beta},\cr
  \href{https://github.com/ericagol/NbodyGradient.jl/blob/0afa7a5fd387b307e3704ded05ec01838b01478c/src/integrator/ahl21/ahl21.jl\#L1124}{c_7} &=& G_2\left(\frac{1}{G_1}+\frac{c_2}{r}\right),\cr
  \href{https://github.com/ericagol/NbodyGradient.jl/blob/0afa7a5fd387b307e3704ded05ec01838b01478c/src/integrator/ahl21/ahl21.jl\#L1125}{c_8} &=& r_0^{-3}\left(k c_6+r r_0+c_3 c_5\right),\cr
  \href{https://github.com/ericagol/NbodyGradient.jl/blob/0afa7a5fd387b307e3704ded05ec01838b01478c/src/integrator/ahl21/ahl21.jl\#L1070}{c_9} &=& 2  h G_2-3  r_0G_3,\cr
  \href{https://github.com/ericagol/NbodyGradient.jl/blob/0afa7a5fd387b307e3704ded05ec01838b01478c/src/integrator/ahl21/ahl21.jl\#L1071}{c_{10}} &=& \frac{k}{r_0^4}\left[-G_2 r_0 h +\frac{k c_9}{\beta}-\frac{c_3 c_{13}}{r}\right],\cr
  \href{https://github.com/ericagol/NbodyGradient.jl/blob/0afa7a5fd387b307e3704ded05ec01838b01478c/src/integrator/ahl21/ahl21.jl\#L1126}{c_{12}} &=& G_0 h-G_1 r_0,\cr
  \href{https://github.com/ericagol/NbodyGradient.jl/blob/0afa7a5fd387b307e3704ded05ec01838b01478c/src/integrator/ahl21/ahl21.jl\#L1067}{c_{13}} &=& G_1 h-G_2 r_0,\cr
  \href{https://github.com/ericagol/NbodyGradient.jl/blob/0afa7a5fd387b307e3704ded05ec01838b01478c/src/integrator/ahl21/ahl21.jl\#L1127}{c_{17}} &=& r_0-r-kG_2,\cr
  \href{https://github.com/ericagol/NbodyGradient.jl/blob/0afa7a5fd387b307e3704ded05ec01838b01478c/src/integrator/ahl21/ahl21.jl\#L1128}{c_{18}} &=& \eta_0 G_1+2kG_2,\cr
  \href{https://github.com/ericagol/NbodyGradient.jl/blob/0afa7a5fd387b307e3704ded05ec01838b01478c/src/integrator/ahl21/ahl21.jl\#L1129}{c_{20}} &=& k(kG_2+r)-G_0r_0\zeta,\cr
  \href{https://github.com/ericagol/NbodyGradient.jl/blob/0afa7a5fd387b307e3704ded05ec01838b01478c/src/integrator/ahl21/ahl21.jl\#L1130}{c_{21}} &=& \frac{(kG_2-r_0)(\beta c_3-k G_1r)}{\beta r^2r_0^3G_1}+\frac{\eta_0 G_1}{rr_0^2}-\frac{2}{r_0^2},\cr
  \href{https://github.com/ericagol/NbodyGradient.jl/blob/0afa7a5fd387b307e3704ded05ec01838b01478c/src/integrator/ahl21/ahl21.jl\#L1131}{c_{22}} &=& \frac{1}{r}\left[-G_1-\frac{G_0G_2}{G_1}+\frac{G_2c_2}{r}\right],\cr
  \href{https://github.com/ericagol/NbodyGradient.jl/blob/0afa7a5fd387b307e3704ded05ec01838b01478c/src/integrator/ahl21/ahl21.jl\#L1072}{c_{24}} &=& \frac{1}{r_0^3}\left[r_0\left(\frac{2k}{\beta r_0}-1\right)-\frac{G_1c_3}{rG_2}\right],\cr
  \href{https://github.com/ericagol/NbodyGradient.jl/blob/0afa7a5fd387b307e3704ded05ec01838b01478c/src/integrator/ahl21/ahl21.jl\#L1132-L1133}{c_{25}} &=& \frac{k}{rr_0^2}\Bigg[-G_2+\frac{k(c_{13}-G_2r_0)}{\beta r_0^2}-\frac{c_{13}}{r_0}-\frac{c_{12}c_3}{rr_0^2}\cr
  &+&\frac{c_{13}c_2c_3}{r^2r_0^2}-\frac{c_{13}(k(kG_2+r)-G_0r_0\zeta)}{\beta r r_0^2}\Bigg],\cr
  \href{https://github.com/ericagol/NbodyGradient.jl/blob/0afa7a5fd387b307e3704ded05ec01838b01478c/src/integrator/ahl21/ahl21.jl\#L1134}{c_{26}} &=& \frac{k}{r^2r_0}\left[-G_2 c_{12}-G_1 c_{13}+\frac{G_2 c_{13}c_2}{r}\right],\cr
  \href{https://github.com/ericagol/NbodyGradient.jl/blob/0afa7a5fd387b307e3704ded05ec01838b01478c/src/integrator/ahl21/ahl21.jl\#L1154}{c_{33}} &=& \frac{D k^2}{r^3 r_0} (h G_2 - r_0 G_3) \cr &+& 
  \frac{k}{\beta r^2 r_0}\Big[-\eta_0 k G_1G_2^2-G_1G_2G_3 k^2\cr &-&r_0\eta_0\beta G_1G_2^2 -r_0 kG_1H_2 - \beta G_2^2G_0r_0^2\Big],\cr
  \href{https://github.com/ericagol/NbodyGradient.jl/blob/0afa7a5fd387b307e3704ded05ec01838b01478c/src/integrator/ahl21/ahl21.jl\#L1139-L1140}{c_{34}} &=& \frac{1}{\beta r^2}\Big[-\beta(\eta_0G_2+r_0G_1)^2-\eta_0 kH_8-H_6k^2\cr&+&(G_2^2-3G_1G_3)\beta k r_0\Big]+\frac{\eta_0 G_2^2}{r G_1} + \frac{k H_8}{\beta r G_1},
  \label{eqn:c1_to_c34}
\end{eqnarray}}
{where we have dropped the $\hat{}$ superscript on the right hand side of equations \ref{eqn:dlnfm1}-\ref{eqn:c1_to_c34} for legibility, but note that all \emph{scalar} quantities in these equations
which are a function of $\mathbf{x}_0$ \emph{must} be evaluated as a function of $\hat{\mathbf{x}}_0$ in lieu of $\mathbf{x}_0$.  In these equations, we have taken care to analytically cancel terms to leading
order in $\gamma$ by defining the following functions}
{\begin{eqnarray}\label{eqn:H3_to_H8}
\href{https://github.com/ericagol/NbodyGradient.jl/blob/0afa7a5fd387b307e3704ded05ec01838b01478c/src/utils.jl\#L256-L269}{H_3} &=& G_1G_2-3G_3,\cr
\href{https://github.com/ericagol/NbodyGradient.jl/blob/0afa7a5fd387b307e3704ded05ec01838b01478c/src/integrator/ahl21/ahl21.jl\#L1083}{H_4} &=& -\beta H_1,\cr
\href{https://github.com/ericagol/NbodyGradient.jl/blob/0afa7a5fd387b307e3704ded05ec01838b01478c/src/utils.jl\#L305-L318}{H_5} &=& G_1G_2-(2+G_0)G_3,\cr
\href{https://github.com/ericagol/NbodyGradient.jl/blob/0afa7a5fd387b307e3704ded05ec01838b01478c/src/utils.jl\#L351-L362}{H_6} &=& 2G_2^2-3G_1G_3,\cr
\href{https://github.com/ericagol/NbodyGradient.jl/blob/0afa7a5fd387b307e3704ded05ec01838b01478c/src/integrator/ahl21/ahl21.jl\#L1261}{H_7} &=& G_1G_2 - 2 G_0G_1G_2 + 3 G_0^2 G_3= \beta G_1 G_2^2-G_0 H_8,\cr
\href{https://github.com/ericagol/NbodyGradient.jl/blob/0afa7a5fd387b307e3704ded05ec01838b01478c/src/integrator/ahl21/ahl21.jl\#L1099}{H_8} &=& G_1 G_2 - 3 G_0 G_3 = -2H_3+3H_5.
\end{eqnarray}}
{These functions are tabulated in Table \ref{tab:G_functions}.  As with $G_3, H_1$ and $H_2$, for small values of $\gamma$ we evaluate these with a series expansion:
\begin{eqnarray}
\href{https://github.com/ericagol/NbodyGradient.jl/blob/0afa7a5fd387b307e3704ded05ec01838b01478c/src/utils.jl\#L271-L303}{H_3(\gamma,\beta)} &=& -\frac{4\gamma^5}{\beta\vert\beta\vert^{1/2}} \sum_{n=0}^\infty \frac{ (\varepsilon \gamma ^2)^{n} (4^{n+1}-1)}{(2n+5)!},\label{eqn:H3}\\
\href{https://github.com/ericagol/NbodyGradient.jl/blob/0afa7a5fd387b307e3704ded05ec01838b01478c/src/utils.jl\#L320-L349}{H_5(\gamma,\beta)} &=& -\frac{2 \gamma^5}{\beta\vert \beta\vert^{1/2}} \sum_{n=0}^\infty \frac{ (\varepsilon\gamma^2)^{n} (n+1)}{(2n+5)!},\label{eqn:H5}\\
\href{https://github.com/ericagol/NbodyGradient.jl/blob/0afa7a5fd387b307e3704ded05ec01838b01478c/src/utils.jl\#L364-L396}{H_6(\gamma,\beta)} &=& \frac{2\gamma^6}{ \beta^{2}} \sum_{n=0}^\infty \frac{(\varepsilon\gamma^2)^{n}\left(4^{n+2}-3n-7\right)}{(2n+6)!},\label{eqn:H6}
\end{eqnarray}
where $\varepsilon=-1$ for $\beta >0$ (elliptic) and $\varepsilon=1$ for $\beta<0$
(hyperbolic) cases.  The coefficients of these series are computed recursively for efficiency. Note that we found expressions for $H_4$, $H_7$, and $H_8$ in terms of the other $G$ and $H$ functions.}

{\subsection{Differential of Kepler then drift step}}

{The Kepler step followed by a negative drift is slightly simpler
as the Gauss functions can be expressed in terms of $\mathbf{x}_0$
rather than $\hat{\mathbf{x}}_0$.  The differential of the scalar
functions are given by}
{\begin{eqnarray}
\href{https://github.com/ericagol/NbodyGradient.jl/blob/0afa7a5fd387b307e3704ded05ec01838b01478c/src/integrator/ahl21/ahl21.jl\#L1270-L1277}{\delta (f\!-h\!\dot f\!-\!1)} &=& \delta k \mathcal{E}_k +
 \delta h \mathcal{E}_h +\delta \mathbf{x}_0 \cdot \mathbf{x}_0 \mathcal{E}_{xx}
   \cr  &+&  \delta (\mathbf{x}_0 \cdot \mathbf{v}_0) \mathcal{E}_{xv} 
   + \delta \mathbf{v}_0 \cdot \mathbf{v}_0 \mathcal{E}_{vv},\cr
   \href{https://github.com/ericagol/NbodyGradient.jl/blob/0afa7a5fd387b307e3704ded05ec01838b01478c/src/integrator/ahl21/ahl21.jl\#L1242-L1243}{\mathcal{E}_k} &=& \frac{k}{r r_0} \Big[-\frac{c_{14} c_{17}}{\beta  r r_0}-\frac{2 G_2 }{\beta }+\frac{4
   H_1 k}{\beta  r_0}\cr&-&\frac{c_1 c_{14} c_{2} }{r^2 r_0}+\frac{c_1 
   (G_1 r_0-H_2 k)}{r r_0}+\frac{c_{14}}{k}-H_1 \Big],\cr 
   \href{https://github.com/ericagol/NbodyGradient.jl/blob/0afa7a5fd387b307e3704ded05ec01838b01478c/src/integrator/ahl21/ahl21.jl\#L1241}{\mathcal{E}_h} &=& \frac{k}{r^2} \left[-\frac{c_{14} c_{2}}{r r_0}+G_1 -\frac{H_2
   k}{r_0}\right],\cr 
   \href{https://github.com/ericagol/NbodyGradient.jl/blob/0afa7a5fd387b307e3704ded05ec01838b01478c/src/integrator/ahl21/ahl21.jl\#L1234}{\mathcal{E}_{xx}} &=& \frac{k}{\beta r^3 r_0^4}\Big[\beta  c_3 (c_{14} c_{2}+c_{23}
   r)+H_1 k r^2 r_0 \left(\beta -\frac{2 k}{r_0}\right)
   \cr&+&c_{14} r \Big(G_0 r_0 \zeta +k
   (r-G_2 k)\Big)\Big],\cr
      \href{https://github.com/ericagol/NbodyGradient.jl/blob/0afa7a5fd387b307e3704ded05ec01838b01478c/src/integrator/ahl21/ahl21.jl\#L1235}{\mathcal{E}_{xv}} &=&  \frac{k }{r^2 r_0}\Big[\frac{c_{14} c_{2} G_2}{r}+k   (G_1 H_1+G_2 H_2)-2 G_1 G_2 r_0\Big],\cr  
      \href{https://github.com/ericagol/NbodyGradient.jl/blob/0afa7a5fd387b307e3704ded05ec01838b01478c/src/integrator/ahl21/ahl21.jl\#L1238-L1240}{\mathcal{E}_{vv}} &=& \frac{k}{\beta r^2 r_0}
    \Big[2\eta_0 k (G_2G_3-G_1H_1)\cr&+&(3G_3H_2-4H_1G_2)k^2 
    +\beta G_2 r_0(3H_1 k-G_2 r_0) \cr &+&\frac{c_{14}}{r}\Big(-\beta (G_2\eta+G_1r_0)^2+\eta_0 k (2G_0G_3-H_2)\cr
     &+&k \beta r_0(H_1-2G_1G_3)-H_6 k^2\Big)\Big],
\end{eqnarray}}
{
\begin{eqnarray}
\href{https://github.com/ericagol/NbodyGradient.jl/blob/0afa7a5fd387b307e3704ded05ec01838b01478c/src/integrator/ahl21/ahl21.jl\#L1271-L1277}{\delta (g-h\dot g)} &=& \delta k \mathcal{F}_k +
 \delta h \mathcal{F}_h
   +  \delta \mathbf{x}_0 \cdot \mathbf{x}_0 \mathcal{F}_{xx} \cr &+&  \delta (\mathbf{x}_0 \cdot \mathbf{v}_0) \mathcal{F}_{xv}
   + \delta \mathbf{v}_0 \cdot \mathbf{v}_0 \mathcal{F}_{vv},\cr
  \href{https://github.com/ericagol/NbodyGradient.jl/blob/0afa7a5fd387b307e3704ded05ec01838b01478c/src/integrator/ahl21/ahl21.jl\#L1258}{\mathcal{F}_k} &=& \frac{k}{r} \left[\frac{c_{15}}{k}-\frac{c_{15} c_{17} }{\beta  r r_0}-\frac{c_{19} }{\beta  r_0}-\frac{c_1 c_{15} c_{2} }{r^2
   r_0}+\frac{c_1 c_{16} }{r r_0}\right],\cr
  \href{https://github.com/ericagol/NbodyGradient.jl/blob/0afa7a5fd387b307e3704ded05ec01838b01478c/src/integrator/ahl21/ahl21.jl\#L1267}{\mathcal{F}_h} &=&\frac{k}{r^3} \left[c_{16} r-c_{15} c_{2}\right],\cr
    \href{https://github.com/ericagol/NbodyGradient.jl/blob/0afa7a5fd387b307e3704ded05ec01838b01478c/src/integrator/ahl21/ahl21.jl\#L1250}{\mathcal{F}_{xx}} &=&\frac{k}{r r_0}\Bigg[-\frac{c_{15} (k (G_2 k+r)-G_0 r_0 \zeta
   )}{\beta  r r_0^2}+\frac{c_{19} k}{\beta  r_0^2}\cr&+&\frac{c_{15} c_{2} c_3}{r^2 r_0^2}-\frac{c_{16} c_3}{r
   r_0^2}+H_2\Bigg],\cr
    \href{https://github.com/ericagol/NbodyGradient.jl/blob/0afa7a5fd387b307e3704ded05ec01838b01478c/src/integrator/ahl21/ahl21.jl\#L1251}{\mathcal{F}_{xv}} &=& \frac{k}{r^2} \left[\frac{c_{15} c_{2} G_2}{r}-c_{15} G_1-c_{16} G_2+H_1r\right],\cr
    \href{https://github.com/ericagol/NbodyGradient.jl/blob/0afa7a5fd387b307e3704ded05ec01838b01478c/src/integrator/ahl21/ahl21.jl\#L1254-L1257}{\mathcal{F}_{vv}} &=&\frac{k}{\beta r^2}    
    \Big[2\eta_0^2(G_1H_1-G_2G_3)+\eta_0 k(4G_2H_1-3H_2G_3)\cr&+&r_0\eta_0(4G_0H_1-2G_1G_3) \cr 
    &+& 3r_0 k\big((G_1+\beta G_3) H_1-G_3 G_2\big)\cr
    &+&\big(G_0 H_8-\beta G_1 (G_2^2+G_1 G_3)\big) r_0^2 \cr 
    &-& \frac{c_{15}}{r} \Big(\beta G_2^2 \eta_0^2+\eta_0 k H_8+H_6 k^2+\beta G_1^2 r_0^2\cr
    &+&\big(2\eta_0 G_1 G_2-k (G_2^2-3G_1 G_3)\big) \beta r_0\Big)\Big],
   \end{eqnarray}}
{\begin{eqnarray}
   \href{https://github.com/ericagol/NbodyGradient.jl/blob/0afa7a5fd387b307e3704ded05ec01838b01478c/src/integrator/ahl21/ahl21.jl\#L1312-L1319}{\frac{\delta \dot f}{\dot f}} &=&\delta k \mathcal{G}_k +
 \delta h \mathcal{G}_h
   +  \delta \mathbf{x}_0 \cdot \mathbf{x}_0 \mathcal{G}_{xx} \cr &+&  \delta (\mathbf{x}_0 \cdot \mathbf{v}_0) \mathcal{G}_{xv}
   + \delta \mathbf{v}_0 \cdot \mathbf{v}_0 \mathcal{G}_{vv},\cr
   \href{https://github.com/ericagol/NbodyGradient.jl/blob/0afa7a5fd387b307e3704ded05ec01838b01478c/src/integrator/ahl21/ahl21.jl\#L1292}{\mathcal{G}_k} &=& \frac{1}{k}+\frac{c_1(r_0-kG_2)}{r_0r^2G_1}-\frac{r+c_{17}}{\beta r r0},\cr 
     \href{https://github.com/ericagol/NbodyGradient.jl/blob/0afa7a5fd387b307e3704ded05ec01838b01478c/src/integrator/ahl21/ahl21.jl\#L1297}{\mathcal{G}_h} &=&  \frac{r_0-kG_2}{r^2G_1},\cr 
   \href{https://github.com/ericagol/NbodyGradient.jl/blob/0afa7a5fd387b307e3704ded05ec01838b01478c/src/integrator/ahl21/ahl21.jl\#L1286}{\mathcal{G}_{xx}} &=& \frac{ 1}{r_0^2} \Bigg[-\frac{k (G_2 k-r_0)}{\beta  r r_0}+\frac{c_{2} c_3}{r^2
   r_0}-\frac{c_3 G_0}{G_1 r r_0}\cr&+&\frac{\eta_0  G_1}{r}-2\Bigg],\cr
  \href{https://github.com/ericagol/NbodyGradient.jl/blob/0afa7a5fd387b307e3704ded05ec01838b01478c/src/integrator/ahl21/ahl21.jl\#L1287}{\mathcal{G}_{xv}} &=&  -\frac{1}{r}\left(\frac{G_0G_2}{G_1}+\frac{r_0G_1+\eta_0 G_2}{r}\right),\cr 
   \href{https://github.com/ericagol/NbodyGradient.jl/blob/0afa7a5fd387b307e3704ded05ec01838b01478c/src/integrator/ahl21/ahl21.jl\#L1290-L1291}{\mathcal{G}_{vv}} &=& \frac{1}{\beta r^2}\Bigg[\frac{(\eta_0 G_2^2 \beta +k H_8) (r_0 G_0 + k G_2)}{G_1}- H_6 k^2 \cr 
    &+& \beta r_0\left(k(H_1-2G_1G_3)-2\eta_0 G_1G_2\right) -
       \beta G_1^2 r_0^2 \Bigg], 
   \end{eqnarray}}
   {\begin{eqnarray} \label{eqn:dgdotm1}
\href{https://github.com/ericagol/NbodyGradient.jl/blob/0afa7a5fd387b307e3704ded05ec01838b01478c/src/integrator/ahl21/ahl21.jl\#L1314-L1319}{\delta(\dot g-1)} &=&\delta k \mathcal{H}_k +
 \delta h \mathcal{H}_h
   +  \delta \mathbf{x}_0 \cdot \mathbf{x}_0 \mathcal{H}_{xx} \cr &+&  \delta (\mathbf{x}_0 \cdot \mathbf{v}_0) \mathcal{H}_{xv}
   + \delta \mathbf{v}_0 \cdot \mathbf{v}_0 \mathcal{H}_{vv},\cr
   \href{https://github.com/ericagol/NbodyGradient.jl/blob/0afa7a5fd387b307e3704ded05ec01838b01478c/src/integrator/ahl21/ahl21.jl\#L1304}{\mathcal{H}_k} &=& \frac{1}{r r_0} \Bigg[\frac{G_2 k (-G_2 k+r+r_0)}{\beta  r}+\frac{c_1 c_{2} G_2
   k}{r^2}\cr 
   &-&\frac{c_1 G_1 k}{r}-G_2 r_0\Bigg],\cr
     \href{https://github.com/ericagol/NbodyGradient.jl/blob/0afa7a5fd387b307e3704ded05ec01838b01478c/src/integrator/ahl21/ahl21.jl\#L1309}{\mathcal{H}_h} &=&\frac{k}{r^3} \left[c_{2} G_2-G_1 r\right], \cr
   \href{https://github.com/ericagol/NbodyGradient.jl/blob/0afa7a5fd387b307e3704ded05ec01838b01478c/src/integrator/ahl21/ahl21.jl\#L1298}{\mathcal{H}_{xx}} &=&
   \frac{k}{r^2 r_0^3}\Bigg[\frac{G_2  (k (G_2 k-r)-G_0 r_0 \zeta )}{\beta}\cr 
   &+&\frac{c_3  (\eta_0  G_2+G_1 r_0)}{r}\Bigg],\cr
   \href{https://github.com/ericagol/NbodyGradient.jl/blob/0afa7a5fd387b307e3704ded05ec01838b01478c/src/integrator/ahl21/ahl21.jl\#L1299}{\mathcal{H}_{xv}} &=& \frac{G_2 k}{r^3} \left[rG_1+r_0G_1+\eta_0 G_2\right],\cr 
 \href{https://github.com/ericagol/NbodyGradient.jl/blob/0afa7a5fd387b307e3704ded05ec01838b01478c/src/integrator/ahl21/ahl21.jl\#L1302-L1303}{\mathcal{H}_{vv}} &=& \frac{k}{\beta r^3}\Big[\eta_0^2\beta G_2^3-\eta_0 k G_2 H_3+3r_0\eta_0\beta G_1G_2^2 \cr 
    &+& r_0 k (3\beta G_1 G_2 G_3-G_0 H_6) \cr
    &+&\beta G_2\big(G_0G_2+G_1^2\big) r_0^2\Big], 
\end{eqnarray}}
{with additional auxiliary definitions,}
{\begin{eqnarray}\label{eqn:c14_c23}
  \href{https://github.com/ericagol/NbodyGradient.jl/blob/0afa7a5fd387b307e3704ded05ec01838b01478c/src/integrator/ahl21/ahl21.jl\#L1221}{c_{14}} &=& r_0G_2-kH_1,\cr
  \href{https://github.com/ericagol/NbodyGradient.jl/blob/0afa7a5fd387b307e3704ded05ec01838b01478c/src/integrator/ahl21/ahl21.jl\#L1222}{c_{15}} &=& \eta_0 H_1+H_2r_0,\cr
  \href{https://github.com/ericagol/NbodyGradient.jl/blob/0afa7a5fd387b307e3704ded05ec01838b01478c/src/integrator/ahl21/ahl21.jl\#L1223}{c_{16}} &=& \eta_0H_2+G_1\gamma r_0\vert \beta \vert^{-1/2},\cr
  \href{https://github.com/ericagol/NbodyGradient.jl/blob/0afa7a5fd387b307e3704ded05ec01838b01478c/src/integrator/ahl21/ahl21.jl\#L1224}{c_{17}} &=& r_0-r-kG_2,\cr
  \href{https://github.com/ericagol/NbodyGradient.jl/blob/0afa7a5fd387b307e3704ded05ec01838b01478c/src/integrator/ahl21/ahl21.jl\#L1226}{c_{19}} &=& 4\eta_0 H_1+3H_2r_0,\cr
  \href{https://github.com/ericagol/NbodyGradient.jl/blob/0afa7a5fd387b307e3704ded05ec01838b01478c/src/integrator/ahl21/ahl21.jl\#L1227}{c_{23}} &=& k H_2-r_0G_1.
\end{eqnarray}
Note that in this case, as the Kepler step takes place first, all of
the scalar quantitites in this equation are defined in terms of $\mathbf{x}_0$.}

{\section{Kepler + Drift Mass derivative expressions} \label{sec:appendix3}}

{Here are the functions $J_1 - J_4$ used in computing the mass derivatives in equations \ref{eqn:dx_DK_mass} and \ref{eqn:dv_DK_mass}:}
{\begin{eqnarray}\label{eqn:J1_J4}
\href{https://github.com/ericagol/NbodyGradient.jl/blob/0afa7a5fd387b307e3704ded05ec01838b01478c/src/integrator/ahl21/ahl21.jl\#L1091}{J_1} &=& \hat r_0 \hat H_4 + k \hat H_6,\cr
\href{https://github.com/ericagol/NbodyGradient.jl/blob/0afa7a5fd387b307e3704ded05ec01838b01478c/src/integrator/ahl21/ahl21.jl\#L1106}{J_2} &=& \hat H_6 \hat G_3 k^2+\hat \eta_0 \hat r_0 (\hat H_6+ \hat G_2 \hat H_4)+\hat r_0^2 \hat G_0 \hat H_5+k\hat \eta_0 \hat G_2 \hat H_6\cr 
&+&(\hat G_1 \hat H_6+\hat G_3 \hat H_4)k\hat r_0,\cr
\href{https://github.com/ericagol/NbodyGradient.jl/blob/0afa7a5fd387b307e3704ded05ec01838b01478c/src/integrator/ahl21/ahl21.jl\#L1146}{J_3} &=& -(\hat G_2 k - \hat r_0)(\hat \beta \hat r_0(\hat G_3-\hat G_1\hat G_2)-\hat \beta\hat \eta_0\hat G_2^2+k\hat H_3),\cr
\href{https://github.com/ericagol/NbodyGradient.jl/blob/0afa7a5fd387b307e3704ded05ec01838b01478c/src/integrator/ahl21/ahl21.jl\#L1159}{J_4} &=& k\Big(-\hat \beta \hat \eta_0^2 \hat G_2^4+\hat \eta_0\hat G_2(\hat G_1\hat  G_2^2+\hat G_1^2\hat G_3-5\hat G_2\hat G_3)k\cr 
&+&\hat G_2\hat G_3 \hat H_3k^2+
        2\hat \eta_0\hat r_0\hat \beta\hat G_2^2(\hat G_3-\hat G_1\hat G_2)  \cr 
        &+&  (4\hat G_3-\hat G_0\hat G_3-\hat G_1\hat G_2)(\hat G_3-\hat G_1\hat G_2)\hat r_0 k\cr
        &-&\hat \beta(\hat G_3 -\hat G_1\hat G_2)^2\hat r_0^2
\Big),
\end{eqnarray}}
{and the functions $J_5 - J_8$ used in computing the mass derivatives in equations   \ref{eqn:dx_KD_mass} and \ref{eqn:dv_KD_mass}:}
{\begin{eqnarray}\label{eqn:J5_J8}
\href{https://github.com/ericagol/NbodyGradient.jl/blob/0afa7a5fd387b307e3704ded05ec01838b01478c/src/integrator/ahl21/ahl21.jl\#L1246-L1248}{J_5} &=&  r\Big(2\eta_0 k (G_1H_1-G_3G_2)+(4G_2H_1-3G_3H_2)k^2\cr 
&-&\eta_0 r_0 \beta G_1 H_1 + (G_3 H_2-4G_2H_1)\beta k r_0 + G_2 H_1\beta^2 r_0^2\Big) \cr
&-& c_{14}\Big(-\eta_0^2\beta G_2^2 - k\eta_0 H_8  - \eta_0 r_0 \beta (G_1G_2 + G_0G_3)\cr 
&+& 2(H_1 - G_1G_3)\beta k r_0 - k^2 H_6 - (G_2 - \beta G_1G_3)\beta r_0^2\Big),\cr
\href{https://github.com/ericagol/NbodyGradient.jl/blob/0afa7a5fd387b307e3704ded05ec01838b01478c/src/integrator/ahl21/ahl21.jl\#L1262-L1265}{J_6} &=&   r_0r\Big(2\eta_0^2(G_3G_2-G_1H_1) + \eta_0 k (3G_3H_2 - 4G_2H_1)\cr 
&+& r_0\eta_0\big(\beta G_3(G_1G_2 + G_0G_3) - 2G_0H_6\big) \cr 
&+& \big(-H_6(G_1 + \beta G_3) + G_2(2G_3 - H_2)\big)r_0 k \cr 
&+&\big(H_7 - \beta^2G_1G_3^2\big)r_0^2\Big)\cr 
&-& r_0 c_{15}\Big(-\beta\eta_0^2G_2^2 + \eta_0 k (-H_2 + 2G_0G_3) - H_6 k^2\cr  
&-& r_0\eta_0\beta\big(H_2 + 2G_0G_3\big) + 2\beta\big(2H_1 - G_2^2\big)r_0 k \cr
&+& \beta\big(\beta G_1G_3 - G_2\big)r_0^2\Big),\cr
\href{https://github.com/ericagol/NbodyGradient.jl/blob/0afa7a5fd387b307e3704ded05ec01838b01478c/src/integrator/ahl21/ahl21.jl\#L1296}{J_7} &=&   (r_0-kG_2)(-\eta_0\beta G_2^2+H_3 k+(G_3-G_1G_2)\beta r_0),\cr
\href{https://github.com/ericagol/NbodyGradient.jl/blob/0afa7a5fd387b307e3704ded05ec01838b01478c/src/integrator/ahl21/ahl21.jl\#L1306-L1308}{J_8} &=&  \beta G_1(G_3 - G_1G_2)r_0^2-\beta\eta_0^2G_2^3+\eta_0 kG_2 H_3\cr 
&+&\eta_0 r_0\beta G_2(G_3-2G_1G_2)+
       (H_6-\beta G_2^3)r_0 k.
\end{eqnarray}}

\section{Table of notation}

Table \ref{tab:symbols} lists the mathematical symbols used throughout this paper.

\onecolumn
\begin{center}
\renewcommand*{\arraystretch}{1.08}
\begin{longtable}{cll}
\caption{Symbols used in this paper} \label{tab:symbols} \\
\hline
\multicolumn{1}{c}{\textbf{Symbol}} &
\multicolumn{1}{c}{\textbf{Definition}} &
\multicolumn{1}{c}{\textbf{Reference}} \\
\hline
\endfirsthead
\multicolumn{3}{c}%
{{\bfseries \tablename\ \thetable{} --} continued from previous page} \\
\hline
\multicolumn{1}{c}{\textbf{Symbol}} &
\multicolumn{1}{c}{\textbf{Definition}} &
\multicolumn{1}{c}{\textbf{Reference}} \\
\hline
\endhead
\hline
\endfoot
\hline
\endlastfoot
$\mathbf{a}_i$ & Instantaneous acceleration of body $i$. & (\ref{eqn:correct_ai})\\
{$\mathbf{a}_{ij}$} & {Difference in acceleration between bodies $i$ and $j$.} & {\S\ref{sec:correction}}\\
$A$ & Set of pairs of bodies interacting via fast kicks. & {\S\ref{sec:kicks}, Alg. \ref{alg:AHL21_algorithm}}\\
$A^C$ & Set of pairs of bodies interacting drift + Keplerian (complement of $A$). & {\S\ref{sec:kicks}, Alg. \ref{alg:AHL21_algorithm}}\\
{$\mathcal{A}_{k}, \mathcal{A}_h, \mathcal{A}_{xx},..., \mathcal{H}_{xv},\mathcal{H}_{vv}$} & {Terms used in derivatives.} & {(\ref{eqn:dlnfm1})-(\ref{eqn:dgdotm1})}\\
{AU} &  {Astronomical unit.}  & {\S \ref{sec:code_units}}\\
$b_{\mathrm{sky},ijk}$ & The impact parameter at the $k$th time of transit of body $i$ in front of body $j$. & \S \ref{sec:transit_times}\\
{$c_1-c_{34}$} & {Quantities defined for computing derivatives of drift+Kepler.} & {(\ref{eqn:c1_to_c34}),(\ref{eqn:c14_c23})}\\
$C$ & Constant in correction term. & \S \ref{sec:correction}\\
$d$ & Distance to observer. & \S \ref{sec:cartesian_coordinates}\\
{$e$} & {Orbital eccentricity.} & {\S \ref{sec:julia}}\\ 
{$D$} & {Constant used in computing derivatives.} & {(\ref{eqn:gamma_D_diff})}\\
{$\Delta E,E_0$}&{Energy error and initial total energy.}&{\S\ref{sec:energy_angmom}}\\
$f,g,\dot f,\dot g$ & Gauss's Kepler propagation functions. & \S \ref{sec:universal_kepler}\\
{$\hat{f},\hat{g},\hat{\dot f}, \hat{\dot g}$} & {Gauss's functions for drift-then-kepler.} & {\S\ref{sec:drift_then_kepler}}\\
{$g_\mathrm{sky,ij}$} &{Sky plane velocity dotted w/ position between bodies $i$ and $j$.} & {(\ref{eqn:gsky})}\\
{$G$} & {Newton's constant.} & {(\ref{eqn:xij})}\\
{$G_i$} & {Functions used in universal Kepler solver.} & {Table \ref{tab:G_functions}}\\
{$\hat G_i$} & {Functions used in universal Kepler solver after initial drift.} & {\S \ref{sec:drift_kepler_details}}\\
$h$ & Symplectic integrator time step (days). & \S \ref{sec:symplectic_integrator}\\
{$H$} & {Hamiltonian.} & {(\ref{eqn:Hamiltonian_splitting})}\\
{$H_A$, $H_B$} & {Integrable Hamiltonians in symplectic splitting.} & {(\ref{eq:hamilt})}\\
{$H_1-H_8$} & {Combinations of $G_i$ functions used in combined Kepler-drift step and derivatives.} &{Table \ref{tab:G_functions}, (\ref{eqn:H1_H2}), (\ref{eqn:H3_to_H8})}\\
{$i,j$} & {Label for bodies $i$ and $j$.} & {(\ref{eqn:Hamiltonian_splitting})}\\
{$J_1-J_8$} & {Intermediate variables used in mass derivatives.} &{(\ref{eqn:J1_J4}),(\ref{eqn:J5_J8})}\\
{$\mathbf{J}_\mathrm{kep}$} & {Jacobian of Kepler/drift step between two bodies (7x7 matrix).} & {\S\ref{sec:kepler_step}}\\
{$\mathbf{J}_\mathrm{substep}$,$\Delta\mathbf{J}_\mathrm{substep}$} & {Jacobian of any substep is $\mathbf{J}_\mathrm{substep}=\mathbf{I}+\Delta\mathbf{J}_\mathrm{substep}$}. & {(\ref{eqn:substep_jacobian})}\\
{$\mathbf{J}_\mathrm{current}$,$\mathbf{J}_\mathrm{prior}$} & {Jacobians of the current and prior steps.} & {(\ref{eqn:substep_jacobian})}\\
{$\Delta\mathbf{J}_\mathrm{D}$} & Jacobian of drift step {is $\mathbf{I}+\Delta\mathbf{J}_\mathrm{D}$}. & \S \ref{sec:drift}\\
{$\mathbf{J}_\mathrm{AHL21}$} & Jacobian of single symplectic step. & \S \ref{sec:Jacobian}\\
{$\mathbf{J}_n$} & {Jacobian after $n$th step.} & {\S \ref{sec:jacobian_total}}\\
{$\Delta\mathbf{J}_\mathrm{4th}$} & Jacobian of fourth-order correction {$\mathbf{I}+\Delta\mathbf{J}_\mathrm{4th}$}. & \S \ref{sec:correction}\\
{$\Delta\mathbf{J}_\mathrm{DK,ij}$, $\Delta\mathbf{J}_\mathrm{KD,ij}$} & {Jacobian of Kepler+drift substep for bodies $i$ and $j$.}  & {\S \ref{sec:kepler_drift_jacobian}}\\
{$k$} & {Central force constant ($=G(m_i+m_j)$).} &  {(\ref{eqn:xij})}\\
{$K_{ij}$} & {Keplerian for pair of bodies $i$ and $j$.} & {(\ref{eqn:Hamiltonian_splitting})}\\
$m_i$ & Mass of $i$th body. & \S \ref{sec:notation}\\
{$m_{ij}$} & {Sum of masses of bodies $i$ and $j$.} & {\S \ref{sec:universal_kepler}}\\
{$\mathbf{m}$} & {Vector of masses.} & {\S \ref{sec:notation}}\\
{$M_\odot$} & {Solar mass.} & {\S \ref{sec:code_units}}\\
{$n$} & {Number of time steps elapsed.} & {\S\ref{sec:transit_times}}\\
$N$ & Number of bodies in system. & {\S \ref{sec:notation}}\\
$N_S$ & Number of time steps & {\S\ref{sec:energy_angmom},\S\ref{sec:transit_precision}}\\
$\mathbf{q}_i(t)$ & Coordinates of the $i$th body at time t & (\ref{eqn:coordinates_body})\\
$\mathbf{q}(t)$ & Position, velocity, mass vector of all bodies (system state at time $t$). & \S \ref{sec:cartesian_coordinates}\\
$\mathbf{q}_n$ & $\mathbf{q}(t)$ at $n$th time step. & \S \ref{sec:transit_times}\\
{$\mathbf{q}_\mathrm{current}$,$\mathbf{q}_\mathrm{prior}$, $\Delta \mathbf{q}$} & {System state at current and prior substep, and the difference:  $\Delta \mathbf{q}= \mathbf{q}_\mathrm{current}-\mathbf{q}_\mathrm{prior}$} & {\S\ref{sec:time_derivative}}\\
$r_0, r, r_{ij}$ & Initial/final separation between bodies $i$ and $j$ for Kepler solver. & \S \ref{sec:universal_kepler}\\
{$\hat r_0, \hat r$} & {Initial/final separation between bodies $i$ and $j$ for drift+Kepler solver.} & {\S \ref{sec:drift_kepler_details}}\\
$s$ & Independent variable in universal Kepler step. & \S \ref{sec:universal_kepler}\\
{$t$} & {Current simulation time.} & {\S \ref{sec:symplectic_integrator}, Alg.\ \ref{alg:DH17_algorithm},\ref{alg:AHL21_algorithm}}\\
$\Delta t_\mathrm{init},\Delta t$ & Initial and final fraction of a timestep to transit time. & \S \ref{sec:transit_times}\\
$t_0$ & Initial time of integration (days). & \S \ref{sec:symplectic_integrator}{, Alg.\ \ref{alg:DH17_algorithm},\ref{alg:AHL21_algorithm}}\\
{$t_n$} & {Time after $n$th step: $t_n = t_0 + n h$.} & {\S \ref{sec:transit_times}}\\
$t_{ijk}$ & The $k$th time of transit of body $i$ in front of body $j$. & \S \ref{sec:transit_times}\\
$t_\mathrm{max}$ & Duration of simulation (days). & \S \ref{sec:symplectic_integrator}{, Alg.\ \ref{alg:DH17_algorithm},\ref{alg:AHL21_algorithm}}\\
$\mathbf{T}_{ij}$ & Correction tensor. & (\ref{eqn:tij})\\
{$T$} & {Total kinetic energy.} & {(\ref{eqn:Hamiltonian_splitting})}\\
{$T_{ij}$} & {Kinetic energy of bodies $i$ and $j$.} & {(\ref{eqn:Hamiltonian_splitting})}\\
$\Delta \mathbf{v}_i$ & Pairwise velocity kick or correction of $i$th body (instead of Kepler step). & (\ref{eqn:kick}), (\ref{eqn:correction}),  {(\ref{eqn:correction_fastkick})}\\
$v_0,v$ & Initial/final relative speed in Kepler problem. & \S \ref{sec:universal_kepler}\\
$v_{\mathrm{sky},ijk}$ & The sky velocity at the $k$th time of transit of body $i$ in front of body $j$. & \S \ref{sec:transit_times}\\
$\mathbf{v}_i=(\dot x_i,\dot y_i,\dot z_i)$ & Cartesian velocities of $i$th body & {(\ref{eqn:drift}),}\S \ref{sec:cartesian_coordinates}\\
{$V$} & {Total potential energy.} & {(\ref{eqn:Hamiltonian_splitting})}\\
{$V_{ij}$} & {Potential energy of bodies $i$ and $j$.} & {(\ref{eqn:Hamiltonian_splitting})}\\
{$\mathbf{w}_{ij}$} & Intermediate quantity for kick derivative. & (\ref{eqn:kick_deriv})\\
$\mathbf{x}_i=(x_i,y_i,z_i)$ & Cartesian coordinates of $i$th body & {(\ref{eqn:drift})}, \S \ref{sec:cartesian_coordinates}\\
$\mathbf{x}_{ij}, \mathbf{v}_{ij}$ & Relative position and velocity of bodies $i$ and $j$ ($\mathbf{x},\mathbf{v}$ in \S \ref{sec:universal_kepler}). & (\ref{eqn:xij})\\
$\mathbf{x}_{0}, \mathbf{v}_{0}$ & Relative position and velocity of bodies $i$ and $j$ at start of universal Kepler step & \S \ref{sec:universal_kepler}\\
{$\hat{\mathbf{x}}_0$} & {Intermediate position in DK step.} & {(\ref{eqn:drift_first})}\\
{$\hat{\mathbf{x}},\hat{\mathbf{v}}$} & {Final position and velocity in DK step.} & {\S \ref{sec:drift_then_kepler}}\\
{$\Delta\mathbf{x}_{DK},\Delta\mathbf{v}_{DK}$} & {Change in position and velocity for DK step between bodies $i$ and $j$.} & {(\ref{eqn:dxv_drift_kepler}),(\ref{eqn:dxv_drift_kepler_redux})}\\
{$\check{\mathbf{x}},\check{\mathbf{v}}$} & {Intermediate position and velocity in KD step.} & {\S \ref{sec:kepler_then_drift}}\\
{$\Delta\mathbf{x}_{KD},\Delta\mathbf{v}_{KD}$} & {Change in position and velocity for KD step between bodies $i$ and $j$.} & {(\ref{eqn:dxv_kepler_drift}),(\ref{eqn:dxv_kepler_drift_redux})}\\
{$\Delta\mathbf{x}_{i,KD},\Delta\mathbf{v}_{i,KD}$} & {Change in position and velocity for KD step for body $i$.} & {(\ref{eqn:kepler_derivatives_ij})}\\
{$\Delta\mathbf{x}_{i,DK},\Delta\mathbf{v}_{i,DK}$} & {Change in position and velocity for DK step for body $i$.} & {(\ref{eqn:kepler_derivatives_ij})}\\
{$x_\mathrm{obs}$} & {Observer position.} & {\S \ref{sec:cartesian_coordinates}}\\
$\alpha_0, \eta_0$ & Quantities used in Kepler solver. & \S \ref{sec:universal_kepler}\\
{$\alpha$} & {Factor from DH17 algorithm(set to zero).} & {\S\ref{sec:symplectic_integrator}, \S \ref{sec:correction} }\\
$\beta$ & Dimensionless energy for Kepler step. & (\ref{eqn:beta})\\
{$\hat\beta$} & {Dimensionless energy for drift+Kepler step.} & {(\ref{sec:drift_kepler_details})}\\
$\gamma$ & Variable used in defining $G_i$ functions. {$=\vert\beta\vert^{1/2}s$} & Table \ref{tab:G_functions}\\
{$\hat\gamma$} & {Value of $\gamma$ computed after an initial drift.} & {\S \ref{sec:drift_kepler_details}}\\
{$\hat\eta_0$} & {Dot product of velocity and position after an initial drift.} & {\S \ref{sec:drift_kepler_details}}\\
{$\epsilon$} & {Small parameter in Hamiltonian splitting for general symplectic integrator.} & {(\ref{eq:hamilt})}\\
{$\epsilon_\mathrm{err}$} & {Numerical error.} & {\S \ref{sec:energy_angmom}}\\
{$\varepsilon_\mathrm{diff}$} & {Fractional change in parameters for finite difference.} & {(\ref{eqn:finite_diff})}\\
{$\varepsilon$} & {Sign of $\beta$.} & {(\ref{eqn:H1}-\ref{eqn:G3}),(\ref{eqn:H3}-\ref{eqn:H6})}\\
{$\zeta$} & {Intermediate variable.} & {(\ref{eqn:zeta})}\\
{$\varpi$} & {Longitude of periastron.} & {\S \ref{sec:julia}} 

\end{longtable}
\end{center}


\bsp    
\label{lastpage}
\end{document}